\documentclass[11pt]{article}

\usepackage{algorithmic} 
\usepackage{algorithm} 
\usepackage{amsfonts}
\usepackage{amssymb}
\usepackage{amsmath}
\usepackage{enumitem}
\usepackage{geometry}
\usepackage{graphicx}
\usepackage{setspace}
\usepackage{booktabs}%
\providecommand{\U}[1]{\protect\rule{.1in}{.1in}}
\newtheorem{theorem}{Theorem}

\newtheorem{result}[theorem]{Result}
\newtheorem{definition}{Definition}
\newtheorem{assumption}[definition]{Assumption}

\usepackage{xr}
\makeatletter
\newcommand*{\addFileDependency}[1]{
\typeout{(#1)}
\@addtofilelist{#1}
\IfFileExists{#1}{}{\typeout{No file #1.}}
}
\makeatother

\newcommand*{\myexternaldocument}[1]{%
\externaldocument{#1}%
\addFileDependency{#1.tex}%
\addFileDependency{#1.aux}%
}
\myexternaldocument{SPC_Nonparametric_Supp}

\usepackage{multirow}
\usepackage{natbib}
\usepackage{xcolor}
\usepackage{hyperref}
\usepackage{bbm}
\newcommand{\cond}{\, \big| \,}
\newcommand{\con}{ ; }
\newcommand{\EXP}{E}
\newcommand{\VAR}{Var}
\newcommand{\potY}[2]{Y_{#2}^{a=#1}}

\newcommand{\true}{*}
\newcommand{\ind}{\mathbbm{1}}
\newcommand{\AVER}{\mathbbm{P}}
\newcommand{\EMP}{\mathbbm{G}}

\DeclareMathOperator*{\argmin}{arg\,min}
\DeclareMathOperator*{\median}{median}

\newcommand{\OutReg}{{\text{OR}}}
\newcommand{\PS}{{\text{PS}}}
\newcommand{\DR}{{\text{DR}}}
\newcommand{\T}{^\intercal}
\newcommand{\sT}{^{*\intercal}}
\newcommand{\GMM}{\text{GMM}}
\newcommand{\bW}{W}
\newcommand{\bw}{w}
\newcommand{\HL}[1]{\hyperlink{(#1)}{(#1)}}
\newcommand{\HT}[1]{\hypertarget{(#1)}{(#1)}}
\newcommand{\InfFt}{\texttt{IF}}
\newcommand{\ETA}{^{(t)}}
\newcommand{\bO}{O}

\newcommand{\bX}{X}

\newcommand{\R}{\mathbbm{R}}
\newcommand{\LSS}{^{(-k)}}
\newcommand{\SSS}{^{(k)}}
\newcommand{\EXPk}{\EXP\LSS}
\newcommand{\VARk}{\text{Var}\LSS}

\newcommand{\EMPk}{\mathbbm{G}_{\mathcal{I}_k}}

\newcommand{\HH}{\mathcal{H}}

\usepackage{tikz}
\usetikzlibrary{arrows, arrows.meta, chains, positioning, quotes, shapes.geometric,
shapes,shapes.arrows,shapes.multipart,decorations.pathmorphing,positioning,shapes.swigs}

\tikzstyle{new edge style 0}=[-{Latex[length=3mm]}]

\newcommand{\indep}{\,\rotatebox[origin=c]{90}{$\models$}\,}
\newcommand{\nindep}{\not\!\!\!\indep}

\hypersetup{
colorlinks=true,
linkcolor=blue,
filecolor=blue,
urlcolor=blue,
citecolor=blue
}

\begin{document}

\setlength{\abovedisplayskip}{8pt}
\setlength{\belowdisplayskip}{8pt}
\setlength{\abovedisplayshortskip}{8pt}
\setlength{\belowdisplayshortskip}{8pt}

\title{\vspace*{-1cm}Single Proxy Control}
 \author{
  Chan Park$^{a}$, David B. Richardson$^{b}$, Eric J. Tchetgen Tchetgen$^{a}$
  \\[0.2cm]
  {\small $a$: Department of Statistics and Data Science, The Wharton School, University of Pennsylvania}\\
  {\small $b$: Department of Environmental \& Occupational Health, University of California-Irvine}\\
    }
 \date{}
  \maketitle
\begin{abstract}
Negative control variables are sometimes used in non-experimental studies to detect the presence of confounding by hidden factors. A negative control outcome (NCO) is an outcome that is influenced by unobserved confounders of the exposure effects on the outcome in view, but is not causally impacted by the exposure. \citet{TT2013_COCA} introduced the Control Outcome Calibration Approach (COCA) as a formal NCO counterfactual method to detect and correct for residual confounding bias. For identification, COCA treats the NCO as an error-prone proxy of the treatment-free counterfactual outcome of interest, and involves regressing the NCO on the treatment-free counterfactual, together with a rank-preserving structural model which assumes a constant individual-level causal effect. In this work, we establish nonparametric COCA identification for the average causal effect for the treated, without requiring rank-preservation, therefore accommodating unrestricted effect heterogeneity across units. This nonparametric identification result has important practical implications, as it provides single proxy confounding control, in contrast to recently proposed proximal causal inference, which relies for identification on a pair of confounding proxies. For COCA estimation we propose three separate strategies: (i) an extended propensity score approach, (ii) an outcome bridge function approach, and (iii) a doubly-robust approach. Finally, we illustrate the proposed methods in an application evaluating the causal impact of a Zika virus outbreak on birth rate in Brazil.

\end{abstract}
\noindent%

 \newpage

\section{Introduction}
Unmeasured confounding is a well-known threat to valid causal inference from observational data. An approach that is sometimes used in practice to assess residual confounding bias, is to check whether \textit{known null effects} can be recovered free of bias, by evaluating whether the exposure or treatment of interest is found to be associated with a so-called negative control outcome (NCO), upon adjusting for measured confounders \citep{Rosenbaum1989, Lipsitch2010, Shi2020}.  An observed variable is said to be a valid \textit{negative control outcome} or more broadly, an \textit{outcome confounding proxy}, to the extent that it is associated with hidden factors confounding the exposure-outcome relationship in view, although not directly impacted by the exposure. Therefore, an NCO which is empirically associated with the exposure, might suggest the presence of residual confounding. In the event such an association is present, a natural question is whether the negative control outcome can be used for bias correction. 

The most well-established NCO approach for debiasing observational causal effect estimates is the difference-in-differences approach (DiD) \citep{CardKrueger1994, DiD_Review2011, Caniglia2020}. In fact, DiD may be viewed as directly leveraging the pre-treatment outcome as an NCO since it cannot logically be causally impacted by the treatment. Identification then follows from an additive equi-confounding assumption that the unmeasured confounder association with the post-treatment outcome of interest matches that with the pre-treatment outcome on the additive scale \citep{Sofer2016}. The baseline outcome in DiD is thus implicitly assumed to be a valid NCO, and equi-confounding is equivalent to the so-called \textit{parallel trends assumption}, that the average trends in treatment-free potential outcomes for treatment and untreated units are parallel. In practice, equi-confounding or equivalently parallel trends may not be reasonable for a number of reasons, including if the outcome trend is also impacted by an unmeasured common cause with the treatment. Furthermore, additive equi-confounding may not be realistic as a broader debiasing method in non-DiD settings where the NCO is not necessarily a pre-treatment measurement of the outcome of interest, but is instead a post-treatment measurement of a different type of outcome (and might therefore have support on a different scale than the outcome of interest has). 

To address these potential limitations of additive equi-confounding, \citet{TT2013_COCA} introduced the Control Outcome Calibration Approach (COCA) as a simple yet formal counterfactual NCO approach to debias causal effect estimates in observational analyses. At its core, COCA essentially treats the NCO variable as a proxy measurement for the treatment-free potential outcome, which therefore is associated with the latter, and which becomes independent of the treatment assignment mechanism, upon conditioning on the treatment-free counterfactual\ outcome. As the treatment-free potential outcome can be viewed as an ultimate source of unmeasured confounding, this assumption formalizes the idea that, as a relevant proxy for the source of residual confounding, the NCO\ would be made irrelevant for the treatment assignment mechanism if one were to hypothetically condition on the underlying potential outcome.

For identification and inference for a continuous outcome, the original COCA approach of \citet{TT2013_COCA}  involves the correct specification of a regression model for the NCO, conditional on the treatment-free potential outcome and measured confounders, together with a rank-preserving structural model which effectively assumes a constant individual-level treatment effect. In this paper, we develop a nonparametric COCA identification framework for the average causal effect for the treated, which equally applies irrespective of the nature of the primary outcome, whether binary, continuous, or polytomous. Importantly, as we show, the proposed COCA\ identification framework completely obviates the need for rank preservation, therefore accommodating an arbitrary degree of effect heterogeneity across units. Relatedly, an alternative counterfactual approach named \textit{proximal causal inference} has recently developed in causal inference literature \citep{Miao2016_arxiv,Miao2018,TT2024_Proximal}, which leverages a pair of negative treatment and outcome control variables, or more broadly treatment and outcome confounding proxies, to nonparametrically identify treatment causal effects subject to residual confounding without invoking a rank-preservation assumption. Importantly, while proximal causal inference relies on two proxies for causal identification, in contrast, COCA is a single-proxy control approach which therefore may present practical advantages. For estimation and inference, we introduce three strategies to implement COCA which improve on prior methods: (i) an \textit{extended propensity score approach}, (ii) a so-called \textit{outcome calibration bridge function approach}, and (iii) \textit{a doubly robust approach} which carefully combines approaches (i) and (ii) and remains unbiased, provided that either approach is also unbiased, without necessarily knowing which method might not be unbiased. Finally, we illustrate the methods with an application evaluating the causal effect of a Zika outbreak on birth rate in Brazil, and we conclude with possible extensions to our methods and a brief discussion.

\section{Notation and Brief Review of COCA} \label{sec:Setup}

\begin{figure}[!htp]
\centering
\scalebox{1}{
\begin{tikzpicture}
\tikzset{line width=1.5pt, outer sep=0pt,
ell/.style={draw,fill=white, inner sep=2pt,
line width=1.5pt},
swig vsplit={gap=5pt,
inner line width right=0.5pt}};
\node[name=A, ell, ellipse] at (0,0){$A$};
\node[name=Y, ell, ellipse] at (4,0) {$Y$};
\node[name=X, ell, ellipse] at (2,1.5) {$X$};
\begin{scope}[>={Stealth[black]}, every edge/.style={draw=black,very thick}]
\path [dashed,<->] (A) edge[bend right=40] (Y);
\path [->] (A) edge (Y);    
\path [->] (X) edge (A);
\path [->] (X) edge (Y);
\end{scope}
\end{tikzpicture} }
\caption{{\small A Graphical Illustration of A Simple Causal Model}}
\label{fig-1}
\end{figure}
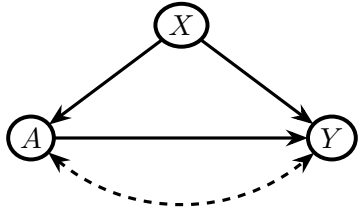

Consider an observational study where, as represented in Figure \ref{fig-1}, one has observed an outcome variable $Y$, a binary treatment $A$ whose causal effect on $Y$ is of interest, and measured pre-treatment covariates $X$. We are concerned that as displayed succinctly in the figure with the bow arc, the association between $A$ and $Y$ is confounded by hidden factors.

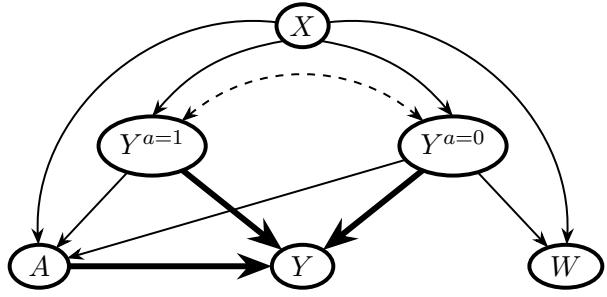
\begin{figure}
\centering 
\scalebox{1}{
\begin{tikzpicture}
\tikzset{line width=1.5pt, outer sep=0pt,
ell/.style={draw,fill=white, inner sep=2pt,
line width=1.5pt},
swig vsplit={gap=5pt,
inner line width right=0.5pt}};
\node[name=A, ell, ellipse] at (0.5,0*0.8){$\, A\, $};
\node[name=Y0, ell, ellipse] at (6,2*0.8) {$\potY{0}{}$}  ;
\node[name=X, ell, ellipse] at (4,4*0.8) {$X$}  ;
\node[name=Y1, ell, ellipse] at (2,2*0.8) {$\potY{1}{}$}  ;
\node[name=W, ell, ellipse] at (7.5,0*0.8) {$\, W\, $};
\node[name=Y, ell, ellipse] at (4,0*0.8) {$\, Y\, $};
\begin{scope}[>={Stealth[black]}, every edge/.style={draw=black}]
\path [dashed,<->] (Y1) edge[bend left=40, line width=0.75pt] (Y0);
\path [->] (Y0) edge[line width=0.75pt] (A);
\path [->] (Y1) edge[line width=0.75pt] (A);
\path [->] (Y0) edge[line width=0.75pt] (W); 
\path [->] (X) edge[bend left=50, line width=0.75pt] (W); 
\path [->] (X.south east) edge[bend left=20, line width=0.75pt] (Y0.north);
\path [->] (X.south west) edge[bend right=20, line width=0.75pt] (Y1.north);
\path [->] (X) edge[bend right=50, line width=0.75pt] (A);

\path [->] (Y1) edge[line width=2.25pt] (Y); 
\path [->] (Y0) edge[line width=2.25pt] (Y); 
\path [->] (A) edge[line width=2.25pt] (Y); 
\end{scope}
\end{tikzpicture} }
\caption{\small A Graphical Illustration of the Assumptions for Control Outcome Calibration Approach. Thick arrows depict the deterministic relationship between $Y$ and $(\potY{1}{},\potY{0}{},A)$, as established by the consistency assumption (Assumption  \ref{assumption:1:consistency}).}
\label{fig: Graphical COCA}
\end{figure}

Throughout, $Y^{a}$ denotes the potential outcome or counterfactual, had possibly contrary to fact, the exposure been set to $a$ by an external hypothetical intervention. Furthermore, throughout, we also make the consistency assumption:
\begin{assumption} \label{assumption:1:consistency}
$Y=Y^{A}$ almost surely.    
\end{assumption}
Hereafter, we aim to make inferences about the causal effect of treatment on the treated (ETT), denoted by $\psi^\true=E\big(  Y^{a=1}-Y^{a=0} \cond A=1\big)$. Under consistency, $\psi_1^* = E\big(  Y^{a=1} \cond A=1\big)$ is identified by $E\big(Y \cond A=1\big)$; to identify the counterfactual mean $\psi_0^* = E\big(  Y^{a=0} \cond A=1\big)$ requires additional assumptions. Standard methods often resort to the \textit{no unmeasured confounding assumption}, i.e., $Y^{a=0} \indep  A \cond X$, a strong assumption we do not make. Instead, we suppose that one has measured a valid NCO $W$, possibly multi-dimensional, which is known a priori to satisfy the following conditions:
\begin{assumption} \label{assumption:NC}
Condition (i): $W^{a}=W$ almost surely for $a=0,1$, where $W^{a}$ is the potential NCO under an external intervention that sets $A=a$; Condition (ii): $W\nindep  Y^{a=0} \cond X$; Condition (iii): $W\indep  A  \cond (Y^{a=0} , X)$.
\end{assumption}
Assumption \ref{assumption:NC}-(i) encodes the key assumption of a known
null causal effect of the treatment on the NCO\ in potential outcome notation. Assumption \ref{assumption:NC}-(ii) encodes that $W$ is  relevant for predicting the treatment-free potential outcome of interest.  Assumption \ref{assumption:NC}-(iii) states that $W$ is independent of $A$ conditional on treatment-free potential outcome and covariates. These conditions formally encode the assumption that $W$ is a valid proxy for the treatment-free potential outcome, a source of residual confounding bias; $W$ is only associated with the treatment mechanism to the extent that it is associated with the confounding mechanism captured by the potential outcome.\ We illustrate these NCO assumptions with the causal graph displayed in Figure \ref{fig: Graphical COCA}. The thick arrows in the graph indicate the deterministic relationship defining the observed outcome in terms of potential outcomes and treatment variables by the consistency assumption. The missing arrows on the graph formally encode the core conditional independence conditions implied by Assumption \ref{assumption:NC}.

\begin{figure}
\centering 
\scalebox{1}{
\begin{tikzpicture}
\tikzset{line width=1.5pt, outer sep=0pt,
ell/.style={draw,fill=white, inner sep=2pt,
line width=1.5pt},
swig vsplit={gap=5pt,
inner line width right=0.5pt}};
\node[name=A, ell, ellipse] at (-0.5,0*0.8){$\, A\, $};
\node[name=Y0, ell, ellipse] at (6.75,2*0.8) {$\potY{0}{}$}  ;
\node[name=U1, ell, ellipse] at (2.5,4*0.8) {$U_1$}  ;
\node[name=U0, ell, ellipse] at (5.5,4*0.8) {$U_0$}  ;
\node[name=Y1, ell, ellipse] at (1.25,2*0.8) {$\potY{1}{}$}  ;
\node[name=W, ell, ellipse] at (8.5,0*0.8) {$\, W\, $};
\node[name=Y, ell, ellipse] at (4,0*0.8) {$\, Y\, $};
\begin{scope}[>={Stealth[black]}, every edge/.style={draw=black}]
\path [->] (Y1) edge[line width=0.75pt] (A);

\path [->] (U0) edge[bend left=50, line width=0.75pt] (W); 
\path [->] (U0.south east) edge[line width=2.25pt] (Y0.north);
\path [->] (U1.south west) edge[line width=0.75pt] (Y1.north);
\path [->] (U0) edge[bend right=70, line width=0.75pt] (A);
\path [->] (U1) edge[bend right=30, line width=0.75pt] (A);
\path [<->] (U1) edge[bend right=30, line width=0.75pt, dashed] (U0);

\path [->] (Y1) edge[line width=2.25pt] (Y); 
\path [->] (Y0) edge[line width=2.25pt] (Y); 
\path [->] (A) edge[line width=2.25pt] (Y); 
\end{scope}
\end{tikzpicture} }
\caption{\small A Graphical Illustration of a Structural Model Compatible with \eqref{eq-SPC model}. Measured covariates $X$ are suppressed for simplicity. The thick arrows depict the deterministic relationships $Y^{a=0} = h_y(U_0)$ and $Y=Y^{A}$.}
\label{fig: Graphical COCA U}
\end{figure}
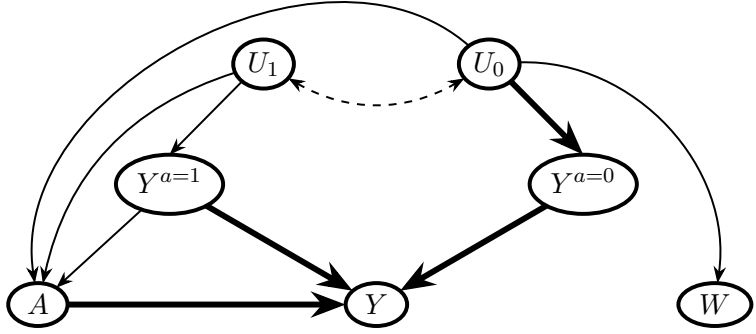

It is enlightening to consider a data generating mechanism that is compatible with Assumption \ref{assumption:NC}. As an example, we consider the following latent variable model for a continuous outcome: 
\begin{subequations} \label{eq-SPC model}
\begin{align} \label{eq-SPC model-1}
&  Y^{a=0} = h_y(U_0,X) \ , \\
&  W \nindep U_0 \cond X \text{ and }
W \indep A \cond (U_0,X) \ ,
\label{eq-SPC model-2}
\end{align}
\end{subequations}
where $U$ is a continuously distributed unobserved variable and $h_y(u,x)$ is a function that is strictly monotone in $u$ for all $x$, but otherwise completely unrestricted. In the Supplementary Material \ref{sec:supp:Latent}, we establish that expressions \eqref{eq-SPC model-1}-\eqref{eq-SPC model-2} imply Assumption \ref{assumption:NC}. Expression \eqref{eq-SPC model-1} means that the treatment-free outcome is a monotonic transformation of an unobserved variable $U_0$, a specific instance of a so-called changes-in-changes model \citep{CiC2006}. Although the latter would also assume under the causal graph in Figure \ref{fig: Graphical COCA} that $W=h_w(U_0,X)$ where $h_w(u,x)$ is strictly monotone in $u$ for all $x$, an assumption we do not make. Expression \eqref{eq-SPC model-2} corresponds to Assumption \ref{assumption:NC}-(ii) and (iii). In fact, our formulation accommodates an additional measurement error in $W$, say $\epsilon_w$, so that $W = h_w(U_0,X,\epsilon_w)$ where $h_w(u,x,\epsilon_w)$ varies in $u$ (potentially non-monotonic) and $\epsilon_w \indep A \cond (U_0,X)$, \eqref{eq-SPC model-2} is satisfied. Figure \ref{fig: Graphical COCA U} provides a graphical representation compatible with, although not necessarily with, expressions \eqref{eq-SPC model-1}-\eqref{eq-SPC model-2}.  

For identification and estimation in the case of a continuous outcome, \citet{TT2013_COCA} further assumed the rank-preserving structural model:%
\begin{align}
Y=Y^{a=0}+\psi^\true  A,\label{rank-preserving}%
\end{align}
which, by consistency, implies a constant individual-level causal effect $\psi^\true =Y^{a=1}-Y^{a=0}$. Under this model, he noted that upon defining $Y(\psi)=Y-\psi A$, then one can deduce from Assumption \ref{assumption:NC} that $W\indep A \cond ( Y(\psi), X)$ if and only if $\psi=\psi^\true $, in which case, given \eqref{rank-preserving}, $Y^{a=0}=Y(\psi^\true )=Y-\psi^\true  A$ which motivates a regression-based implementation of COCA, that entails searching for the parameter value of $\psi=\psi^\true $ such that $E \big\{  W \cond A,Y(\psi),X \big\}  =E \big\{  W \cond Y(\psi),X \big\}$. A straightforward implementation of the approach uses linear models whereby for each value of $\psi$ on a sufficiently fine grid$,$ one obtains an estimate of the regression model $E \big\{  W \cond Y(\psi),X \big\}  =\beta_{1}+\beta_{2}A+\beta_{3}Y(\psi) + \beta_{4} X$  using ordinary least squares (OLS), with estimated coefficients $\big(  \widehat{\beta}_{1}(\psi),\widehat{\beta}_{2}(\psi),\widehat{\beta}_{3}(\psi), \widehat{\beta}_4(\psi) \big)$. Then a 95\% confidence interval for $\psi^\true $ consists of all values of $\psi$ for which a valid test of the null hypothesis $\beta_{2}(\psi)=0$ fails to reject at the 0.05 type 1 error level. Such hypothesis test might be performed by verifying whether the interval $\widehat{\beta}_{2}(\psi)\pm1.96\widehat{\mathrm{SE}} \big(  \widehat{\beta}_{2}(\psi)\big)$ covers $0,$ with $\widehat{\mathrm{SE}} \big(  \widehat{\beta}_{2}(\psi)\big)$ the OLS estimate of the standard error of $\widehat{\beta}_{2}(\psi)$. \citet{TT2013_COCA} also describes a potentially simpler one-shot approach which fits a single regression $E\big( W \cond A,Y,X\big)  =\beta_{1}+\beta_{2}^*A+\beta_{3}Y+\beta_{4}X$ via OLS where $\beta_{2}^*=-\beta_{3}/\psi_{0},$ in which case $\widehat{\psi} =-\widehat{\beta}_{2}^*/\widehat{\beta}_{3}$ where $\big(\widehat{\beta}_{2}^*,\widehat{\beta}_{3}\big)  $ are OLS estimates; a corresponding standard error estimator of $\widehat{\psi}$ is given in the Supplementary Material \ref{sec:supp:COCA_SE} for convenience.  Though practically convenient, validity of either approach relies on both correct specification of the linear model for $W$ given $(Y^{a=0},X)$, and on the rank-preserving structural model, which may be biologically implausible. In the following, we describe alternative methods aimed at addressing these limitations.

\section{Identification of the Effect of Treatment on the Treated}

\subsection{Identification via Extended Propensity Score Weighting} \label{sec:3}

In order to establish identification, consider the extended propensity score function (EPS):
\begin{align*}
\pi^* \big(  y, x\big)  =\Pr\big(  A=1 \cond Y^{a=0}=y, X=x\big)
\end{align*} 
which makes explicit the fact that, in the presence of unmeasured confounding, the treatment mechanism will generally depend on the treatment-free potential outcome even after conditioning for all observed confounders. For notational brevity, we denote the treatment odds by $\omega^*(y,x) = \pi^*(y,x) / \{1-\pi^*(y,x)\}$. Since $\omega^*$ and $\pi^*$ have a one-to-one relationship, we use $\omega^*$ throughout to model the exposure mechanism. We assume that positivity holds, i.e.,
\begin{assumption}\label{assumption-positivity} $\mathcal{S}(1) \subseteq \mathcal{S}(0)$ where $\mathcal{S}(a)$ is the support of $(\potY{0}{},X) \cond (A=a)$ for $a=0,1$.
\end{assumption}

Next, we note that were $\omega^*$ known, the average treatment-free potential outcome in the treated would then be empirically identified by the expression
\begin{align}
\psi_0^*  =
\frac{E\big\{  \big(  1-A\big)
Y \omega^* \big(  Y, X\big) 
\big\}  }{E\big\{  \big(  1-A\big)
\omega^* \big(  Y, X\big)
\big\} }  \ .\label{EPS ID}%
\end{align} 
See the Supplementary Material \ref{sec:supp:proof:EPSID} for the details. Therefore, the ETT would be identified by%
\begin{align*}
\psi^*  =E\big( Y \cond A=1\big)
-
\frac{E\big\{  \big(  1-A\big)
Y \omega^* \big(  Y, X\big) 
\big\}  }{E\big\{  \big(  1-A\big)
\omega^* \big(  Y, X\big)
\big\} }  \ . 
\end{align*}
As $\omega^*$ is unknown, we next demonstrate how the NCO assumption can be leveraged to identify the latter quantity.  Let $p^*\big(  w, x\big)  =\Pr\big(  A=1 \cond W=w, X=x\big)$.  The proposed approach to identify $\omega^* $ is based on the following equality, which we prove in the Supplementary Material \ref{sec:supp:proof:EPSeq}:
\begin{result}      \label{result-1}
Under Assumptions \ref{assumption:1:consistency}-\ref{assumption-positivity}, the following result holds almost surely:
\begin{align}
\frac{p^*\big(  W,X\big)  }{1-p^*\big(  W,X\big)  } &  =E\left\{  \omega^* (Y,X)  \cond W,X,A=0\right\}
=\sum_{y}\frac{\pi^* \big(  y,X\big)  }{1-\pi^* \big(  y,X\big)  }
f^*\big(y \cond W,X,A=0\big)  \label{EPS equation}
\end{align}
where $f^*\big(  y \cond W,X,A=0\big)  $ is the conditional law of $Y$ given $(W,X,A)$ evaluated at $Y=y$.
\end{result}
\noindent Result \ref{result-1} provides an expression relating the exposure mechanism of interest to the observed data distribution, as of the three quantities involved in the expression $p^*$, $f^*$, and $\omega^*$; two are uniquely determined by the observed data, mainly $p^*$ and $f^*$, which are then related to the unknown function of interest $\omega^* $ in Result \ref{result-1}; see the Supplementary Material \ref{sec:supp:Result-1} for a graphical illustration. Equation  \eqref{EPS equation} is known as an integral equation, more precisely a Fredholm integral equation of the first kind. In slight abuse of notation, the sum may be interpreted as an integral if $Y$ is Lebesgue measurable. We then have the following identification result.

\begin{result}      \label{result-2}
If Assumptions \ref{assumption:1:consistency}-\ref{assumption-positivity} hold, and the integral equation \eqref{EPS equation} in Result \ref{result-1} admits a unique solution, then $\omega^*$ and $\pi^*$ are nonparametrically identified from the observed data by solving \eqref{EPS equation}, and
$\psi_0^*$ is nonparametrically identified by \eqref{EPS ID}.
\end{result}

Sufficient conditions for the existence and uniqueness of a solution to such an equation are well-studied; see Section \ref{sec:supp:Exist h} for details. Such conditions were recently discussed in the context of proximal causal inference \citep{Miao2016_arxiv,Miao2018,TT2024_Proximal}. A detailed comparison of COCA with proximal causal inference is relegated to Section \ref{sec:Discussion}. Intuitively, the condition that the integral equation admits a unique solution essentially requires that $W$ is sufficiently relevant for $Y^{a=0}$ in the sense that for any variation in the latter, there is corresponding variation in the former. This assumption is akin to the assumption of relevance in the context of instrumental variable methodology, which states that variation in the instrument should induce variation in the treatment. Importantly, Result \ref{result-2} applies whether $Y$ is binary, continuous or polytomous, provided that $W$ is sufficiently relevant for $Y^{a=0}$, to ensure that equation \eqref{EPS ID} admits a solution. Also, the result obviates the need for a rank-preserving structural model, and delivers fully nonparametric identification of the causal effect of treatment on the treated.  

\subsection{Identification via COCA Confounding Bridge Function} \label{sec:3:bridge}

In this Section, we introduce an alternative nonparametric identification and estimation approach which does not rely on modeling the EPS, but instead relies on the existence of a so-called confounding bridge function formalized below.

\begin{assumption}   \label{condition-1}
For all $(y,x)$, there exists a function (possibly nonlinear) $b^*(w,x)$ that satisfies the following equation%
\begin{align}
y=E\left\{  b^*\big(  W,X\big)  \cond Y=y,X=x,A=0\right\}  .\label{Outcome COCA}%
\end{align}
\end{assumption}

Intuitively, upon noting that under Assumptions \ref{assumption:1:consistency}-\ref{assumption:NC}, Assumption \ref{condition-1} can equivalently be stated in terms of potential outcomes%
\begin{align}
Y^{a=0}=E\left\{  b^*\big(  W,X\big)  \cond Y^{a=0},X\right\}
\label{Counterfactual bridge}%
\end{align}
which essentially formalizes the idea that $W$ is a sufficiently relevant proxy for the potential outcome $Y^{a=0}$ if there exist a (potentially nonlinear) transformation of $(W,X)$ whose conditional expectation given $(Y^{a=0},X)$ recovers $Y^{a=0}$. As $b^*(W,X)$ provides a bridge between the observed data equation \eqref{Outcome COCA}, and its potential outcome counterpart \eqref{Counterfactual bridge}, we aptly refer to $b^*\big(W,X\big)  $ as a \textit{COCA confounding bridge function}. Note that classical measurement error is a special case of the equation in the display above in which case $b^*$ is the identity map and $W=Y^{a=0}+e$ where $e$ is an independent mean zero error.  The condition can therefore be viewed as a nonparametric generalization of classical measurement error which allows $W$ and $Y$ to be of arbitrary nature and does not assume the error to be unbiased on the additive scale.  We further  illustrate the assumption in the case of binary $W$ and $Y$. As shown in the Supplementary Material \ref{sec:supp:proof:Binary}, in this case with suppressing covariates, the following $b^*(W)$ satisfies \eqref{Outcome COCA}:
\begin{align*}
b^*(W)
=
\frac{ -(1-W) \cdot \Pr\big(  W=1 \cond Y=0,A=0\big) + W \cdot  \Pr\big(  W=0 \cond Y=0,A=0\big)  }{\Pr\big(
W=1 \cond Y=1,A=0\big)  -\Pr\big(  W=1 \cond Y=0,A=0\big)  };
\end{align*} 
provided that $\Pr\big(  W=1 \cond Y^{a=0}=1\big)  \not =\Pr\big(  W=1 \cond Y^{a=0}=0\big)$, encoding the requirement that $W$ cannot be independent of $Y^{a=0}$, i.e., Assumption \ref{assumption:NC}. Beyond the binary case, for more general outcome types, Assumption \ref{condition-1} likewise formally defines a Fredholm integral equation of the first kind, for which sufficient conditions for existence of a solution are well characterized in functional analysis textbooks; we again refer the reader to \citet{Miao2016_arxiv}.   We are now ready to state our result, which we prove in the Supplementary Material \ref{sec:supp:proof:OutcomeBridge}:
\begin{result}  \label{result-3}
Suppose that Assumptions \ref{assumption:1:consistency}-\ref{condition-1} hold, and $b^*$ satisfies \eqref{Outcome COCA}. Then,
\begin{align}
\psi_0^*  = 
E \big\{  b^*\big(  W, X\big) \cond A=1 \big\}  \label{Outcome bridge id} \ \text{ and } \ 
\psi^* =
E \big\{ Y - b^*(W,X) \cond A=1 \big\} \ .
\end{align} 
\end{result}

In the binary example discussed above where $b^{*}(w)$ was uniquely identified, we have that
\begin{align*}
\psi_0^*   & =E\left\{   b^*\big(  W\big)
 \cond A=1\right\}
=\frac{\Pr\big(  W=1 \cond A=1\big)  -\Pr\big(  W=1 \cond Y=0,A=0\big)  }%
{\Pr\big(  W=1 \cond Y=1,A=0\big)  -\Pr\big(  W=1 \cond Y=0,A=0\big)  }%
\ .
\end{align*}

We briefly highlight a key feature of the above result reflected in its proof, which is that $b^*\big(  W, X\big)  $ need not be uniquely identified by equation  \eqref{Outcome COCA}, and that any such solution leads to a unique value for $E\big(  Y^{a=0} \cond A=1\big)  .$ Interestingly, the identifying formula in the display above was also obtained by \cite{TT2013_COCA} in the binary case, although he did not emphasize the key role of the bridge function as a general framework for identification beyond the binary case. 

\subsection{Semiparametric Efficiency Theory}

Let $\mathcal{M}$ denote a semiparametric model defined as a collection of observed data laws that admit a solution to \eqref{Outcome COCA}, i.e., 
\begin{align*}
\mathcal{M}
=
\Big\{
P \, \Big| \,  
\text{$P(O)$ is regular and Assumption \ref{condition-1} holds, i.e.,  there exists $b^*$ solving \eqref{Outcome COCA}} 
\Big\} \ .
\end{align*}
We further consider the following surjectivity condition:
\begin{itemize}[leftmargin=0cm]
\item[] \HT{Surjectivity}: Let $T:\mathcal{L}_2(W,X) \rightarrow \mathcal{L}_2(Y,A=0,X)$ denote the operator given by  $T(g)  =  \EXP \big\{ g(W,X) \cond Y,A=0,X \big\}$. At the true data law, $T$ is surjective.
\end{itemize} 
The surjectivity condition states that the Hilbert space $\mathcal{L}_2(W,X)$ is sufficiently rich so that any element in $\mathcal{L}_2(Y,A=0,X)$ can be recovered from an element in $\mathcal{L}_2(W,X)$ via the conditional expectation mapping; see \citet{Cui2023}, \citet{Dukes2023_ProxMed}, and \citet{Ying2023} for related discussions. In addition, we consider a submodel $\mathcal{M}_{\text{sub}}$:
\begin{align*}
&
\mathcal{M}_{\text{sub}}
=
\Bigg\{
P \in \mathcal{M} \, \Bigg| \,  
\begin{array}{l}         
\text{$\omega^*$ and $b^*$ are unique solutions to \eqref{EPS  equation} and \eqref{Outcome COCA}, respectively, } \\
\text{and \HL{Surjectivity} is satisfied}  
\end{array}
\Bigg\}
\end{align*}
We then establish the semiparametric local efficiency bound for $\psi^*$ under $\mathcal{M}$ at the submodel $\mathcal{M}_{\text{sub}}$.

\begin{result} \label{result-IF}
Suppose that Assumptions \ref{assumption:1:consistency}-\ref{condition-1} hold. Then, the following results hold.
\begin{itemize}

\item[(i)] The following function $\InfFt(O \con \omega^*, b^*)$ is an influence function for $\psi^*$ under $\mathcal{M}$.
\begin{align} \label{EIF id}
\InfFt(O \con \omega^*, b^*) = 
\frac{ A \big\{ Y -  b^*\big(W, X \big) - \psi^* \big\} - (1-A) \omega^*(Y,X)
\left\{  Y-b^*\big(
W,X\big)  \right\} }{\Pr \big(  A=1\big)}   
\ .
\end{align}

\item[(ii)] The influence function $\InfFt(O \con \omega^*, b^*)$ is the efficient influence function for $ \psi^*  $ under $\mathcal{M}$ at the submodel $\mathcal{M}_{\text{sub}}$. Therefore, the corresponding semiparametric local efficiency bound for $\psi^*$ is $\VAR \big\{ \InfFt (\bO \con \omega^*,b^*) \big\}$.

\end{itemize}

\end{result}
The influence function $\InfFt$ shares similarity with an influence function for $\psi^*$ in the proximal causal inference framework; see Section G of \citet{Cui2023} for details. Interestingly, the influence function has the following doubly robust property \citep{Scharfstein1999, Lunceford2004, Bang2005}; see the Supplementary Material \ref{sec:supp:proof:DR} for the proof:
\begin{result} \label{result-DR}
Suppose that Assumptions \ref{assumption:1:consistency}-\ref{condition-1} are satisfied. In addition, suppose that either (i) $E\big\{  b^\dagger \big(  W,X\big)  \cond A=0,Y=y, X\big\}  =y$ or (ii) 
$\omega^\dagger \big(  y,x\big)  = \omega^*\big(  y,x\big)  $, but not necessarily both, is satisfied. Then, we have that $\EXP \big\{ \InfFt(O \con \omega^\dagger, b^\dagger) \big\}=0$.

\end{result}
In words, if either the COCA confounding bridge function or the EPS, but not necessarily both, is correctly specified, the influence function is an unbiased estimating function of $\psi^*$.

Using expressions \eqref{EPS ID}, \eqref{Outcome bridge id}, and \eqref{EIF id}, one can construct parametric estimators of $\psi^*$. Specifically, the first estimator using \eqref{EPS ID} entails a priori specifying a parametric model for the EPS, say a logistic regression model. The second estimator based on \eqref{Outcome bridge id} entails a priori specifying a parametric model for the COCA bridge function, say a linear model. Lastly, the third estimator based on \eqref{EIF id} entails parametric models for both EPS and COCA bridge functions. The first two estimators rely on the correct exposure and COCA bridge function specifications, respectively. Thus, misspecification of either model will likely result in biased inferences about the ETT. On the other hand, the last estimator has a doubly-robust property \citep{Scharfstein1999, Lunceford2004, Bang2005} in that it can be used for unbiased inference about the ETT if either EPS or COCA bridge function is correct, without a priori knowledge of which model,  if any, is incorrect. In Section \ref{sec:supp:GMM}, we provide details on constructing these three parametric estimators and their large sample behavior. 

A significant limitation of the three parametric estimators is their dependence on specific parametric specifications of nuisance components, which can lead to biased inference if the model specifications are incorrect. To address this concern, a potential solution is to develop an estimator where nuisance components are estimated using nonparametric methods, drawing on advancements in recent learning theory. In the following Section, we construct such an estimator and study its statistical properties.

\section{A Semiparametric Locally Efficient Estimator} \label{sec:estimator}

Our estimator is derived from the influence function $\InfFt$ in Result \ref{result-DR} and adopts the cross-fitting approach \citep{Schick1986, Victor2018}, which is implemented as follows. We randomly split $N$ study units, denoted by $\mathcal{I}=\{1,\ldots,N\}$, into $K$ non-overlapping folds, denoted by $\{ \mathcal{I}_1, \ldots,\mathcal{I}_K \}$. For each $k=1,\ldots,K$, we estimate the EPS and COCA confounding bridge functions using observations in $\mathcal{I}_k^c = \mathcal{I} \setminus \mathcal{I}_k$, and then evaluate the estimated nuisance functions using observations in $\mathcal{I}_k$ to obtain an estimator of $\psi^*$. We refer to $\mathcal{I}_k^c$ and $\mathcal{I}_k$ as the estimation and evaluation folds, respectively. To use the entire sample, we take the simple average of the $K$ estimators. 

We introduce the following additional notation in order to facilitate the discussion. Let $\mathcal{H}(V)$ be the Reproducing Kernel Hilbert Space (RKHS) of $V$ endowed with a universal kernel function $\mathcal{K}$, such as the Gaussian kernel $\mathcal{K}$, i.e., $\mathcal{K}(v,v') = \exp \big\{ - \big\| v- v' \|_2^2 / \kappa \big\} $ where $\kappa \in (0,\infty)$ is a bandwidth parameter; see Chapter 4 of \citet{SVM2008} for the definition and examples of the universal kernal function. For each $k=1,\ldots,K$, let $\AVER \LSS (V) = |\mathcal{I}_k^c|^{-1} \sum_{i \in \mathcal{I}_k^c} V_i$ and $\AVER \SSS (V) = |\mathcal{I}_k|^{-1} \sum_{i \in \mathcal{I}_k} V_i$. For a function $g (\bO)$, let $\big\| g \big\|_{P,2} = \big[ \EXP \big\{ g^2(\bO) \big\} \big]^{1/2}$ be the $L_2(P)$-norm of $g$. 

We estimate the EPS and COCA bridge functions by adopting a recently developed minimax estimation approach \citep{Ghassami2022}. We remark that other approaches (e.g., \citet{PMMR2021}) can also be adopted with minor modification. Note that $\omega^*(Y,X)$ and $b^*(W,X)$ satisfy
\begin{align*}
&
E\left[  \big\{ (1-A) \omega^*(Y,X) - A \big\}  p \big(  W,X\big)  \right]  =0 \ , \ \forall p \in \mathcal{L}_2(W,X)
\ ,
\\
&
E\left[  (1-A) \big\{ Y - b^*(W,X) \big\} q(Y,X)   \right]  =0 \ , \ \forall q \in \mathcal{L}_2(Y,X)
\ .
\end{align*}
Therefore, following \citet{Ghassami2022}, minimax estimators of $\omega^*$ and $b^*$ are given by
\begin{align*}
& 
\widehat{\omega}\LSS (\cdot)
\\
&
=
\argmin_{\omega \in \mathcal{H}(Y,X)}
\bigg[
\displaystyle{ \max_{p \in \mathcal{H}(W,X)}  }
\bigg[
\AVER\LSS \bigg[
\begin{array}{l}
p (W,X)
\big\{
(1-A) \omega(Y,X)
-A
\big\} \\
- p^2(W,X)
\end{array}
\bigg]
-
\lambda_{p} \big\| p \big\|_{\mathcal{H}}^2
\bigg] 
+
\lambda_{\omega} \big\| \omega \big\|_{\mathcal{H}}^2 
\bigg] 
\\
& 
\widehat{b}\LSS (\cdot)
\\
&
=
\argmin_{h \in \mathcal{H}(\bW,\bX)}
\bigg[ 
\displaystyle{ \max_{q \in \mathcal{H}(Y,X)}  }
\bigg[
\AVER\LSS \bigg[
\begin{array}{l}         
q (Y,X)
(1-A)
\big\{
Y - b(W,X)
\big\} \\
- q^2(Y,X) 
\end{array} \bigg]
-
\lambda_{q} \big\| q \big\|_{\mathcal{H}}^2
\bigg]
+
\lambda_{b} \big\| b \big\|_{\mathcal{H}}^2 
\bigg] 
\end{align*}
where $\big\| \cdot \big\|_{\mathcal{H}}$ is an RKHS norm and $\lambda_p$, $\lambda_{\omega}$, $\lambda_q$, and $\lambda_b$ are positive regularization parameters. 

We make a few remarks about the minimax estimation approach, of which details are relegated to the Supplementary Material \ref{sec:supp:MMEstimation}. First, despite the complicated formulas, closed-form representations of $\widehat{\omega}\LSS$ and $\widehat{b}\LSS$ are available from the representer theorem \citep{KW1970, SHS2001}. Second, the bandwidth and regularization parameters can be selected via cross-validation. Lastly, $\omega^*$ may vary widely because it is a ratio of two probabilities. In such cases, the proposed minimax estimator may result in significantly small or negative estimates. To mitigate this issue, one may consider a practical approach to regularize the minimax estimator when it appears to be ill-behaved. 

Using the minimax estimators of the nuisance functions, a semiparametric estimator $\widehat{\psi}$ of $\psi^*$ is then obtained as follows:
\begin{align*}
&
\widehat{\psi}
=
\frac{1}{K}
\sum_{k=1}^{K}
\widehat{\psi}^{(k)}
\ , \\
&
\widehat{\psi}^{(k)}
=
\frac{ 
\frac{1}{N}
\sum_{i=1}^{N} A_i Y_i
-
\AVER\SSS \big[  
(1-A) \widehat{\omega}\LSS(Y,X) \big\{ Y - \widehat{b}\LSS(W,X) \big\}
+ A \widehat{b}\LSS(W,X) 
\big] }{ \frac{1}{N} \sum_{i=1}^{N} A_i}    \ .
\end{align*}

Under regularity conditions, the semiparametric estimator $\widehat{\psi}$ is consistent and asymptotically normal for $\psi^*$.  
\begin{assumption} \label{reg}
Suppose that the following conditions hold for all $k=1,\ldots,K$:
\begin{itemize}
\item[(i)] (\textit{Boundedness}) There exists a finite constant $C>0$ such that 
\begin{align*}
&
\big| \omega^*(y,x) \big| \in [C^{-1},C]
\ , \quad 
\big| \widehat{\omega}\LSS(y,x) \big| \in [C^{-1},C]
\quad \text{ for all $(y,x)$,}
\\
&
\big| b^*(w,x) \big| \leq C
\ , \quad 
\big| \widehat{b}\LSS(w,x) \big| \leq C
\quad \text{ for all $(w,x)$,} 
\quad
\EXP(Y^4 ) \leq C \ .
\end{align*} 

\item[(ii)] (\textit{Consistency}) As $N \rightarrow \infty$, we have $\big\| \widehat{\omega}\LSS  - \omega^*  \big\|_{P,2}=o_P(1)$ and $\big\| \widehat{b}\LSS  - b^*  \big\|_{P,2}=o_P(1)$.

\item[(iii)] (\textit{Cross-product Rates}) As $N \rightarrow \infty$, we have 
\begin{align*}
    \min 
\Bigg[
\begin{array}{l}    
    \big\| \widehat{\omega}\LSS - \omega^* \big\|_{P,2}  
    \big\| 
    \EXP\LSS\big[ 
    (1-A)
    \big\{ b^*(W,X) - \widehat{b}\LSS(W,X) \big\}
    \cond Y,X 
    \big]
\big\|_{P,2}
    ,
    \\
    \big\| \widehat{b}\LSS - b^* \big\|_{P,2}
\big\| 
    \EXP\LSS\big[ 
    (1-A)
    \big\{ \omega^*(Y,X) - \widehat{\omega}\LSS(Y,X) \big\}
    \cond W,X 
    \big]
\big\|_{P,2} 
\end{array}
\Bigg]
= o_P(N^{-1/2}) \ .
\end{align*}
\end{itemize}
\end{assumption}
Assumption \ref{reg}-(i) states that nuisance functions and the corresponding estimators are uniformly bounded. Assumption \ref{reg}-(ii) states that the estimated nuisance functions are consistent for the true nuisance functions in the $L_2(P)$ norm sense. Assumption \ref{reg}-(iii) states that the cross-product rate of nuisance function estimators are $o_P(N^{-1/2})$. Assumption \ref{reg}-(iii) is satisfied if $b^*$ and $\omega^*$ are sufficiently smooth, the conditional expectation operators $f(W,X) \mapsto \EXP\{ f(W,X) \cond Y,A=0,X \}$ and $f(Y,X) \mapsto \EXP\{ f(Y,X) \cond W,A=0,X \}$ are sufficiently smooth, and $\widehat{\omega}\LSS$ and $\widehat{b}\LSS$ are estimated over an RKHS with fast enough eigendecay; see Section 5 of \citet{Ghassami2022} for details. Importantly, if one nuisance function is estimated at sufficiently fast rates, the other nuisance function is allowed to converge at a substantially slower rate provided that the cross-products remain $o_P(N^{-1/2})$.  This is an instance of the mixed-bias property described by \citet{Rotnitzky2020} and \citet{Ghassami2022}. It is also worth highlighting the structure of the mixed bias in the current context which is the minimum of two product biases, each containing a bias term for a nuisance function and a projected bias term for the other nuisance function; this property was first reported in \citet{Ghassami2022} for a large class of functionals including ours.  

Result \ref{result-CAN} establishes that $\widehat{\psi}$ is consistent and asymptotically normal (CAN) for $\psi^*$.
\begin{result} \label{result-CAN}
Suppose that Assumptions \ref{assumption:1:consistency}-\ref{reg} hold. Then, we have $\sqrt{N} 
\big( \widehat{\psi} - \psi^* \big)
\stackrel{D}{\rightarrow} N(0, \sigma^2)$ where $\sigma^2 = \VAR \big\{ \InfFt(O \con \omega^*,b^*) \big\}$, and a consistent estimator of $\sigma^2$ is $\widehat{\sigma}^2
=
\sum_{k=1}^{K} \widehat{\sigma}^{2,(k)} / K$ where
\begin{align*}
&
\widehat{\sigma}^{2,(k)}
=
\frac{ \AVER\SSS \big[ 
\big[ A \big\{ Y -  \widehat{b}\LSS\big(W, X \big) - \widehat{\psi}\SSS \big\} - \big(  1-A\big) \widehat{\omega}\LSS(Y,X)
\big\{  Y-\widehat{b}\LSS\big(W,X\big)  \big\}   \big]^2
\big] } { \big( \sum_{i=1}^{N} A_i/N \big) ^2 }
\ .
\end{align*}
\end{result}
Using the variance estimator $\widehat{\sigma}^2$, valid $100(1-\alpha)$\% confidence intervals for the ETT are given by  $ \big( \widehat{\psi} + z_{\alpha/2} \widehat{\sigma} / \sqrt{N} ,  \widehat{\psi} + z_{1-\alpha/2} \widehat{\sigma} / \sqrt{N} \big)$  where $z_\alpha$ is the $100\alpha$th percentile of the standard normal distribution. Alternatively, one may construct confidence intervals using the multiplier bootstrap \citep[Chapter 2.9]{VW1996}; see Section \ref{sec:supp:MMEstimation} for details. 

Lastly, the cross-fitting estimator depends on a specific sample split, and thus, may produce outlying
estimates if some split samples do not represent the entire data. To mitigate this issue, \citet{Victor2018} proposes to use median adjustment from multiple cross-fitting estimates; the detail can be found in the Supplementary Material \ref{sec:supp:MMEstimation}.

\section{Data Application: Zika Virus Outbreak in Brazil}            \label{sec:Data}

The Zika virus, which can be transmitted from a pregnant woman to her fetus, can cause serious brain abnormalities, including microcephaly (i.e., an abnormally small head) \citep{Zika2016}. Brazil is one of the countries hardest hit by the Zika virus. In particular, the outbreak in 2015 resulted in  over 200,000 cases in Brazil by 2016 \citep{Zika2018}. As a result, many prior works \citep{Zika2018_3, Zika2018_2, Zika2022, TTPR2023} asked whether the Zika virus outbreak caused a drop in birth rates. 

We re-analyzed the dataset analyzed in \citet{Zika2022} and \citet{TTPR2023}. In the dataset, we focused on 673 municipalities in two states of Brazil, Pernambuco and Rio Grande do Sul, which are northeastern and southernmost states. Out of the 1248 cases of microcephaly that occurred in Brazil by November 28, 2015, 51.8\% (646 cases) were reported in Pernambuco (PE), less than 10 cases of Zika-related microcephaly were reported in Rio Grande do Sul (RS) \citep{Zika2017_Brazil2}, which shows that PE was severely impacted by the Zika virus outbreak, while RS was minimally affected. Based on their epidemiologic histories, we defined 185 and 488 municipalities in PE and RS as treated and control groups, respectively. 

For each municipality, we included the following variables in the analysis. As pre-treatment covariates, we included municipality-level population size, population density, and proportion of females measured in 2014. We used the post-epidemic municipality-level birth rate in 2016 as the outcome $Y$, where the birth rate is defined as the total number of live human births per 1,000 persons. We used the pre-epidemic municipality-level birth rates in 2013 and 2014 as the outcome proxies (i.e., NCO), denoted by $W_1$ and $W_2$, respectively. To be valid proxies, the birth rates in 2013 and 2014 must satisfy Assumption \ref{assumption:NC}: (i) birth rates in 2013 and 2014 cannot be causally impacted by the Zika virus epidemic which occurred in 2015, (ii) birth rates in 2013 and 2014 are correlated with what the birth rate in 2016 would have been had there not been a Zika virus epidemic, and (iii) birth rates in 2013 and 2014 are independent of a municipality's Zika epidemic status, upon conditioning on its Zika virus epidemic-free potential birth rate in 2016. The first two conditions are uncontroversial, while the third condition largely relies on the extent to which pre-epidemic birth rates can accurately be viewed as a proxy for the counterfactual birth rate had the pandemic not occurred, and as such would not further be predictive of whether the municipality experienced a high rate of Zika virus incidence, conditional on the region's epidemic-free counterfactual birth rate in 2016. Although one might consider this last assumption reasonable, ultimately, it is empirically untestable without making an alternative assumption. Nevertheless, in the Supplementary Material \ref{sec:supp:sensitivity}, we describe a straightforward sensitivity analysis to evaluate the extent to which violation of the assumption might impact inference. 

Using the dataset, we estimated $\psi^* = E \big(Y^{a=1}-Y^{a=0} \cond A=1 \big)$, i.e., the difference between the observed average birth rate of Pernambuco and a forecast of what it would have been had the Zika outbreak been prevented. Therefore, the ETT quantifies the average treatment effect of the Zika outbreak on the birth rate within the Pernambuco region. Of note, the crude estimand $E(Y \cond A=1) - E(Y \cond A=0)$ was estimated to be equal to $3.384$, suggesting that municipalities in the PE region (with higher incidence of Zika virus) experienced a higher birth rate than RS regions in 2016 during the Zika virus outbreak. An immediate concern is that this crude association between $A$ and $Y$ might be subject to significant confounding bias, leading us to conduct two separate analyses geared at addressing residual confounding bias; the proposed COCA methods, which we compared with a standard difference-in-differences analysis. Thus, we estimate the ETT using the approach outlined in Section \ref{sec:estimator} where the NCOs are specified as either (i) $W_1$, birth rate in 2013, or (ii) $W_2$, birth rate in 2014, or (iii) $(W_1,W_2)$. For comparison, we also obtained doubly-robust parametric estimators of the ETT using the three NCO specifications; see the Supplementary Material \ref{sec:supp:data} for details on how these estimators were constructed.

Table \ref{Tab-1} summarizes corresponding results. We find that the six COCA estimates vary between $-1.833$ and $-2.410$, meaning between $1.833$ and $2.410$ birth per 1,000 persons were reduced in PE due to the Zika virus outbreak, an empirical finding better aligned with the scientific hypothesis that Zika may likely adversely impact the birth rates of exposed populations. Compared to the crude estimate of 3.384, the negative effect estimates indeed provide compelling evidence of potential confounding. We also obtain an estimate using the difference-in-difference estimator under a standard parallel trends assumption (e.g., \citet{CardKrueger1994, Angrist2009}), which yields a considerably smaller effect estimate varying between $-1.156$ and $-1.041$; noting that the DiD estimator requires the assumption of equi-confounding of the $A-Y$ association in the pre- and post-periods, while our proposed estimator does not (but instead requires conditions (i)-(iii) outlined above). Regardless of the estimator, all estimates appear to be consistent with the anticipated adverse causal impact of the Zika Virus epidemic. Consequently, we conclude that based on inferences aimed at accounting for confounding (DiD and COCA), the Zika virus outbreak likely led to a decline in the birthrate of affected regions in Brazil, which agrees with similar findings in the literature \citep{Zika2018_3,Zika2018_2,Zika2022,TTPR2023}.

\begin{table}[!htp]
\renewcommand{\arraystretch}{1.05} \centering
\setlength{\tabcolsep}{7pt}
\footnotesize
\begin{tabular}{|c|c|ccc|}
\hline
\multirow{2}{*}{Estimator} & \multirow{2}{*}{Statistic} & \multicolumn{3}{c|}{NCO} \\ \cline{3-5} 
&  & \multicolumn{1}{c|}{$W_1$} & \multicolumn{1}{c|}{$W_2$} & $(W_1,W_2)$ \\ \hline
 \multirow{3}{*}{Semiparametric COCA} & Estimate & \multicolumn{1}{c|}{$-2.410$} & \multicolumn{1}{c|}{$-2.182$} & $-2.180$ \\ \cline{2-5}
 & SE & \multicolumn{1}{c|}{$0.356$} & \multicolumn{1}{c|}{$0.503$} & $0.342$ \\ \cline{2-5}
 & 95\% CI & \multicolumn{1}{c|}{($-3.107$,$-1.713$)} & \multicolumn{1}{c|}{($-3.168$,$-1.196$)} & ($-2.850$,$-1.510$) \\ \hline
 \multirow{3}{*}{Doubly-robust parametric COCA} & Estimate & \multicolumn{1}{c|}{$-2.235$} & \multicolumn{1}{c|}{$-1.833$} & $-2.182$ \\ \cline{2-5}
 & SE & \multicolumn{1}{c|}{$0.502$} & \multicolumn{1}{c|}{$0.519$} & $0.415$ \\ \cline{2-5}
 & 95\% CI & \multicolumn{1}{c|}{($-3.220$,$-1.250$)} & \multicolumn{1}{c|}{($-2.850$,$-0.816$)} & ($-2.996$,$-1.368$) \\ \hline
 \multirow{3}{*}{Standard DiD under parallel trends} & Estimate & \multicolumn{1}{c|}{$-1.156$} & \multicolumn{1}{c|}{$-1.041$} & $-1.041$ \\ \cline{2-5}
 & SE & \multicolumn{1}{c|}{$0.199$} & \multicolumn{1}{c|}{$0.195$} & $0.195$ \\ \cline{2-5}
 & 95\% CI & \multicolumn{1}{c|}{($-1.546$,$-0.767$)} & \multicolumn{1}{c|}{($-1.424$,$-0.658$)} & ($-1.424$,$-0.658$) \\ \hline
\end{tabular}
\vspace*{0.5cm}
\caption{Summary of Data Analysis. Values in ``Estimate'' row represent the estimates of the ETT. Values in ``SE'' and ``95\% CI'' rows represent the standard errors (SEs) associated with the estimates and the corresponding 95\% confidence intervals (CIs), respectively. The reported values are expressed as births per 1,000 persons.}
\label{Tab-1}

\end{table}

\section{Discussion and Possible Extensions}        \label{sec:Discussion}
We have described a COCA nonparametric identification framework, therefore extending previous results of \citet{TT2013_COCA} to a more general setting accommodating outcomes of arbitrary nature and obviating the need for an assumption of constant treatment effects, i.e. rank preservation. We have proposed three estimation strategies including a doubly robust method which has appealing robustness properties. Interestingly, the COCA central identifying assumption, that conditioning on the treatment-free counterfactual would in principle shield the treatment assignment from any association with the NCO is isomorphic to an analogous assumption in the missing data literature where an outcome might be missing not at random; however, a fully observed so-called \emph{shadow variable} (the missing data analog of an NCO) reasonably assumed to be conditionally independent of the missing data process given the value of the potentially missing outcome. For example, \cite{Zahner1992} considered a study of the children’s mental health evaluated through their teachers’ assessments in Connecticut. However, the data for the teachers’ assessments are subject to nonignorable missingness. As a proxy of the teacher’s assessment, a separate parent report is available for all children in this study. The parent report is likely to be correlated with the teacher’s assessment, but is unlikely to be related to the teacher’s response rate given the teacher’s assessment and fully observed covariates. Hence, the parental assessment is regarded as a shadow variable for the teacher's assessment in this study.  The literature on shadow variables is fast-growing \citep{DH2010, Kott2014, Wang2014, MTT2016, LMTT2023, MLTTG2024}, the methods developed in this paper have close parallels to shadow variable counterparts in this literature. This connection to shadow variables is particularly salient when the COCA confounding bridge function is not uniquely defined, which can easily occur for instance when the shadow variable (or analogously the NCO) is multivariate, therefore significantly complicating inference. Fortunately, the methods developed by \citet{LMTT2023} for the analogous shadow variable setting directly apply to the corresponding COCA setting and thus provide a complete solution for identification and inference for the average treatment effect for the treated without relying on completeness conditions nor on unique identification of either the EPS or the COCA confounding bridge function. We refer the interested reader to this latter work for further details. It is worth noting that the doubly robust estimator proposed in this paper appears to be completely new, and different from those of \citet{MTT2016}, \citet{LMTT2023}, and \citet{MLTTG2024}  and therefore may also be of use in shadow variable applications. Likewise, the doubly robust estimators proposed in the latter works can equally be applied to the current COCA setting as an alternative inferential approach. 

Additionally, as mentioned in the previous Section, the key assumption that conditioning on the treatment-free potential outcome, would in principle make the NCO or outcome proxy irrelevant to treatment mechanism is ultimately untestable, and may in certain settings not hold exactly. In fact, this would be the case if the NCO were in fact explicitly used in assigning the treatment in which case the assumption might be violated. In order to address such eventuality, the analyst might consider several candidate proxies/NCOs when available, and may even perform an over-identification test, by inspecting the extent to which the estimated causal effect depends on the choice of proxy. Alternatively, a sensitivity analysis might also be performed to evaluate the potential impact of a hypothesized departure from the assumption. In the context of the Zika virus application, an over-identification test and a sensitivity analysis were carried out as illustrative examples, with the corresponding results and discussion provided in the Supplementary Material \ref{sec:supp:sensitivity} and \ref{sec:supp:over-identification}.

Finally, as previously mentioned in the Introduction, COCA offers an alternative approach to proximal causal inference for debiasing observational estimates of the ETT by leveraging negative control or valid confounding proxies. A key difference highlighted earlier between these two frameworks is that COCA relies on a single valid NCO which directly proxies the treatment-free potential outcome, while proximal causal inference requires both valid NCO and negative control treatment variables that proxy an underlying unmeasured confounder. Importantly, COCA takes advantage of the fact that the treatment-free potential outcome is observed in the untreated, while in proximal causal inference, the unmeasured confounders for which proxies are available, are themselves never observed, arguably a more challenging identification task. Despite the practical advantage of needing one rather than two proxies, it is important to note that though COCA identifies the ETT, it fails to nonparametrically identify the population average treatment effect (ATE), without an additional assumption. In contrast, proximal causal inference provides nonparametric identification of both causal parameters and thus can be interpreted as providing richer identification opportunities. A key reason for this difference in the scope of identification is the fact that in the current paper, we have emphasized an interpretation of the NCO as a proxy for the treatment-free potential outcome but not for the potential outcome under treatment, in the sense that under our conditions, it must be that the treatment-free potential outcome does not only shield the treatment from the NCO, but also shields the potential outcome under treatment from the latter. Two potential strategies to recover COCA identification of the population ATE, might be either (i) evoke a rank-preservation assumption which if appropriate would imply that the ETT and the ATE are equal (this is the assumption made in \cite{TT2013_COCA}); or (ii) identify a second proxy NCO which is a valid proxy for the counterfactual outcome under treatment. The second condition would be needed if $Y^{a=1}$ can also be viewed as a hidden confounder.  By a symmetry argument, one can show that (ii) in fact would provide identification of the average counterfactual outcome under treatment for the untreated. A weighted average of both counterfactual means would then provide identification of the ATE. Details are not provided, but can easily be deduced from the presentation.

\section*{Acknowledgment}

The authors would like to thank James Robins, Thomas Richardson, and Ilya Shpitser for helpful discussions.

%

%
%
%

\section*{Data Availability}

The data and the analysis R code are accessible on the GitHub repository located at \url{http://github.com/qkrcks0218/SingleProxyControl}.

\newpage

\appendix 

\section*{Supplementary Material}

This document provides details of ``Single Proxy Control.''	 In Section \ref{sec:supp:Detail}, we provide details of the main paper. In Section \ref{sec:supp:Proof}, we provide proofs of the results in the main paper.

\section{Details of the Main Paper} \label{sec:supp:Detail}

\subsection{Details of the Latent Variable Model \eqref{eq-SPC model}} \label{sec:supp:Latent}

Recall that model \eqref{eq-SPC model} is given by
\begin{align} 
&  Y^{a=0} = h_y(U_0,X) \ , \ h_y(u,x): \text{strictly increasing in $u$ for all $x$}
\tag{\ref{eq-SPC model-1}}
\\
&  W \nindep U_0 \cond X \text{ and }
W \indep A \cond (U_0,X)
\tag{\ref{eq-SPC model-2}}
\end{align}

Under these assumptions, we establish that
\begin{align*}
& 
W \nindep U_0 \cond X
\quad
\Rightarrow
\quad
W \nindep h_y(U_0,X) \cond X
\quad
\Rightarrow
\quad
W \nindep \potY{0}{} \cond X
\end{align*}
justifying Assumption \ref{assumption:NC}-(ii): $W\nindep  Y^{a=0} \cond X$, and
\begin{align*}
&
W \cond (U_0=u,A=1,X) \stackrel{D}{=}
W \cond (U_0=u,A=0,X)
\\
\Rightarrow
\quad
&
W \cond (h_y^{-1}(\potY{0}{},X)=u,A=1,X) \stackrel{D}{=}
W \cond (h_y^{-1}(\potY{0}{},X)=u,A=0,X)
&& (\because \ \text{$h$ is monotonic})
\\
\Rightarrow
\quad
&
W \cond (\potY{0}{} = h_y(u,X),A=1,X) \stackrel{D}{=}
W \cond (\potY{0}{} = h_y(u,X),A=0,X) 
&& (\because \ \text{Definition of $\potY{0}{}$})
\\
\Rightarrow
\quad
&
W \cond (\potY{0}{} = y,A=1,X) \stackrel{D}{=}
W \cond (\potY{0}{} = y,A=0,X) 
\end{align*} 
justifying Assumption \ref{assumption:NC}-(iii): $W\indep  A  \cond (Y^{a=0},X)$.

\begin{figure}[!htp]
\centering 
\scalebox{1}{
\begin{tikzpicture}
\tikzset{line width=1.5pt, outer sep=0pt,
ell/.style={draw,fill=white, inner sep=2pt,
line width=1.5pt},
swig vsplit={gap=5pt,
inner line width right=0.5pt}};
\node[name=A, ell, ellipse] at (-0.5,0*0.8){$\, A\, $};
\node[name=Y0, ell, ellipse] at (6.25,2*0.8) {$\potY{0}{}$}  ;
\node[name=U1, ell, ellipse] at (2.5,4*0.8) {$U_{y1}$}  ;
\node[name=U0, ell, ellipse] at (4.75,4*0.8) {$U_{y0}$}  ;
\node[name=Uw, ell, ellipse] at (7,4*0.8) {$U_{w}$}  ;

\node[name=Y1, ell, ellipse] at (1.75,2*0.8) {$\potY{1}{}$}  ;
\node[name=W, ell, ellipse] at (8.5,0*0.8) {$\, W\, $};
\node[name=Y, ell, ellipse] at (4,0*0.8) {$\, Y\, $};
\begin{scope}[>={Stealth[black]}, every edge/.style={draw=black}]

\path [->] (Y1) edge[line width=0.75pt] (A);

\path [->] (Uw) edge[bend left=10, line width=2.25pt] (W); 
\path [->] (U0.south) edge[line width=2.25pt] (Y0.north);
\path [->] (U1.south) edge[line width=0.75pt] (Y1.north);
\path [->] (U0) edge[bend right=70, line width=0.75pt] (A);
\path [->] (U1) edge[bend right=30, line width=0.75pt] (A);
\path [<->] (U1) edge[bend right=30, line width=0.75pt, dashed] (U0);
\path [<->] (U0) edge[bend right=30, line width=0.75pt, dashed] (Uw);

\path [->] (Y1) edge[line width=2.25pt] (Y); 
\path [->] (Y0) edge[line width=2.25pt] (Y); 
\path [->] (A) edge[line width=2.25pt] (Y); 
\end{scope}
\end{tikzpicture} }
\caption{\small A Graphical Illustration of a Structural Model Compatible with \eqref{eq-SPC model}. Measured covariates $X$ are suppressed for simplicity. The thick arrows depict the deterministic relationships $Y^{a=0} = h_y(U_{y0})$, $W = h_w(U_w)$, and $Y=Y^{A}$.}
\label{fig: Graphical COCA U Special}
\end{figure}
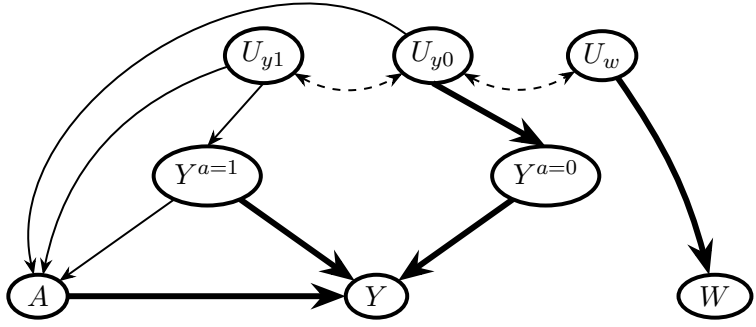

It is instructive to consider a data generating mechanism that can be formulated in a manner analogous to the changes-in-changes model \citep{CiC2006}. To this end, we now consider a special case of model \eqref{eq-SPC model}:
\begin{subequations} 
\begin{align} 
&  Y^{a=0} = h_y(U_{y0},X) \ , \ h_y(u,x): \text{strictly increasing in $u$ for all $x$} 
\tag{SPC-a}
\label{eq-SPC model-Special-1}
\\
&  W = h_w(U_{w},X)
\tag{SPC-b}
\label{eq-SPC model-Special-2}
\\
&
U_{y0} \nindep U_{w} \cond X \quad \text{ and } \quad
U_{w} \indep A \cond U_{y0}, X 
\tag{SPC-c}
\label{eq-SPC model-Special-3}
\end{align} 
\end{subequations}
Figure \ref{fig: Graphical COCA U Special} provides a graphical representation compatible with expressions \eqref{eq-SPC model-Special-1}-\eqref{eq-SPC model-Special-3}.  

It is straight forward to show that \eqref{eq-SPC model-Special-2} and \eqref{eq-SPC model-Special-3} imply \eqref{eq-SPC model-2} as follows:

Under these assumptions, we establish that
\begin{align*}
& 
U_{y0} \nindep U_w \cond X
\quad
\Rightarrow
\quad
U_{y0} \nindep h_w(U_w,X) \cond X
\quad
\Rightarrow
\quad
U_{y0} \nindep W \cond X
\end{align*}
and
\begin{align*}
&
U_w \cond (U_{y0}=u,A=1,X) \stackrel{D}{=}
U_w \cond (U_{y0}=u,A=0,X)
\\
\Rightarrow
\quad
&
h(U_w,X) \cond (U_{y0}=u,A=1,X) \stackrel{D}{=}
h(U_w,X) \cond (U_{y0}=u,A=0,X)
\\
\Rightarrow
\quad
&
W \cond (U_{y0}=u,A=1,X) \stackrel{D}{=}
W \cond (U_{y0}=u,A=0,X) 
\end{align*}

Models \eqref{eq-SPC model-Special-1}-\eqref{eq-SPC model-Special-3} share similarity with a changes-in-changes model in panel data setting \citep{CiC2006}. Specifically, when a pre-treatment outcome is viewed as $W$, the changes-in-changes model in panel data setting is represented as follows:
\begin{subequations} 
\begin{align} 
&  Y^{a=0} = h_y(U_{y0},X) \ , 
&& h_y(u,x): \text{strictly increasing in $u$ for all $x$} \label{eq-CiC-1}
\tag{CiC-a}
\\
&  W = h_w(U_{w},X) \ , 
&& h_w(u,x): \text{strictly increasing in $u$ for all $x$} 
\label{eq-CiC-2}
\tag{CiC-b}
\\
&
U_{y0} \cond A, X \stackrel{D}{=} U_{w} \cond A,X
\label{eq-CiC-3}
\tag{CiC-c}
\end{align}
\end{subequations}
Expression \eqref{eq-SPC model-Special-1} means that the treatment-free outcome is a monotonic transformation of an unobserved variable $U_{y0}$. We remark that the same model for $\potY{0}{}$ is used in the changes-in-changes model \eqref{eq-CiC-1}. Expression \eqref{eq-SPC model-Special-2} means that the negative control outcome is a transformation of an unobserved variable $U_{w}$. It is important to highlight that in model \eqref{eq-SPC model-Special-2}, the function $h_w$ is not necessarily monotonic in $U_w$, unlike \eqref{eq-CiC-2}. Consequently, \eqref{eq-SPC model-Special-2} is strictly weaker than \eqref{eq-CiC-2}. Lastly, expressions \eqref{eq-SPC model-Special-3} and \eqref{eq-CiC-3} share a common goal of restricting the distributions of the unmeasured confounders, but have significant differences. Specifically, \eqref{eq-CiC-3} states that $U_{y0}$ and $U_w$ have the identical distribution given $(A,X)$ and do not place any restriction on the correlation between $U_{y0}$ and $U_w$. On the other hand, \eqref{eq-SPC model-Special-3} states that (i) $U_{y0}$ and $U_w$ must be correlated given $X$ and (ii) $U_{w}$ is conditionally independent of $A$ given $(U_{y0},X)$, but it does not necessitate that $U_{y0}$ and $U_w$ have the identical distribution given $(A,X)$. Consequently, $U_{y0}$ and $U_{w}$ may have different supports under expression \eqref{eq-SPC model-Special-3}, a condition that does not apply to expression \eqref{eq-CiC-3}. Lastly, we note that both \eqref{eq-SPC model-Special-3} and \eqref{eq-CiC-3} hold if $U_{y0}=U_{w}$ almost surely, yet perfect correlation between the two unmeasured confounders may not be reasonable.

\subsection{Standard Error of the COCA Estimator in \citet{TT2013_COCA}}       \label{sec:supp:COCA_SE}

We provide details on the standard error of the COCA estimator obtained from the ordinary least squares estimators of the regression model $\EXP ( W | A,Y,X) = \beta_1 + \beta_2^*A + \beta_3 Y + \beta_4 X$. It is well known that the variance estimator of the OLS estimator $(\widehat{\beta}_1 , \widehat{\beta}_2^* , \widehat{\beta}_3, \widehat{\beta}_4)$ is given as
\begin{align*}
\widehat{V}
:=
\widehat{\sigma}^2
\left[
\sum_{i=1}^{N}
\begin{pmatrix}
1 & A_i & Y_i & X_i
\\
A_i & A_i^2 & A_i Y_i  & A_i X_i 
\\
Y_i & A_i Y_i & Y_i^2 & Y_i X_i
\\
X_i & A_i X_i & Y_i X_i & X_i^2
\end{pmatrix}
\right]^{-1}
\ , \ 
\widehat{\sigma}^2
:=
\frac{ \sum_{i=1}^{N}
\big(
W_i - \widehat{\beta}_1 - \widehat{\beta}_2^* A_i - \widehat{\beta}_3 Y_i - \widehat{\beta}_4 X_i
\big)^2 }{N-4}
\end{align*}
where $N$ is the number of observed units, indexed by subscript $i = 1, \ldots N$. 
Let $f(a,b)=a/b$. Then, the first-order Taylor expansion around $(a_0,b_0)$ is
\begin{align*}
\frac{a}{b} 
=
f(a,b) 
\simeq f(a_0,b_0) + f_1 (a_0,b_0) (a-a_0) + f_2 (a_0,b_0) (b-b_0)
=
\frac{a_0}{b_0} + \frac{a-a_0}{b_0} - \frac{ a_0 (b-b_0) } {b_0^2}
\end{align*}
where $f_1(a,b) = \partial f(a,b) / \partial a$ and $f_2(a,b) = \partial f(a,b) / \partial b$. 
Therefore, we find
\begin{align*}
\EXP \big\{ ( \widehat{\psi}_0 - \widehat{\psi}_0)^2 \big\}
& =
\EXP \bigg\{
\bigg(
\frac{ \widehat{\beta}_2^*}{\widehat{\beta}_3}
-
\frac{ {\beta}_2^*}{{\beta}_3}
\bigg)^2
\bigg\}
\\
& \simeq
\EXP \bigg\{
\bigg(
\frac{ \widehat{\beta}_2^* - \beta_2^* }{{\beta}_3}
-
\frac{ \beta_2^* (\widehat{\beta}_3 - \beta_3) }{{\beta}_3^2}
\bigg)^2
\bigg\}
\\
& =
\frac{ \VAR (\widehat{\beta}_2^*) }{\beta_3^2}
-
2
\frac{ \beta_2^* Cov (\widehat{\beta}_2^*, \widehat{\beta}_3) }{\beta_3^3}
+
\frac{ (\beta_2^*)^2 \VAR (\widehat{\beta}_3) }{\beta_3^4}
\end{align*}
Consequently, the standard error estimator of $\widehat{\psi}$ is given as $SE_{\psi_0}
=
\widehat{\beta}_3^{-1} 
\Big( 
\widehat{V}_{22}  - 
2 \widehat{\psi}_0 \widehat{V}_{23}
+
\widehat{\psi}_0^2 \widehat{V}_{33}
\Big)^{1/2}$ where $\widehat{V}_{ij}$ is the $(i,j)$th element of $\widehat{V}$.  

\subsection{A Graphical Illustration of Result \ref{result-1}} \label{sec:supp:Result-1}

\begin{figure}[!htp]
\centering
\includegraphics[width=0.75\linewidth]{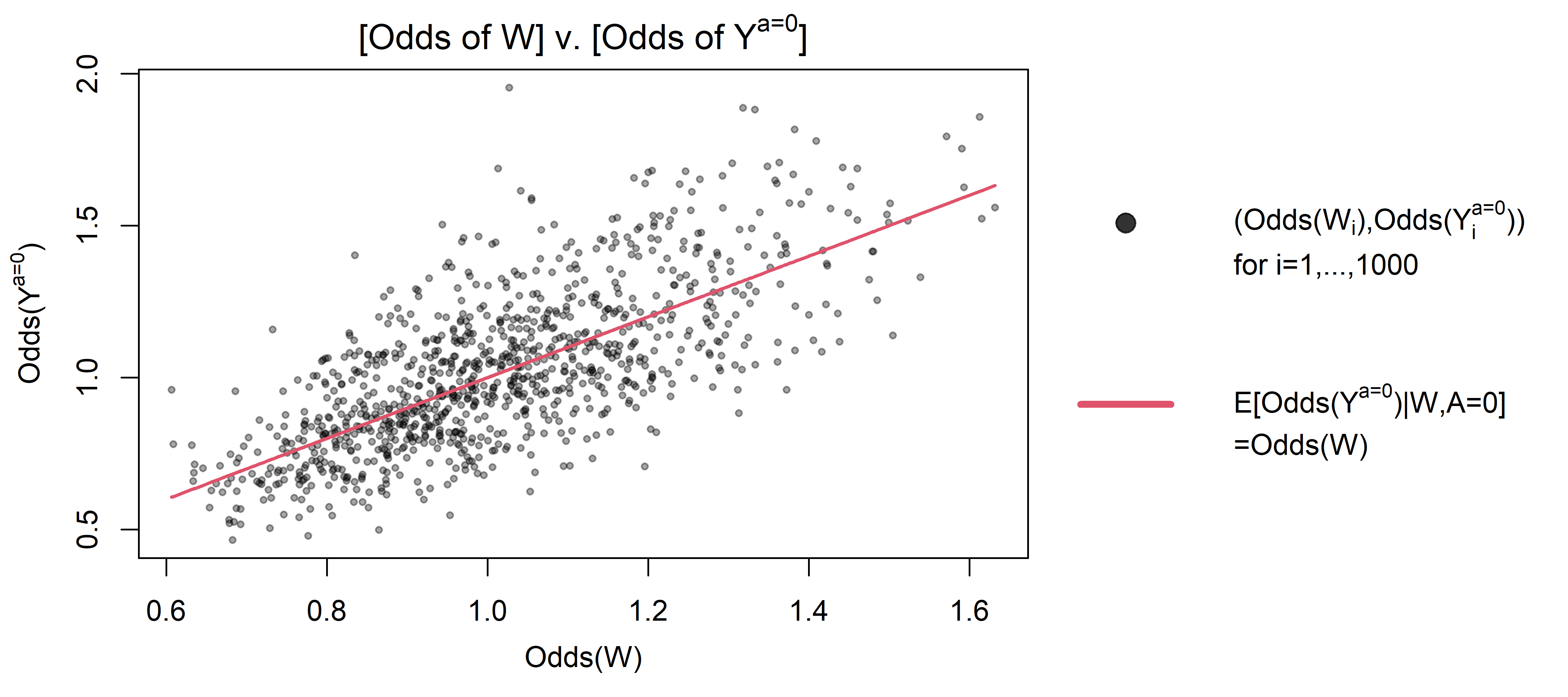}
\caption{A scatterplot of $\text{Odds}(W_i)$ and $\text{Odds}(Y_i^{a=0})$ for $i=1,\ldots,N$. The red solid line shows the deterministic relationship $\text{Odds}(W=w) = \EXP \{ \text{Odds}(Y^{a=0}) \cond W=w,A=0 \}$.}
\label{fig-scatter}
\end{figure}

Recall that Result \ref{result-1} is given by
\begin{align}
\frac{p^*\big(  W,X\big)  }{1-p^*\big(  W,X\big)  } &  =E\left\{  \omega^* (Y,X)  \, \big| \, W,X,A=0\right\}
=\sum_{y}\frac{\pi^* \big(  y,X\big)  }{1-\pi^* \big(  y,X\big)  }f^*\big(
y|W,X,A=0\big)  \tag{\ref{EPS equation}}
\end{align}
where $f^*\big(  y \cond W,X,A=0\big)  $ is the conditional law of $Y$ given $(W,X,A)$ evaluated at $Y=y$. To illustrate, we consider the following data generating process:
\begin{itemize}
\item $A \sim \text{Ber}(0.5)$
\item $Y^{a=0} \sim N(0.25 A,1)$
\item $W=Y^{a=0} + \epsilon$ where $\epsilon \sim N(0,1)$ with $\epsilon \indep (Y^{a=0},A)$
\end{itemize}
We find the odds function relating $W$ and $A$ and that relating $Y^{a=0}$ and $A$ are given by
\begin{align*}
&
\text{Odds}(W=w)
=
\frac{p^*(w)}{1-p^*(w)}
=
\frac{ \Pr(A=1 \cond W=w) }{\Pr(A=0 \cond W=w) }
=
\frac{ \phi(w \con 0.25,2) }{\phi(w \con 0,2 ) }
\\
&
\text{Odds}(Y^{a=0}=y)
=
\frac{ \pi^*(y) }{ 1-\pi^*(y) }
=
\frac{ \Pr(A=1 \cond Y^{a=0} = y) }{ \Pr(A=0 \cond Y^{a=0} = y) }
=
\frac{ \phi(y \con 0.25,1) }{\phi(y \con 0,1 ) }
\end{align*}
where $\phi(v \con \mu,\sigma^2)$ is the density of $N(\mu,\sigma^2)$. In addition, $\potY{0}{} \cond (W,A=0) \sim N(W/2,0.5)$. From straightforward algebra, we find $E \big\{ \text{Odds}(Y^{a=0}) \cond W,A=0 \big\} = \text{Odds}(W)$. We then generate $N=1000$ observations from the data generating process, and we draw a scatterplot of $\text{Odds}(W_i)$ and $\text{Odds}(Y_i^{a=0})$ in Figure \ref{fig-scatter}.

\subsection{Details of the Construction of Parametric Estimators via GMM}       \label{sec:supp:GMM}

In this Section, we provide details of the generalized method of moments (GMM,  \citet{Hansen1982_GMM}), which is implemented in many publicly available software programs, such as \texttt{gmm} package in R \citep{GMM_Rpackage}. 
Before we present details, we introduce additional notations. 
Let $N$ be the number of observed units, indexed by subscript $i = 1, \ldots N$. 
Let $O_i=(Y_i,A_i,W_i)$ be the observed data for the $i$th unit. Let $\AVER (h) $ denote the average of function $h(O)$ across $N$ units, i.e., $\AVER(g) = N^{-1} \sum_{i=1}^{N} g( O_i )$. 
For random variables $V$ and $Z$, let   $V \stackrel{D}{\rightarrow} Z$ denote $V$ converges to $Z$  in distribution. 

Let $\psi_1^\true = E ( \potY{1}{} \cond A=1 )$ and $ \psi_0^\true = E ( \potY{0}{} \cond A=1 )$. The additive and multiplicative additive average causal effects of treatment on the treated are represented as $\psi^\true := \psi_1^\true - \psi_0^\true$. Let $\theta$ be a vector of parameters of interest and $g$ be a vector-valued function that is restricted by the mean-zero moment restriction, i.e. $\EXP \big\{ g(O \con \theta^\dagger) \big\} = 0$ where $\theta^\dagger$ is the unique parameter that achieves the mean-zero moment restriction. 

\subsubsection{A Moment Restriction based on the EPS} \label{sec:para EPS}

The first approach entails a priori specifying a parametric logistic regression
model for the EPS, say:%
\begin{align*}
\log \omega(y,x \con \alpha)
=\log\frac{\pi\big(  y,x; \alpha\big)  }{1-\pi\big(y,x;\mathbf{\alpha}\big)  }= \mathbf{\alpha}\T S_{y}(y,x) \ , 
\end{align*}
where  $S_{y}\big(  y,x \big)  $ is a user-specified sufficient statistic for the log-odds ratio ratio association between $Y^{a=0}$ and $A$ evaluated at $Y^{a=0}=y$ given $X=x$. Options for specifying $S_{y}$ includes both more parsimonious
specifications, e.g., $S_{y}\big(  y ,x \big) = (1, y,x\T)\T$, as well as more flexible
specifications, say  $S_{y}\big(  y,x\big)  =\big(  1, y,y^{2},\ldots,y^{v},x\T,y x\T,\ldots,y^{v} x\T\big)\T
$ or 
\begin{align*}
&
S_{y}\big(  y,x\big)
=\big\{ 1, \ind( y\leq y_{1})  ,\ind(
y_{1}<y\leq y_{2})  ,\ldots,\ind(  y_{K}<y )
,
\ind( y\leq y_{1}) x\T 
,\ldots, 
\ind( y_{K}<y) x\T
\big\}\T  
\end{align*}
where
$y_{k}$ is the $k^{th}$ user-specified percentile of $Y$.  Alternative
nonparametric smoothing techniques, e.g. nearest neighbor, kernel smoothing, splines or wavelets might also be considered. We assume that $\omega^*(Y,X) = \omega(Y,X \con \alpha^*)$ for an unknown unique value $\alpha^*$.

While such logistic regression model specifications for the EPS might look familiar, estimation and inference about its unknown parameter
$\mathbf{\alpha}^*$ via maximum likelihood is complicated by the fact
that $Y^{a=0}$ $\ $is only observed among untreated units with $A=0.$ Instead,
to find an empirical estimate of $\mathbf{\alpha}$, we observe the following property of $\omega^*$ and $\pi^*$:
\begin{align*}
& 
\text{For any function } p(W,X) , \text{ we have }
\\
&
E\left[  \left\{ (1-A) \omega^*(Y,X) - A  \right\}  p \big(  W,X\big)  \right] 
=
E\left[  \left\{  \frac{  1-A  }{1-\pi^* ( Y,X )  }-1\right\}  p \big(  W,X\big)  \right]
=0 \ .
\end{align*}
The identity is trivial from Result \ref{result-1}, the law of iterated expectations, and some straightforward algebra:
\begin{align*}
E \big[  \big\{ (1-A) \omega^*(Y,X) \big\} p (W,X) \big]
&
=
E \big[ \Pr(A=0 \cond W,X) E \big\{  \omega^*(Y,X) \cond W,X,A=0 \big\} p (W,X) \big]
\\
&
=
E \big[ \Pr(A=0 \cond W,X) \frac{ \Pr(A=1 \cond W,X)  } { \Pr(A=0 \cond W,X) 
} p (W,X) \big]
\\
&
=
E \big[ \Pr(A=1 \cond W,X) p (W,X) \big]
\\
&
=
E \big[ A p (W,X) \big] \ , 
\end{align*}
and
\begin{align*}
\frac{(1-A) \pi^*(Y,X)}{1-\pi^*(Y,X)} - A
&
=
\frac{\pi^*(Y,X) - A \pi^*(Y,X) - A + A\pi^*(Y,X)}{1-\pi^*(Y,X)}
=
\frac{1 - A }{1-\pi^*(Y,X)}    -1 \ .
\end{align*}
Consequently, one can use the following moment restriction to estimate $\alpha^*$:
\begin{align} \label{eq-supp-EPS Moment}
E\left[  \left\{  \frac{  1-A  }{1-\pi ( Y,X \con \alpha )  }-1\right\}  r_{w}\big(  W,X\big)  \right]
=
0
\text{ for a user-specified function }r_{w}(W,X) 
\quad
\Leftrightarrow
\quad
\alpha = \alpha^*
\ .
\end{align}
We need to choose $r_{w}(w,x)$ so that $\dim(r_{w}) \geq \dim(\alpha)$, which is required to admit a unique $\alpha^*$.

\subsubsection{A Moment Restriction based on the COCA Confounding Bridge Function}

We consider a parametric model of 
$b\big(  W, X;\eta\big)  =\eta\T S_{w}\big(  W, X\big)  $ for user-specified function $S_{w}\big(  W, X\big)$, assuming that $b^*\big(  W, X\big)  =\eta\sT S_{w}\big(W,X\big)  $ for an unknown unique value $\eta^*$. Similar to $S_{y}\big(  y,x\big)  $ in the previous Section, $S_{w}\big(  W, X\big)  $
can likewise account for nonlinearities by specifying specific basis
functions, such as polynomials, splines, or wavelets. We then note that a
standard least-squares regression $Y$ on $(W,X)$ would generally fail to recover
$\eta^*$ even if the model is correctly specified. In other words, under
our assumptions it will generally be the case that $b^*\big(  W, X \big)
\neq E\big(Y|A=0,W,X\big)$, as this would in fact imply that  $E\left\{
E\big(  Y \cond A=0,W,X\big)  \cond A=0,Y=y,X\right\}  =y$ which does not follow from our assumptions.
A correct approach in fact is based on the following consequence of our
assumptions:%
\begin{align} \label{eq-supp-OR Moment}
E\big[  \big(  1-A\big)  \big\{  \eta\T S_{w}\big(  W, X\big)
-Y \big\}  r_{y}\big(  Y, X\big)  \big]  =0 \text{ for a user-specified
function }r_{y} 
\quad \Leftrightarrow \quad \eta = \eta^* \ .
\end{align} 
Again, we need to choose $r_{y}(Y,X)$ so that $\dim(r_{y}) \geq \dim (\eta)$ to admit a unique $\eta^*$. If only one NCO is available and $r_{y}$ is chosen as $r_{y}\big(  Y,X\big)  =S_{w}\big(Y,X\big)$, $\eta^*$ is represented as
\begin{align*}
\eta^*=E\big\{  S_{w}\big(  Y,X\big)  S_{w}\T\big(  W,X\big)
\cond A=0 \big\}  ^{-1}
E\big\{  YS_{w}\big(  Y,X\big)  \cond A= 0 \big\}    
\end{align*} 
assuming $E\big\{  S_{w}\big(  Y,X\big)  S_{w}\T\big(  W,X\big)
\cond A=0 \big\}$ is invertible.

\subsubsection{Construction of the GMM Estimators}

Based on the moment restrictions \eqref{eq-supp-EPS Moment} and \eqref{eq-supp-OR Moment}, we consider the following three moment functions $g$ and the corresponding parameters $\theta$:
\begin{itemize}[leftmargin=0cm]
\item (Extended Propensity Score)
\begin{align}       \label{eq:MomentEq:EPS}
\theta 
=
\begin{pmatrix}
\psi_1 \\ \psi_0 \\ \alpha
\end{pmatrix}
\in \R^{\dim(\alpha)+2}
\ , \
g_{\PS}(O \con \theta ) 
=
\begin{pmatrix}
A \big( Y - \psi_1 \big)
\\
(1-A)  \frac{ \pi (Y, X \con \alpha) }{1-\pi(Y, X \con \alpha)} (Y - \psi_0)
\\
\big\{ \frac{1-A}{1-\pi(Y, X \con \alpha)} - 1 \big\} r_{w}(W, X)
\end{pmatrix}
\in \R^{\dim(r_{w})+2}
\end{align}
where $ \dim(\theta) \leq  \dim(g_{\PS})$.
\item (COCA Confounding Bridge Function)
\begin{align}      \label{eq:MomentEq:OBF}
\theta 
=
\begin{pmatrix}
\psi_1 \\ \psi_0 \\ \eta
\end{pmatrix}
\in \R^{\dim(\eta)+2}
\ , \
g_{\OutReg} (O \con \theta ) 
=
\begin{pmatrix}
A \big( Y - \psi_1 \big)
\\
A \big\{ b(W, X \con \eta) - \psi_0 \big\}
\\
(1-A) \big\{ b(W, X \con \eta) - Y \big\} r_{y}(Y, X) 
\end{pmatrix}
\in \R^{\dim(r_{y})+2}
\end{align}
where $ \dim(\theta) \leq  \dim(g_{\OutReg})$.
\item (Doubly Robust)
\begin{align}     \label{eq:MomentEq:DR}
&
\theta 
=
\begin{pmatrix}
\psi_1 \\ \psi_0 \\ \alpha \\ \eta
\end{pmatrix}
\in \R^{\dim(\alpha)+\dim(\eta)+2}
\ , 
\nonumber
\\
&
g_{\DR}(O \con \theta ) 
=
\begin{pmatrix}
A \big( Y - \psi_1 \big)
\\
\frac{(1-A) \pi(Y,X \con \alpha)}{1-\pi(Y,X \con \alpha)} \{ Y - b(W,X \con \eta) \}
+
A \{ b(W,X \con \eta) - \psi_0 \}
\\
\big\{ \frac{1-A}{1-\pi(Y,X \con \alpha)} - 1 \big\} r_{w}(W,X)
\\
(1-A) \big\{ b(W,X \con \eta) - Y \big\} r_{y}(Y,X) 
\end{pmatrix}
\in \R^{\dim(r_{y})+\dim(r_{y})+2}
\end{align}
where $ \dim(\theta) \leq  \dim(g_{\DR})$.

\end{itemize}

The unique parameter $\theta^\dagger$ can be represented as the minimizer of the weighted squared norm of $\EXP \{ g(O \con \theta)\}$. Specifically, with a chosen weighting matrix $\Omega$ (which can depend on the observed data), $\theta^\dagger$ is represented as
\begin{align} \label{eq:GMM-population}
\theta^\dagger
=
\argmin_{\theta} 
\big[ \EXP \big\{ g(O \con \theta) \big\} \big] \T 
\Omega 
\big[ \EXP \big\{ g(O \con \theta) \big\} \big]
\end{align}
From the law of large numbers, we find $\psi_1^\dagger=\psi_1^\true$. In section \ref{sec:supp:proof:EPSID} of the Appendix, we show that $\psi_0^\dagger=\psi_0^\true$ for each estimation strategy if
\begin{itemize}
\item  (Extended Propensity Score) the EPS is correctly specified, i.e. 
\begin{align}   \label{eq:supp:EPS}
\pi(y , x \con \alpha^\dagger) = \Pr(A=1 \cond Y^{a=0} = y, X=x)
\end{align}
\item  (COCA Confounding Bridge Function) the COCA confounding bridge function is correctly specified, i.e. 
\begin{align}   \label{eq:supp:OBF}
y = \EXP \big\{ b(W, X \con \eta^\dagger) \cond Y=y, X=x, A=0 \big\}
\end{align}
\item (Doubly Robust) either the EPS or the COCA confounding bridge function is correctly specified; i.e. either \eqref{eq:supp:EPS} or \eqref{eq:supp:OBF} is satisfied.
\end{itemize}
For efficiency, $\Omega$ is chosen as $\Sigma^{-1}$, the inverse of the variance matrix of $g(O \con \theta^\dagger)$, i.e. $\Omega = \Sigma^{-1} = \big[ \EXP \big\{ g(O \con \theta^\dagger) g(O \con \theta^\dagger)\T \big\} \big]^{-1}$.

The GMM estimator is obtained from the empirical analog of the minimization task in \eqref{eq:GMM-population}. Specifically, we get
\begin{align}       \label{eq:supp:GMMsolve}
\widehat{\theta}_{\GMM}
=
\argmin_{\theta} 
\big[ \AVER \big\{ g(O \con \theta) \big\} \big] \T 
\widehat{\Omega}
\big[ \AVER \big\{ g(O \con \theta) \big\} \big]
\end{align}
where $\widehat{\Omega}$ is often chosen as a matrix that is consistent for $\Sigma^{-1}$; see equation \eqref{eq:supp:OptOmega} below for the exact form.

Under regularity conditions (see \citet{NMcF1994} for details), the GMM estimator is consistent and asymptotically normal (CAN) as follows:
\begin{align*}
& 
\sqrt{N}
\big(
\widehat{\theta}_{\GMM} - \theta^\dagger
\big)
\stackrel{D}{\rightarrow}
N\big( 0, 
\Sigma_{\GMM}
\big)
\\
& 
\Sigma_{\GMM}
=
(G\T \Omega G)^{-1} ( G \T \Omega \Sigma \Omega\T G) (G\T \Omega G)^{-1}
\ , \
G = 
\EXP 
\bigg\{
\frac{\partial g(O \con \theta) }{\partial \theta\T}
\bigg|_{\theta=\theta^\dagger}
\bigg\}
\in \mathbbm{R}^{ \dim(g) \times \dim(\theta) }
\end{align*}
In particular, $\Sigma_{\GMM}$ reduces to $\Sigma_{\GMM} = (G\T \Sigma^{-1} G)^{-1}$ if $\Omega= \Sigma^{-1}$, which is the most efficient in the class of all GMM estimators. Therefore, we obtain the consistency and asymptotic normality of the GMM-based estimator of $\psi^\true = \psi_1^\true-\psi_0^\true$ as follows:
\begin{align*}
&
\sqrt{N} \big( \widehat{\psi}_{\GMM} - \psi^\true \big)
\stackrel{D}{\rightarrow}
N \big(
0 , v\T \Sigma_{\GMM} v
\big)
\ , \\
&
v = \big(
-1 , 1  , 0  , \cdots , 0
\big)\T
\ , \
\widehat{\psi}_{\GMM} = v\T \widehat{\theta}_{\GMM}
\ , \ 
\psi^\true = v\T \theta^\true
\end{align*}
The variance estimator of $\widehat{\psi}_{\GMM}$ can be obtained from the empirical analog, i.e. $\widehat{\VAR}(\widehat{\psi}_{\GMM}) = v\T \widehat{\Sigma}_{\GMM} v /N $ where
\begin{align}
&
\widehat{\Sigma}_{\GMM}
=
(\widehat{G}\T \widehat{\Omega} \widehat{G})^{-1} 
( \widehat{G} \T \widehat{\Omega} \widehat{\Sigma} \widehat{\Omega}\T \widehat{G}) 
(\widehat{G}\T \widehat{\Omega} \widehat{G})^{-1}
\ , 
\nonumber
\\
&
\widehat{G} = 
\AVER
\bigg\{
\frac{\partial g(O \con \theta) }{\partial \theta}
\bigg|_{\theta=\widehat{\theta}_{\GMM}}
\bigg\}
\ , 
\quad \quad 
\widehat{\Sigma}
=
\AVER \big\{ g(O \con \widehat{\theta}_{\GMM}) g(O \con \widehat{\theta}_{\GMM})\T \big\}
\label{eq:supp:OptOmega}
\end{align}
If we choose $\widehat{\Omega} = \widehat{\Sigma}^{-1}$, $\widehat{\Sigma}_{\GMM}$ reduces to $\widehat{\Sigma}_{\GMM} =  (\widehat{G}\T \widehat{\Omega} \widehat{G})^{-1} $.

Since $\widehat{\Omega}$ depends on the parameter $\theta$, it may be difficult to obtain the GMM estimator by directly working on \eqref{eq:supp:GMMsolve}. Therefore, a popular procedure is the following two-step approach in that:
\begin{itemize}
\item[(1)] Compute the preliminary GMM estimator $\widehat{\theta}_{\GMM}^{(pre)}$ from \eqref{eq:supp:GMMsolve} where the weight matrix $\widehat{\Omega}$ is chosen as the identity matrix or other positive definite fixed matrix. 
\item[(2)] Compute the GMM estimator $\widehat{\theta}_{\GMM}$ from \eqref{eq:supp:GMMsolve} where the weight matrix $\widehat{\Omega}^{(pre)}$ is chosen as $\widehat{\Omega}^{(pre)} = \big[ \AVER \big\{ g(O \con \widehat{\theta}_{\GMM}^{(pre)}) g(O \con \widehat{\theta}_{\GMM}^{(pre)})\T \big\} \big]^{-1}$. 
\end{itemize}
In the second step, $\widehat{\Omega}^{(pre)}$ is consistent for $\Sigma^{-1}$ because $\widehat{\theta}_{\GMM}^{(pre)}$ is consistent (even though it is inefficient). One can repeat the iterative procedure multiple times until convergence criteria are satisfied.

\subsubsection{Local Efficiency of the Doubly-robust GMM Estimator}

Recall that $\mathcal{M}$ is defined as 
\begin{align*}
\mathcal{M}
=
\Bigg\{
P \, \Bigg| \,  
\begin{array}{l}
\text{$P(O)$ is regular and Assumption \ref{condition-1} holds,}\\
\text{i.e.,  there exists $b^*(W,X)$ solving integral equation \eqref{Outcome COCA}}  
\end{array}
\Bigg\} \ .
\end{align*} 
We show that the doubly robust GMM estimator is semiparametric locally efficient in $\mathcal{M}$ at the intersection model $\mathcal{M}_{\pi} \cap \mathcal{M}_{b}$ where 
\begin{align*}
\mathcal{M}_{\pi}: & \quad \text{$\pi(Y,X \con \alpha)$ is correctly specified, i.e., $\pi(Y,X \con \alpha^*) = \pi^*(Y,X)$}
\\
\mathcal{M}_{b}: & \quad \text{$b(W,X \con \eta)$ is correctly specified, i.e., $b(Y,X \con \eta^*) = b^*(W,X)$} \ .
\end{align*}
Moreover, we choose $r_a(W , X)$ and $r_w(Y,X)$ as 
\begin{align*}
&
r_a(W,X) 
=
\EXP \bigg[ 
\frac{(1-A) \nabla_{\alpha} \pi(Y,X \con \alpha)}{ \{ 1-\pi(Y,X \con \alpha) \}^2} 
\, \bigg| \, 
W,X \bigg] \in \mathbbm{R}^{\dim(\alpha)}
\ ,
\\
&
r_w(Y,X) 
=
\EXP \big\{ 
(1-A) \nabla_{\eta} b(W,X \con \eta)
\, \big| \, 
Y,X \big\} \in \mathbbm{R}^{\dim(\eta)} \ .
\end{align*}
Then, the moment equation $g_{\text{DR}}$ in \eqref{eq:MomentEq:DR} is just identified, i.e., $\dim(g_{\text{DR}}) = \theta$. Therefore, under regularity conditions, (see \citet{NMcF1994} for details), we obtain
\begin{align*}
\frac{1}{\sqrt{N}}
\big(
\widehat{\theta}_{\GMM} - \theta^*
\big)
=
\frac{1}{N}
\sum_{i=1}^{N}
\Big\{ 
- G^{-1} g_{\text{DR}}(O_i \con \theta^*)
\Big\} + o_P(1) \ ,
\end{align*} 
where $G$ is
{\footnotesize
\begin{align*}
G
=
\EXP 
\bigg\{
\frac{\partial g_{\text{DR}} (O \con \theta) }{\partial \theta\T}
\bigg|_{\theta=\theta^*}
\bigg\}
=
\EXP 
\left(
\begin{array}{cccc}
-A & 0 & 0 & 0  \\
0 & -A & \HT{g1} & \HT{g2} \\
0 & 0 & * & 0 \\
0 & 0 & 0 & * 
\end{array}
\right)
\end{align*}    }
Here, $\HL{g1}$ and $\HL{g2}$ are zero as follows:
\begin{align*}
\HL{g1}
& 
=
E \Bigg[ 
(1-A) \nabla_{\alpha} \bigg\{ \frac{\pi(Y,X \con \alpha)}{1-\pi(Y,X \con \alpha)} \bigg\}
\big\{ Y - b(W,X \con \eta) \big\}
\Bigg] 
\\&
=
E \Bigg[ 
(1-A) \nabla_{\alpha} \bigg\{ \frac{\pi(Y,X \con \alpha)}{1-\pi(Y,X \con \alpha)} \bigg\}
\underbrace{ E \big\{ Y - b(W,X \con \eta) \cond Y,A=0,X \big\} }_{=0}
\Bigg] 
=
0
\end{align*}
and
\begin{align*}
\HL{g2}
& =
E \Bigg[ 
-
(1-A)  \frac{\pi(Y,X \con \alpha)}{1-\pi(Y,X \con \alpha)}
\nabla_{\eta} b(W,X \con \eta)
+ A \nabla_{\eta} b(W,X \con \eta)
\Bigg]
\\
&
=
E \Big[ 
-
\pi(Y^{a=0},X \con \alpha)
E \big\{ \nabla_{\eta} b(W,X \con \eta) \cond Y^{a=0},X \big\}
+ A \nabla_{\eta} b(W,X \con \eta)
\Big]
\\
&
=
E \Big[ 
-
A \nabla_{\eta} b(W,X \con \eta)
+ A \nabla_{\eta} b(W,X \con \eta)
\Big]
=
0 \ .
\end{align*}
Consequently, the influence function of $\widehat{\theta}_{\GMM}$ reduces to
{\footnotesize
\begin{align*}
& 
- G^{-1} g_{\text{DR}}(O_i \con \theta^*)
\\
&
= 
\left(
\begin{array}{cccc}
\frac{1}{\Pr(A=1)} & 0 & 0 & 0  \\
0 & \frac{1}{\Pr(A=1)} & 0 & 0 \\
0 & 0 & * & 0 \\
0 & 0 & 0 & * 
\end{array}
\right)
\begin{pmatrix}
A \big( Y - \psi_1^* \big)
\\
\frac{(1-A) \pi(Y,X \con \alpha)}{1-\pi(Y,X \con \alpha)} \{ Y - b(W,X \con \eta) \}
+
A \{ b(W,X \con \eta) - \psi_0^* \}
\\
\big\{ \frac{1-A}{1-\pi(Y,X \con \alpha)} - 1 \big\} r_a(W,X)
\\
(1-A) \big\{ b(W,X \con \eta) - Y \big\} r_{w}(Y,X) 
\end{pmatrix}
=
\begin{pmatrix}
\frac{A ( Y - \psi_1^* )}{\Pr(A=1)}
\\
\InfFt(O \con \pi^*, \eta^*)
\\
*
\\
*
\end{pmatrix} \ .
\end{align*}}%
Therefore, the estimator $\widehat{\psi}_1-\widehat{\psi}_0$ achieves the semiparmetric local efficiency bound for the ETT $\psi^*$ under $\mathcal{M}$ at the submodel $\mathcal{M}_{\text{sub}}$ so long as the posited parametric models belong to the intersection model $\mathcal{M}_{\pi} \cap \mathcal{M}_{b}$.

\subsubsection{Regularized GMM Estimators}

Finding the optimal solution in \eqref{eq:supp:GMMsolve} can be quite challenging when dealing with finite samples, especially if the objective function is not convex. To mitigate the issue, one may regularize \eqref{eq:supp:GMMsolve} and solve a penalized version of GMM, e.g.,
\begin{align} \label{eq-penalized GMM}
&
\widehat{\theta}_{\GMM} (\lambda)
=
\argmin_{\theta} 
\big[ \AVER \big\{ g(O \con \theta) \big\} \big] \T 
\widehat{\Omega}
\big[ \AVER \big\{ g(O \con \theta) \big\} \big] 
+
\lambda \mathcal{R}(\theta)
\ .
\end{align}    
where $\mathcal{R}(\theta)$ is a regularization term and $\lambda \in (0,\infty)$ is a regularization parameter, which is chosen from cross-validation; see Algorithm \ref{alg:CV lambda} below for details.
\begin{algorithm}[!htb]
\begin{algorithmic}[1]
\REQUIRE Candidates of Regularization Parameters $\{ \lambda_1,\ldots,\lambda_L\}$
\FOR{$\ell=1,\ldots,L$}
\FOR{$i=1,\ldots,N$}
\STATE Obtain $\widehat{\theta}_{\ell}^{(-i)}$ from \eqref{eq-penalized GMM} using a regularization parameter $\lambda_\ell$ and $N-1$ observations $\{1,\ldots,i-1,i+1,\ldots,N\}$.
\STATE Evaluate $g(O_i \con \widehat{\theta}_\ell^{(-i)})$
\ENDFOR    
\STATE Obtain the average moment $\overline{g}_\ell = \big\| N^{-1} \sum_{i=1}^{N} g(O_i \con \widehat{\theta}_\ell^{(-i)}) \big\|_2^2$
\ENDFOR
\STATE Choose $\ell$ that minimizes $\overline{g}_\ell$, i.e., $\ell^* = \argmin_{\ell} \overline{g}_\ell$ 
\RETURN Regularization parameter $\lambda = \lambda_{\ell^*}$
\end{algorithmic}
\caption{Cross-validation for Choosing $\lambda$}
\label{alg:CV lambda}
\end{algorithm}

\subsection{Sufficient Conditions for the Existence and Uniqueness of the COCA Confounding Bridge Function} \label{sec:supp:Exist h}

In this Section, we discuss sufficient conditions for the existence and uniqueness of the solutions to the integral equations in equations \eqref{EPS equation} and \eqref{Outcome COCA}:
\begin{align*}
&y=E\left\{  b^*\big(  W,X\big)  \cond Y=y,X=x,A=0\right\}
\\
&\frac{p^*\big(  w,x\big)  }{1-p^*\big(  w,x\big)  } =E\left\{  \omega^* (Y,x)  \, \big| \, W=w,X=x,A=0\right\}
\end{align*}
Since both equations share a similar structure, our focus here is primarily on the case of the COCA bridge function $b^*$. 

\subsubsection{Conditions for Existence}

We begin by providing sufficient conditions for the existence of the COCA confounding bridge function $b^*$. In brief, we follow the approach in \citet{Miao2018}. The proof relies on Theorem 15.18 of \citet{Kress2014}, which is stated below for completeness.\\[0.25cm]
\noindent
\textbf{Theorem 15.18.} \citep{Kress2014}
Let $A:X \rightarrow Y$ be a compact operator with singular system $\big\{ \mu_n,\phi_n,g_n \big\}_{n=1,2,\ldots}$. The integral equation of the first kind $A\phi = f$ is solvable if and only if 
\begin{align*}
& 1. \quad
\text{$f \in \mathcal{N}(A^{\text{adjoint}})^\perp = \big\{ f \, \big| \, A^{\text{adjoint}}(f) = 0 \big\}^{\perp}$}
\ , 
&&
2. \quad
\text{$\sum_{n=1}^{\infty} \mu_n^{-2} \big| \langle f,g_n \rangle |^2 < \infty$}
\end{align*}

To apply the Theorem, we introduce some additional notations. For a fixed $X=x$, let $\mathcal{L}_{Wx}$ and $\mathcal{L}_{Yx}$ be the spaces of square-integrable functions of $W$ and $Y$, respectively, which are equipped with the inner products 
\begin{align*}
&
\langle h_1, h_2 \rangle_{Wx} = \int h_{1x}(\bw) h_{2x}(\bw) \, f_{W|XA} (\bw \cond X=x,A=0) \, d\bw = \EXP \big\{ h_{1x}(\bW) h_{2x}(\bW) \cond X=x, A=0 \big\} 
\\
&
\langle g_{1x}, g_{2x} \rangle_{Yx} = \int g_{1x}(y) g_{2x}(y) \, f_{Y|XA} (y \cond X=x,A=0) \, dy = \EXP \big\{ g_{1x}(Y) g_{2x}(Y) \cond X=x,A=0 \big\} \ .
\end{align*}
Let $\mathcal{K}_x: \mathcal{L}_{Wx} \rightarrow \mathcal{L}_{Yx}$ be the conditional expectation of $h( W ) \in \mathcal{L}_{Wx}$ given $(Y,X=x,A=0)$, i.e.,
\begin{align*}
\mathcal{K}_x (h) \in \mathcal{L}_{Yx} 
\text{ satisfying }
\big( \mathcal{K}_x(h) \big) (y) 
=
\EXP \big\{ h(W) \cond Y=y,X=x,A=0 \big\}
\text{ for } h \in \mathcal{L}_{Wx} \ .
\end{align*}
Then, the COCA confounding bridge function evaluated at $X=x$, i.e., $b_x^* \in \mathcal{L}_{W}$, solves $\mathcal{K}_x ( b_x^* ) = [\text{identity map}:Y \rightarrow Y] \in \mathcal{L}_{Yx} $, i.e., 
\begin{align*}
E\left\{  b_x^*\big(  W \big)  \cond Y=y,X=x,A=0\right\}  
=
\int b_x^*(\bw) f_{W |YXA} (\bw \cond y,x,0 ) \, d\bw = y , \ \forall y \ .
\end{align*}

Now, we assume the following conditions for each $x$:
\begin{itemize}[leftmargin=0.4cm, itemsep=0cm]
\item[] \HT{Bridge-1} $\iint
f_{W |YXA } (\bw \cond y,x,0 )
f_{Y|WXA} (y \cond \bw,x,0 )
\, d\bw \, d y
< \infty$;
\item[] \HT{Bridge-2} For $g_x \in \mathcal{L}_{Yx}$, $\EXP \big\{ g_x(Y) \cond W,X=x,A=0 \big\} = 0$ implies $g_x(Y)= 0$ almost surely;
\item[] \HT{Bridge-3} $\EXP \big( Y^2 \cond X=x,A=0 \big) < \infty$;
\item[] \HT{Bridge-4} Let the singular system of $\mathcal{K}_x$ be $\big\{ \mu_{n,x},\phi_{n,x},g_{n,x} \big\}_{n=1,2,\ldots}$. Then, we have 
\begin{align*}
\sum_{n=1}^{\infty} \mu_{n,x}^{-2} \big| \langle Y ,g_{n,x} \rangle_{Yx} |^2 < \infty \text{ almost surely. }
\end{align*}
\end{itemize}

First, we show that $\mathcal{K}_x$ is a compact operator under Condition \HL{Bridge-1} for each $x$. Let $\mathcal{K}_x^{\text{adjoint}} : \mathcal{L}_{Yx} \rightarrow \mathcal{L}_{Wx}$ be the conditional expectation of $g_x(Y) \in \mathcal{L}_{Yx}$ given $(W,X,A=0)$, i.e., 
\begin{align*}
\mathcal{K}_x^{\text{adjoint}} (g) \in \mathcal{L}_{Wx} 
\text{ satisfying }
\big( \mathcal{K}_x(g) \big) (\bw) 
=
\EXP \big\{ g(Y) \cond \bW=\bw,X=x,A=0 \big\}
\text{ for } g \in \mathcal{L}_{Yx} \  .
\end{align*}
Then, $\mathcal{K}_x$ and $\mathcal{K}_x^{\text{adjoint}}$ are the adjoint operator of each other as follows:
\begin{align*}
\langle \mathcal{K}_x(h) , g \rangle_{Yx}
& =
\EXP
\big[
\EXP \big\{ h(W) \cond Y,X,A=0 \big\}
g(Y)
\big]	
\\
& 
=
\EXP
\big\{
h(W) g(Y)
\cond X,A=0
\big\}	
\\
&
=
\EXP
\big[
h(W) \EXP \big\{ g(Y) \cond W,X,A=0 \big\}
\big]	
=
\langle h , \mathcal{K}_x^{\text{adjoint}}(g) \rangle_{Wx} \ .
\end{align*}
Additionally, as shown in page 5659 of \citet{Carrasco2007}, $\mathcal{K}_x$ and $\mathcal{K}_x^{\text{adjoint}}$ are compact operators under Condition \HL{Bridge-1}. Moreover, by Theorem 15.16 of \citet{Kress2014}, there exists a singular value decomposition of $\mathcal{K}_x$ as $\big\{ \mu_{n,x},\phi_{n,x},g_{n,x} \big\}_{n=1,2,\ldots}$. 

Second, we show that $\mathcal{N}(\mathcal{K}_x^{\text{adjoint}})^\perp = \mathcal{L}_{Yx}$, which suffices to show $\mathcal{N}(\mathcal{K}_x^{\text{adjoint}}) = \big\{ 0 \big\} \subseteq \mathcal{L}_{Yx}$. Under Condition \HL{Bridge-2}, we have 
\begin{align*}
g \in \mathcal{N}(\mathcal{K}_x^{\text{adjoint}})
\quad 
\Rightarrow 
\quad 
\EXP \big\{ g (Y) \cond W = w,X,A=0 \big\}
=
0, \ \forall \bw
\quad \Rightarrow
\quad
g(Y) = 0 \text{ almost surely.}
\end{align*}
where the first arrow is from the definition of the null space $\mathcal{N}$, and the second arrow is from Condition \HL{Bridge-2}. Therefore, any $g \in \mathcal{N}(\mathcal{K}_x^{\text{adjoint}})$ must satisfy $g(y) = 0 $ almost surely, i.e., $\mathcal{N}(\mathcal{K}_x^{\text{adjoint}})= \big\{ 0 \big\} \subseteq \mathcal{L}_{Yx}$ almost surely. 

Third, from the definition of $\mathcal{L}_{Wx}$, $g(Y) = Y \in \mathcal{L}_{Yx} = \mathcal{N}(\mathcal{K}_x^{\text{adjoint}})^\perp $ under Condition \HL{Bridge-3}.

Combining the three results, we establish that $Y$ satisfies the first condition of Theorem 15.18 of \citet{Kress2014}. The second condition of the Theorem is exactly the same as Condition \HL{Bridge-4}. Therefore, we establish that the Fredholm integral equation of the first kind $\mathcal{K}_x ( h ) = [\text{identity map}:Y \rightarrow Y] $ is solvable under Conditions \HL{Bridge-1}-\HL{Bridge-4}. Consequently, for each $x$, there exists a function $b_x^*(W)$ satisfying
\begin{align*}
\EXP \big\{ b_x^*(W) \cond Y=y,X=x,A=0 \big\} = y \ , \ \forall y \ .
\end{align*}
We define $b^*(W,X)$ as a function satisfying $b^*(W=w,X=x) = b_x^*(w)$, which then solves
\begin{align*}
\EXP \big\{ b^*(W,X) \cond Y=y,X=x,A=0 \big\} = y \ , \ \forall y \ .
\end{align*}

\subsubsection{Conditions for Uniqueness} 

The COCA confounding bridge function is unique under the following completeness condition:
\begin{itemize}[leftmargin=0cm]
\item[] \HT{Completeness}: Suppose that $\EXP \big\{ g(W) \cond Y=y,X=x,A=0 \big\} = 0$ for any $(y,x)$. Then, $g(W)=0$ almost surely. 
\end{itemize}
The proof is given below. Suppose that $b_1^*(W,X)$ and $b_2^*(W,X)$ satisfy the bridge function condition \eqref{Outcome COCA}. Then, we find
\begin{align*}
0 = Y - Y = \EXP \big\{ b_1^*(W,x) - b_2^*(W,x) \cond Y=y,X=x,A=0 \big\} \ .
\end{align*}
From \HL{Completeness}, $\Delta_{x} (W) = b_1^*(W,x) - b_2^*(W,x) = 0$ almost surely for any $x$. This implies $b_1^*(W,X)=b_2^*(W,X)$ almost surely. Therefore, the COCA confounding bridge function is unique.

\subsection{Details of the Minimax Estimation} \label{sec:supp:MMEstimation}

\subsubsection{Closed-form Representations of the Minimax Estimators}

For notational brevity, let $M_k = |\mathcal{I}_k^c|$. Recall $\omega^*$ and $b^*$ satisfy
\begin{align*}
&
E\left[  \left\{ (1-A) \omega^*(Y,X) - A  \right\}  p \big(  W,X\big)  \right] = 0 \quad , \quad \forall p \ ,
\\
&
E\big[  \big(  1-A\big)  \big\{  Y - b^*(W,X) \big\}  q \big(  Y, X\big)  \big]  =0
\quad , \quad \forall q \ .
\end{align*}
Following \citet{Ghassami2022}, minimax estimators of $\omega^*$ and $b^*$ are given by
\begin{align*}
& 
\widehat{\omega}\LSS (\cdot)
=
\argmin_{\omega \in \mathcal{H}(Y,X)}
\Bigg[
\displaystyle{ \max_{p \in \mathcal{H}(W,X)}  }
\Bigg[
\AVER\LSS \bigg[
\begin{array}{l}
p (W,X)
\big\{
(1-A) \omega(Y,X)
-A
\big\} \\
- p^2(W,X)
\end{array}
\bigg]
-
\lambda_{p} \big\| p \big\|_{\mathcal{H}}^2
\Bigg] 
+
\lambda_{\omega} \big\| \omega \big\|_{\mathcal{H}}^2 
\Bigg] 
\ ,
\\
& 
\widehat{b}\LSS (\cdot)
=
\argmin_{h \in \mathcal{H}(\bW,\bX)}
\Bigg[ 
\displaystyle{ \max_{q \in \mathcal{H}(Y,X)}  }
\Bigg[
\AVER\LSS \bigg[
\begin{array}{l}         
q (Y,X)
(1-A)
\big\{
Y - b(W,X)
\big\} \\
- q^2(Y,X) 
\end{array} \bigg]
-
\lambda_{q} \big\| q \big\|_{\mathcal{H}}^2
\Bigg]
+
\lambda_{b} \big\| b \big\|_{\mathcal{H}}^2 
\Bigg] 
\ .
\end{align*} 
Note that all true and estimated functions are defined in terms of the following representations:
\begin{align} \label{eq-Ghassami}
&
f^*(v_f)
\text{ satisfies }
E \big[ \big\{ S - D \cdot f^*(V_f) \big\} g(V_g) \big] = 0 
\text{ for any } g \ ,
\nonumber 
\\
& 
\widehat{f} \LSS (v_f)
\nonumber
\\
&
=
\argmin_{f \in \mathcal{H}(V_f)}
\Big[
\max_{g \in \mathcal{H}(V_g)}
\Big[
\AVER\LSS \big[  \big[ g (V_g) \cdot \big\{ S - D \cdot f (V_f) \big) \big\} - g^2 (V_g) \big] - \lambda_g \big\| g \big\|_{\mathcal{H}_g}^2
\Big]
+ \lambda_f \big\| f \big\|_{\mathcal{H}_f}^2
\Big] \ ,
\end{align}
where $S,D,V_f,V_g$ are appropriately chosen. Speicifcally, for the weight function $\omega^*$, we choose $V_f=(Y,X)$, $V_g=(W,X)$, $S=A$, $D=1-A$; for the COCA confounding bridge function $b^*$, we choose $V_f = (W,X)$, $V_g = (Y,X)$, $S = (1-A)Y$, $D=1-A$. 

A closed-form representation of the solution to \eqref{eq-Ghassami} is available from the represented theorem \citep{KW1970, SHS2001}. Specifically, we have $\widehat{f} \LSS (v_f) = \sum_{i \in \mathcal{I}_k^c} \widehat{\gamma}_i \mathcal{K} (V_{f,i}, v_{f})$ where $\widehat{\gamma} = \big( \widehat{\gamma}_i \big)_{i \in \mathcal{I}_k^c}$ is equal to
\begin{align*}
& \widehat{\gamma}
= 
\big(
\mathcal{F} \mathcal{D}_{T} \Gamma \mathcal{D}_{T}   \mathcal{F}
+ M_k^2 \lambda_f  \mathcal{F} \big)^{\dagger}
\mathcal{F} \mathcal{D}_{T} \Gamma \mathcal{S} \ , 
\end{align*}
where 
\begin{align*}
& 
\Gamma = 0.25 \mathcal{G} \big\{  M_k^{-1}   \mathcal{G} + \lambda_g I_{M_k \times M_k} \big\}^{-1}
\in \R^{M_k \times M_k}
\\
& \mathcal{F}
=
\Big[
\mathcal{K} \big( V_{f,i}, V_{f,j} \big)
\Big]_{i,j \in \mathcal{I}_k^c}  
\in \R^{M_k \times M_k}
\quad , 
&&
\mathcal{G}
=
\Big[
\mathcal{K} \big( V_{g,i}, V_{g,j} \big)
\Big]_{i,j \in \mathcal{I}_k^c}  
\in \R^{M_k \times M_k}
\\
&
\mathcal{D} = \text{diag} \Big[ 
D_i
\Big]_{i \in \mathcal{I}_k^c}  \in \R^{M_k \times M_k}
\quad , 
&&
\mathcal{S} = \Big[ S_i \Big]_{i \in \mathcal{I}_k^c}  \in \R^{M_k} \ .
\end{align*}
Therefore, minimax estimators of the nuisance functions are readily available by appropriately choosing $(V_f,V_g,S,D)$. 

\subsubsection{Cross-Validation}

We use cross-validation to select the hyperparameters $ \mathfrak{h} = (\kappa_f,\kappa_g, \lambda_f,\lambda_g)$ that minimize an empirical risk evaluated over a validation set $\mathcal{V}$, denoted by $\mathcal{R}_(\mathcal{V}) (\mathfrak{h})$. For the empirical risk, we can either use (i) the projected risk \citep{Dikkala2020, Ghassami2022}, or (ii) the V-statistic \citep{PMMR2021}. To motivate (i), we remark that \citet{{Dikkala2020, Ghassami2022}} defined the population-level projected risk of \eqref{eq-Ghassami} as $ \EXP \big[ \big\| \EXP \big[ D \big\{ \widehat{f}\LSS(V_f) - f^*(V_f) \big\} \cond V_g \big] \big\|_2^2 \big] $. In addition, they showed that its empirical counterpart evaluated over a validation set $\mathcal{V}$ is given by $
\widehat{R}_{\mathcal{V}} (\mathfrak{h}) = \{ \widehat{\epsilon}_{\mathcal{V}}\LSS \} \T \Gamma_{\mathcal{V}} \{ 	\widehat{\epsilon}_{\mathcal{V}}\LSS \}$
where
\begin{align*}
&
\widehat{\epsilon}_{\mathcal{V}}\LSS
=
\big[ S_{j} - D_{j} \cdot  \widehat{f}\LSS(V_{f,j}) \big]_{j \in \mathcal{V} } \in \R^{|\mathcal{V}|} \ , \\
&
\Gamma_{\mathcal{V}} = 0.25 \mathcal{G}_{\mathcal{V}} \big\{  |\mathcal{V}|^{-1}   \mathcal{G}_{\mathcal{V}} + \lambda_g I_{|\mathcal{V}| \times |\mathcal{V}|} \big\}^{-1}
\in \R^{|\mathcal{V}| \times |\mathcal{V}|}
\ , 
&&
\mathcal{G}_{\mathcal{V}}
= \big[ \mathcal{K} (V_{g,i},V_{g,j}) \big]_{ i , j \in \mathcal{V} } \in \R^{|\mathcal{V}| \times |\mathcal{V}|} 
\ .
\end{align*} 

To motivate the V-statistic, we begin by introducing Lemma 2 of \citet{PMMR2021}:\\
\textbf{Lemma 2} \citep{PMMR2021}: Suppose that
\begin{align}	\label{eq-kernel bound}
\EXP
\big[
\big\{
S - D \cdot f(V_f)
\big\}^2 \mathcal{K} (V_g, V_g )
\big] < \infty \ .
\end{align}
Then, we have
\begin{align*}
&
\max_{ 
\substack{ 
g \in \mathcal{H}(V_g)
\\ 
g: \| g  \| \leq 1} }
\Big[
\EXP
\big[
g (V_g)
\big\{
S - D \cdot f(V_f)
\big\}
\big] \Big]^2
&
=
\EXP \left[
\big\{ S - D \cdot f(V_f) \big\}
\big\{ S' - D' \cdot f(V_f') \big\}
\times
\mathcal{K} ( V_g, V_g' )
\right] \ ,
\end{align*}
where $(S',D',V_f',V_g')$ are independent copies of $(S,D,V_f,V_g)$. 
The result is also reported in other works, e.g., Theorem 3.3 of \citet{Muandet2020} and Lemma 1 of \citet{Zhang2023}. The condition \eqref{eq-kernel bound} implies that $\EXP \big[ 
\big\{
S - D \cdot f(V_f)
\big\} \mathcal{K} (V_g, \cdot ) \big]$ is Bochner integrable \citep[Definition A.5.20]{SVM2008}. One important property of the Bochner integrability is that an integration and a linear operator can be interchanged. Therefore, we find 
\begin{align*}
\max_{\substack{ g \in \HH(V_g) \\ g: \| g  \| \leq 1}}
\Big[
{\EXP}
\big[
g (V_g)
\big\{
S - D \cdot f(V_f)
\big\}
\big] \Big]^2
&
=
\max_{\substack{ g \in \HH(V_g) \\ g: \| g  \| \leq 1}}
\Big[
{\EXP}
\big[
\big\{
S - D \cdot f(V_f)
\big\}
\langle g, \mathcal{K} ( V_g, \cdot ) \rangle
\big] \Big]^2
\\
&
=
\max_{\substack{ g \in \HH(V_g) \\ g: \| g  \| \leq 1}}
\Big[ 
\Big \langle g, 
{\EXP}
\big[ 
\big\{
S - D \cdot f(V_f)
\big\} \mathcal{K} ( V_g , \cdot ) 
\big]
\Big \rangle
\Big]^2
\\
&
=
\Big\|
{\EXP}
\big[ 
\big\{
S - D \cdot f(V_f)
\big\}
\mathcal{K} ( V_g, \cdot ) 
\big] 
\Big\|_{\HH(V_g) }^2
\\
&
=
\Big\langle
{\EXP}
\big[ 
\big\{
S - D \cdot f(V_f)
\big\}
\mathcal{K} ( V_g , \cdot ) 
\big] 
, 
{\EXP}
\big[ 
\big\{
S - D\cdot f(V_f)
\big\}
\mathcal{K} ( V_g , \cdot ) 
\big] 
\Big\rangle
\\
&
=
{\EXP}
\left[ 
\Big \langle 
\big\{
S - D \cdot f(V_f)
\big\}
\mathcal{K} ( V_g , \cdot )  
, 
{\EXP} 
\big[
\big\{
S' - D \cdot f(V_f')
\big\}
\mathcal{K} ( V_g', \cdot )  
\big] \Big \rangle
\right] 
\\
&
=
{\EXP}
\left[ 
\Big \langle 
\big\{
S - D \cdot f(V_f)
\big\}
\mathcal{K} ( V_g , \cdot )  
,  
\big\{
S' - D \cdot f(V_f')
\big\}
\mathcal{K} ( V_g', \cdot )   \Big \rangle
\right] 
\\
&
=
\EXP \left[
\big\{ S - D \cdot f(V_f) \big\}
\big\{ S' - D' \cdot f(V_f') \big\}
\times
\mathcal{K} ( V_g, V_g' )
\right]  \ .
\end{align*}
The first line holds from $g \in \HH(V_g)$, implying that $g(V_g)	 = \langle g , \mathcal{K}( V_g, \cdot ) \rangle$. 
The second line holds from the Bochner integrability.
The third line holds from the fact that $\HH(V_g)$ is a vector space, and ${\EXP}
\big[ 
\big\{
S - D \cdot f(V_f)
\big\} \mathcal{K} ( V_g, \cdot ) 
\big] \in \HH(V_g)$ from the Bochner integrability. Therefore, by choosing $g \propto {\EXP}
\big[ 
\big\{
S - D \cdot f(V_f)
\big\} \mathcal{K} ( V_g, \cdot )  \big]$, we obtain the result. 
The fourth line is trivial from the definition of the norm $\| \cdot \|_{\HH(V_g)}$. 
The fifth and sixth lines are from the Bochner integrability. 
The last line is trivial. Using this risk function, one may use the following empirical risk evaluated over a validation set $\mathcal{V}$ to find hyperparameters:
\begin{align} \label{eq-Vstat}
\mathcal{R}_{\mathcal{V}}(\mathfrak{h}) = 
\sum_{j,j' \in \mathcal{V}}
\big\{ S_{j} - D_{j} \cdot  \widehat{f}\LSS(V_{f,j}) \big\}
\big\{ S_{j'} - D_{j'} \cdot  \widehat{f}\LSS(V_{f,j'}) \big\}
\mathcal{K}( V_{f,j}, V_{f,j'}) \ . 
\end{align}
Using these empirical risks, we may choose the hyperparameters by following Algorithm \ref{alg:CV MM} below:
\begin{algorithm}[!htb]
\begin{algorithmic}[1]
\REQUIRE Candidates of Hyperparameters $\{ \mathfrak{h}_1,\ldots,\mathfrak{h}_L \}$
\STATE Split $\mathcal{I}_k^c$ into non-overlapping $K$ folds $\{ \mathcal{V}_1,\ldots,\mathcal{V}_K\}$
\FOR{$\ell=1,\ldots,L$}
\FOR{$c=1,\ldots,C$}
\STATE Estimate the nuisance functions $\omega$ and $b$ using hyperparameters $\mathfrak{h}_\ell$ and observations in $\big\{ \mathcal{V}_1,\ldots,\mathcal{V}_{c-1},\mathcal{V}_{c+1},\ldots,\mathcal{V}_{C}\}$
\STATE Evaluate the empirical risk $\mathcal{R}_{\mathcal{V}_c}(\mathfrak{h}_\ell)$ over the remaining set $\mathcal{V}_c$
\ENDFOR
\STATE Evaluate the average empirical risk $\overline{\mathcal{R}} (\mathfrak{h}_\ell) = C^{-1} \sum_{c=1}^{C} \mathcal{R}_{\mathcal{V}_c}(\mathfrak{h}_\ell)$
\ENDFOR    
\STATE Choose $\ell$ that minimizes $\overline{\mathcal{R}}$, i.e., $\ell^* = \argmin_{\ell} \overline{\mathcal{R}}(\mathfrak{h}_{\ell})$
\RETURN Hyperparameter $\mathfrak{h}=\mathfrak{h}_{\ell^*}$
\end{algorithmic}
\caption{Cross-validation for Choosing Hyperparameters}
\label{alg:CV MM}
\end{algorithm}

\subsubsection{A Remedial Strategy to Address Widely Varying Nuisance Functions} \label{sec:supp:remedial}

We present a remedial strategy to address widely varying $\omega^*$. First, following Section \ref{sec:para EPS}, we may obtain a GMM estimator of $\omega^*$, denoted by $\widetilde{\omega}\LSS(y,x)$, using observations in $\mathcal{I}_k^c$. If the range of $\widetilde{\omega}\LSS$ is excessively wide (e.g., the maximum is larger than ten times the minimum), we recommend estimating $r^*(x,x) = \omega^*(y,x)/ \widetilde{\omega}\LSS(y,x)$ instead of $\omega^*$; note that $r$ stands for the ratio. If $\widetilde{\omega}\LSS(y,x)$ is a good estimator, $r^*(y,x)$ would be close to 1, and thus, the minimax estimator of $r^*$ would vary less than the minimax estimator of $\omega^*$. Motivated by this phenomenon, we obtain a minimax estimator of $\omega^*$ as $\widehat{\omega}\LSS (y,x) 
=
\widetilde{\omega}\LSS(y,x) \cdot \widehat{r}(y,x)$
where $\widehat{r}\LSS(y,x)$ is estimated from the following minimax estimation strategy:
\begin{align*}
& \widehat{r}\LSS(y,x)
\\
&
=
\argmin_{r \in \mathcal{H}(Y,X)}
\left[ 
\begin{array}{l}
\displaystyle{ \max_{p \in \mathcal{H}(W,X)}  }
\Big[
\AVER\LSS \big[
p (W,X)
\big\{
(1-A) r(Y,X) \widetilde{\omega}\LSS(Y,X)
-A
\big\} - p^2(W,X) \big]
-
\lambda_{p} \big\| p \big\|_{\mathcal{H}(W,X)}^2
\Big]
\\
+
\lambda_{r} \big\| r \big\|_{\mathcal{H}(Y,X)}^2 
\end{array} 
\right]  \ .
\end{align*} 

\subsubsection{Multiplier Bootstrap}

One can obtain a standard error of the minimax estimator $\widehat{\psi}$ and confidence intervals for the ETT $\psi^*$ based on the multiplier bootstrap procedure; see Algorithm \ref{alg:MB} for details:
\begin{algorithm}[!htb]
\begin{algorithmic}[1]
\REQUIRE Number of bootstrap estimates $B$
\STATE For $i = 1 ,\ldots, N$, obtain the estimated influence functions
\begin{align*}
\widehat{\Psi}(O_i)
& =
\frac{ 
A_i \big\{ Y_i -  \widehat{b}\LSS\big(W_i, X_i \big) - \widehat{\psi}\LSS \big\} - \big(  1-A_i\big) \widehat{\omega}\LSS(Y_i,X_i)
\big\{  Y_i-\widehat{b}\LSS\big(
W_i,X_i\big)  \big\}  }{ \sum_{i=1}^{N} A_i/N } 
\end{align*}
where $k$ is the index that satisfies $ i \in \mathcal{I}_k$
\FOR{$b=1,\ldots,B$}
\STATE Generate i.i.d. random variables $\epsilon_{i}^{(b)} \sim N(0,1)$ for $i=1,\ldots,N$
\STATE Calculate $\widehat{e}^{(b)} = N^{-1} \sum_{i=1}^{N} \epsilon_{i}^{(b)} \widehat{\Psi} (\bO_i)$
\ENDFOR
\STATE Let $\widehat{\sigma}_{\text{boot}}^{2}$ be the empirical variance of $\big\{ \widehat{e}^{(b)} \cond b=1,\ldots,B \big\}$
\STATE Let $\widehat{q}_{\text{boot},\alpha}$ be the $100\alpha$-th percentile of $\big\{ \widehat{e}^{(b)} \cond b=1,\ldots,B \big\}$
\RETURN Variance estimate $\widehat{\sigma}_{\text{boot}}^{2}$; $100(1-\alpha)$\% confidence interval $[\widehat{q}_{\text{boot},\alpha/2}, \widehat{q}_{\text{boot},1-\alpha/2}]$
\end{algorithmic}
\caption{Multiplier Bootstrap Procedure}
\label{alg:MB}
\end{algorithm}

\subsubsection{Median Adjustment} \label{sec:supp:median}

Lastly, the cross-fitting estimator depends on a specific sample split, and thus, may produce outlying estimates if some split samples do not represent the entire data. To mitigate this issue, \citet{Victor2018} proposes to use so-called median adjustment from multiple cross-fitting estimates, which is implemented as follows. First, let $\widehat{\psi}_{s}$ $(s=1,\ldots,S)$ be the $s$th cross-fitting estimate with a variance estimate $\widehat{\sigma}_{s}^2$. Then, the median-adjusted cross-fitting estimate and its variance estimate are defined as follows: 
\begin{align} \label{eq-median}
& \widehat{\psi}_{\median}
:=
\median_{s=1,\ldots,S} \widehat{\psi}_{s}
\ , \quad \widehat{\sigma}_{\median}^2
:=
\median_{s=1,\ldots,S} \big\{ \sigma_s^2 + (\widehat{\psi}_s - \widehat{\psi}_{\median} )^2 \big\} \ .
\end{align}
These estimates are more robust to the particular realization of sample partition.

\subsection{Details of the Specifications for the Zika Virus Application}       \label{sec:supp:data}

In this Appendix, we present a detailed explanation of the Zika virus application. We remark that the replication code is available at \url{https://github.com/qkrcks0218/SingleProxyControl}, and the code implements the estimation process outlined below. 

Recall that the variables are defined as follows:
\begin{itemize}[leftmargin=0.25cm]
\item $N=617$: Number of municipalities
\item $A_i=1$: municipality $i$ belongs to Pernambuco (treated)
\item $A_i=0$: municipality $i$ belongs to Rio Grande do Sul (control)
\item $Y_i$: Birth rate of municipality $i$ in 2016
\item $W_{1i}$: Birth rate of municipality $i$ in 2014
\item $W_{2i}$: Birth rate of municipality $i$ in 2013
\item $X_i=(X_{1i},X_{2i},X_{3i})\T$: A vector of municipality $i$'s log population, log population density, proportion of females in 2014
\end{itemize}

\subsubsection{GMM Estimators} \label{sec:supp:GMM-Zika}

We first provide details of GMM estimators. We consider three specifications according to how NCOs are used:
\begin{itemize}[leftmargin=0.25cm]
\item (NCO 1) We use $W_1$, birth rate in 2013, as an NCO. 
\item (NCO 2) We use $W_2$, birth rate in 2014, as an NCO. 
\item (NCO 3) We use $(W_1,W_2)$, birth rates in 2013 and 2014, as NCOs. 
\end{itemize}

According to the specifications, we use penalized GMM in \eqref{eq-penalized GMM} to estimate the ETT, where the nuisance functions and basis functions in the moment function $g$ (\eqref{eq:MomentEq:EPS},\eqref{eq:MomentEq:OBF},\eqref{eq:MomentEq:DR}) are specified is specified as follows:
\begin{itemize}[leftmargin=0.25cm]
\item (NCO 1) 
\begin{itemize}[leftmargin=0.25cm]

\item $\pi(Y,X \con \alpha)
=
\text{expit}
\big\{ (1,Y,X\T)\alpha \big\}$ $\Leftrightarrow$ $\omega(Y,X \con \alpha) = \exp \big\{ (1,Y,X\T)\alpha \big\}$, $ \alpha = (\alpha_0,\alpha_1,\ldots,\alpha_4)\T \in \R^{5}$

\item $r_{y}(Y,X)=(1,Y,X\T,Y X\T)\T \in \R^{8}$

\item $b(W_1,X \con \eta) = (1,W_1,X\T) \eta$, $ \eta = (\eta_0,\eta_1,\ldots,\eta_4) \in \R^{5}$

\item $r_{w}(W_1,X) = (1,W_1,X\T,W_1 X\T)\T \in \R^{8}$    
\end{itemize}

\item (NCO 2)
\begin{itemize}[leftmargin=0.25cm]

\item $\pi(Y,X \con \alpha)
=
\text{expit}
\big\{ (1,Y,X\T)\alpha \big\}$ $\Leftrightarrow$ $\omega(Y,X \con \alpha) = \exp \big\{ (1,Y,X\T)\alpha \big\}$, $ \alpha = (\alpha_0,\alpha_1,\ldots,\alpha_4)\T \in \R^{5}$

\item $r_{y}(Y,X)=(1,Y,X\T,Y X\T)\T \in \R^{8}$

\item $b(W_2,X \con \eta) = (1,W_2,X\T) \eta$, $ \eta = (\eta_0,\eta_1,\ldots,\eta_4) \in \R^{5}$

\item $r_{w}(W_2,X) = (1,W_2,X\T,W_1 X\T)\T \in \R^{8}$

\end{itemize}

\item (NCO 3)
\begin{itemize}[leftmargin=0.25cm]
\item $\pi(Y,X \con \alpha)
=
\text{expit}
\big\{ (1,Y,X\T)\alpha \big\}$ $\Leftrightarrow$ $\omega(Y,X \con \alpha) = \exp \big\{ (1,Y,X\T)\alpha \big\}$, $ \alpha = (\alpha_0,\alpha_1,\ldots,\alpha_4)\T \in \R^{5}$

\item $r_{y}(Y,X)=(1,Y,X\T,Y X\T)\T \in \R^{8}$

\item $b(W_1,W_2,X \con \eta) = (1,W_1,W_2,X\T) \eta$, $ \eta = (\eta_0,\eta_1,\ldots,\eta_6) \in \R^{5}$

\item $r_{w}(W_2,X) = (1,W_1, W_2,X\T,W_1 X\T,W_2 X\T)\T \in \R^{12}$

\end{itemize}
\end{itemize}
The penalization term $\mathcal{R}(\theta)$ in equation \eqref{eq-penalized GMM} is determined based on the specific moment function used:
\begin{itemize}
\item ($g_{\PS}$) $\mathcal{R}(\theta) = \| \alpha_{-0} \|_{2}^{2}$ where $\alpha_{-0} = (\alpha_1,\ldots,\alpha_4)$
\item ($g_{\OutReg}$) $\mathcal{R}(\theta) = \| \eta_{-0} \|_{2}^{2}$ where $\eta_{-0} = (\eta_1,\ldots,\eta_{\dim(W)+3})$ 
\item ($g_{\DR}$) $\mathcal{R}(\theta) = \| \alpha_{-0} \|_{2}^{2} + \| \eta_{-0} \|_{2}^{2}$
\end{itemize}
The regularization parameter $\lambda$ is chosen from cross-validation in Algorithm \ref{alg:CV lambda}. Table summarizes the choice of $\lambda$:
\begin{table}[!htp]
\renewcommand{\arraystretch}{1.2} \centering
\setlength{\tabcolsep}{7pt}
\begin{tabular}{|c|ccc|}
\hline
\multirow{2}{*}{Moment Function} & \multicolumn{3}{c|}{NCO}                                              \\ \cline{2-4} 
& \multicolumn{1}{c|}{$W_1$} & \multicolumn{1}{c|}{$W_2$} & $(W_1,W_2)$ \\ \hline
$g_{\PS}$                        & \multicolumn{1}{c|}{$10^{-3.5}$}      & \multicolumn{1}{c|}{$10^{-4}$}      &     $10^{-4}$        \\ \hline
$g_{\OutReg}$                    & \multicolumn{1}{c|}{$10^{-4}$}      & \multicolumn{1}{c|}{$10^{-4}$}      &    $10^{-3.5}$         \\ \hline
$g_{\DR}$                        & \multicolumn{1}{c|}{$10^{-3.5}$}      & \multicolumn{1}{c|}{$10^{-3.5}$}      &  $10^{-4.5}$           \\ \hline
\end{tabular}
\caption{Choice of $\lambda$ Obtained from Cross-validation in Algorithm \ref{alg:CV lambda}} 
\label{tab:Lambda}
\end{table}

\subsubsection{Minimax Estimators}

We next provide details of minimax estimators. First, we employ the cross-fitting procedure with $K=2$ split samples. Second, the hyperparameters are chosen based on 5-fold cross-validation in Algorithm \ref{alg:CV MM} using the V-statistic in \eqref{eq-Vstat} as a criterion. Third, we employ the remedial strategy in Section \ref{sec:supp:remedial} where $\widetilde{\omega}$ is estimated from GMM using $g_{\PS}$ in Section \ref{sec:supp:GMM-Zika} where the regularization parameter $\lambda$ is set to $10^{-3}$. Lastly, we implement median adjustment in Section \ref{sec:supp:median} based on 500 cross-fitting estimates.

\subsubsection{Summary of the Result}

Table \ref{Tab-Zika Full} summarizes corresponding results. We find that the twelve COCA estimates vary between $-1.833$ and $-3.599$, meaning between $1.833$ and $3.599$ birth per 1,000 persons were reduced in PE due to the Zika virus outbreak, an empirical finding better aligned with the scientific hypothesis that Zika may likely adversely impact the birth rates of exposed populations. We remark that parametric estimators utilizing the COCA bridge function yield larger effect sizes compared to the other three COCA estimators. However, the estimates show minimal variability across different NCO specifications. Regardless of the estimator used, all the estimates support the conclusion that the Zika virus outbreak likely led to a decline in the birthrate in the affected regions of Brazil.

\begin{table}[!htp]
\renewcommand{\arraystretch}{1.1} \centering
\setlength{\tabcolsep}{7pt}
\small
\begin{tabular}{|cc|c|ccc|}
\hline
\multicolumn{2}{|c|}{\multirow{2}{*}{Estimator}}                                                                                       & \multirow{2}{*}{Statistic} & \multicolumn{3}{c|}{NCO}                                              \\ \cline{4-6} 
\multicolumn{2}{|c|}{}                                                                                                                 &                            & \multicolumn{1}{c|}{$W_1$} & \multicolumn{1}{c|}{$W_2$} & $(W_1,W_2)$ \\ \hline

\multicolumn{2}{|c|}{\multirow{3}{*}{Semiparametric}} & Estimator & \multicolumn{1}{c|}{$-2.410$} & \multicolumn{1}{c|}{$-2.182$} & $-2.180$ \\ \cline{3-6} 
\multicolumn{1}{|c}{} & & SE & \multicolumn{1}{c|}{$0.356$} & \multicolumn{1}{c|}{$0.503$} & $0.342$ \\ \cline{3-6} 
\multicolumn{1}{|c}{} & & 95\% CI & \multicolumn{1}{c|}{$(-3.107,-1.713)$} & \multicolumn{1}{c|}{$(-3.168,-1.196)$} & $(-2.850,-1.510)$ \\ \hline 
\multicolumn{1}{|c|}{\multirow{9}{*}{Parametric}} & \multirow{3}{*}{EPS} & Estimator & \multicolumn{1}{c|}{$-2.334$} & \multicolumn{1}{c|}{$-2.298$} & $-2.462$ \\ \cline{3-6} 
\multicolumn{1}{|c|}{} & & SE & \multicolumn{1}{c|}{$0.365$} & \multicolumn{1}{c|}{$0.492$} & $0.457$ \\ \cline{3-6} 
\multicolumn{1}{|c|}{} & & 95\% CI & \multicolumn{1}{c|}{$(-3.050,-1.618)$} & \multicolumn{1}{c|}{$(-3.261,-1.334)$} & $(-3.357,-1.567)$ \\ \cline{2-6} 
\multicolumn{1}{|c|}{} & \multirow{3}{*}{\begin{tabular}[c]{@{}c@{}}COCA \\ bridge\\ function\end{tabular}} & Estimator & \multicolumn{1}{c|}{$-3.560$} & \multicolumn{1}{c|}{$-3.446$} & $-3.599$ \\ \cline{3-6} 
\multicolumn{1}{|c|}{} & & SE & \multicolumn{1}{c|}{$0.520$} & \multicolumn{1}{c|}{$0.545$} & $0.703$ \\ \cline{3-6} 
\multicolumn{1}{|c|}{} & & 95\% CI & \multicolumn{1}{c|}{$(-4.579,-2.542)$} & \multicolumn{1}{c|}{$(-4.516,-2.377)$} & $(-4.976,-2.222)$ \\ \cline{2-6} 
\multicolumn{1}{|c|}{} & \multirow{3}{*}{Doubly-robust} & Estimator & \multicolumn{1}{c|}{$-2.235$} & \multicolumn{1}{c|}{$-1.833$} & $-2.182$ \\ \cline{3-6} 
\multicolumn{1}{|c|}{} & & SE & \multicolumn{1}{c|}{$0.502$} & \multicolumn{1}{c|}{$0.519$} & $0.415$ \\ \cline{3-6} 
\multicolumn{1}{|c|}{} & & 95\% CI & \multicolumn{1}{c|}{$(-3.220,-1.250)$} & \multicolumn{1}{c|}{$(-2.850,-0.816)$} & $(-2.996,-1.368)$ \\ \hline
\multicolumn{2}{|c|}{\multirow{3}{*}{DiD under parallel trends}} & Estimator & \multicolumn{1}{c|}{$-1.156$} & \multicolumn{1}{c|}{$-1.041$} & $-1.041$ \\ \cline{3-6} 
\multicolumn{1}{|c}{} & & SE & \multicolumn{1}{c|}{$0.199$} & \multicolumn{1}{c|}{$0.195$} & $0.195$ \\ \cline{3-6} 
\multicolumn{1}{|c}{} & & 95\% CI & \multicolumn{1}{c|}{$(-1.546,-0.767)$} & \multicolumn{1}{c|}{$(-1.424,-0.658)$} & $(-1.424,-0.658)$ \\ \hline 

\end{tabular}
\vspace*{0.5cm}
\caption{Summary of Data Analysis. Values in ``Estimate'' row represent the estimates of the ETT $\psi^*$. Values in ``SE'' and ``95\% CI'' rows represent the standard errors (SEs) associated with the estimates and the corresponding 95\% confidence intervals (CIs), respectively. The reported values are expressed as births per 1,000 persons.}
\label{Tab-Zika Full}

\end{table}

\subsection{Sensitivity Analysis}       \label{sec:supp:sensitivity}

We provide details of the sensitivity analysis described in Section \ref{sec:Discussion} of the main paper. To perform such sensitivity analysis, one might consider a parametric approach in Section \ref{sec:supp:GMM} where the EPS model is slightly modified by including the NCO into the EPS to encode a departure from the assumption. For example, we may consider:  
\begin{align*}
& \log\frac{\Pr\big(  A=1|Y^{a=0}=y, X=x, W=w\big)  }{\Pr\big(  A=0|Y^{a=0}%
=y, X=x, W=w\big)  } 
= \alpha \sT S_{y}(y,x) + \alpha_w S_{w}\big(  w\big);
\end{align*}
where $S_{w}$ is specified by the user and  $\mathbf{\alpha}_{w}$ is a pre-specified sensitivity parameter encoding a hypothetical violation of a valid NCO assumption in the direction of  $S_{w}$. For instance, one might let  $S_{w}(W)=W$ and vary $\mathbf{\alpha}_{w}$ over a grid of values in the neighborhood of zero (which corresponds to the valid NCO assumption). For each hypothetical value of $\mathbf{\alpha}_{w}$, one would then re-estimate the EPS using methods described in the paper with the sensitivity function specified as an offset. 

We implement a sensitivity analysis in the context of the Zika virus application. 
We focus on the GMM estimator based on the EPS model using the following specifications:
\begin{itemize}[leftmargin=0.25cm]
\item (NCO 1) 
\begin{itemize}[leftmargin=0.25cm]

\item $\pi_{SA}(Y,W,X \con \alpha, \alpha_w)
=
\text{expit}
\big\{ (1,Y,X\T)\alpha + \alpha_w W_1 \big\}$ \\
$\Leftrightarrow$ $\omega_{SA} (Y,X \con \alpha, \alpha_w) = \exp \big\{ (1,Y,X\T)\alpha + \alpha_w W_1 \big\}$, $ \alpha = (\alpha_0,\alpha_1,\ldots,\alpha_4)\T \in \R^{5}$

\item $r_{y}(Y,X)=(1,Y,X\T,Y X\T)\T \in \R^{8}$

\item $b(W_1,X \con \eta) = (1,W_1,X\T) \eta$, $ \eta = (\eta_0,\eta_1,\ldots,\eta_4) \in \R^{5}$

\item $r_{w}(W_1,X) = (1,W_1,X\T,W_1 X\T)\T \in \R^{8}$    
\end{itemize}

\item (NCO 2)
\begin{itemize}[leftmargin=0.25cm]

\item $\pi_{SA}(Y,W,X \con \alpha, \alpha_w)
=
\text{expit}
\big\{ (1,Y,X\T)\alpha + \alpha_w W_2  \big\}$\\
$\Leftrightarrow$ $\omega_{SA} (Y,X \con \alpha, \alpha_w) = \exp \big\{ (1,Y,X\T)\alpha + \alpha_w W_2  \big\}$, $ \alpha = (\alpha_0,\alpha_1,\ldots,\alpha_4)\T \in \R^{5}$

\item $r_{y}(Y,X)=(1,Y,X\T,Y X\T)\T \in \R^{8}$

\item $b(W_2,X \con \eta) = (1,W_2,X\T) \eta$, $ \eta = (\eta_0,\eta_1,\ldots,\eta_4) \in \R^{5}$

\item $r_{w}(W_2,X) = (1,W_2,X\T,W_1 X\T)\T \in \R^{8}$

\end{itemize}

\item (NCO 3)
\begin{itemize}[leftmargin=0.25cm]
\item $\pi_{SA}(Y,W,X \con \alpha, \alpha_w)
=
\text{expit}
\big\{ (1,Y,X\T)\alpha  + \alpha_w(W_1+W_2) /2 \big\}$ \\
$\Leftrightarrow$ $\omega_{SA} (Y,X \con \alpha, \alpha_w) = \exp \big\{ (1,Y,X\T)\alpha   + \alpha_w(W_1+W_2) /2 \big\}$, $ \alpha = (\alpha_0,\alpha_1,\ldots,\alpha_4)\T \in \R^{5}$

\item $r_{y}(Y,X)=(1,Y,X\T,Y X\T)\T \in \R^{8}$

\item $b(W_1,W_2,X \con \eta) = (1,W_1,W_2,X\T) \eta$, $ \eta = (\eta_0,\eta_1,\ldots,\eta_6) \in \R^{5}$

\item $r_{w}(W_2,X) = (1,W_1, W_2,X\T,W_1 X\T,W_2 X\T)\T \in \R^{12}$

\end{itemize}
\end{itemize}
Therefore, using the EPS $\pi_{SA} (Y,W,X \con \alpha, \alpha_w)$ (here, subscript $SA$ emphasizes that the EPS is for the sensitivity analysis) with a specified sensitivity parameter value $\alpha_w \in \mathbbm{R}$, we redefine the moment equation used in the GMM procedure as follows:
\begin{align*}
g_{\PS}(O \con \theta ) 
& =
\begin{pmatrix}
A \big( Y - \psi_1 \big)
\\
(1-A) \frac{ \pi_{SA} (Y,W,X \con \alpha, \alpha_w) }{1-\pi_{SA}(Y,W,X \con \alpha, \alpha_w)} ( Y -\psi_0  )
\\
\big\{ \frac{1-A}{1-\pi_{SA}(Y,W,X \con \alpha, \alpha_w)} - 1 \big\} r_{w}(W,X)
\end{pmatrix} 
\ .
\end{align*}
Using this $g$ function, we use penalized GMM in \eqref{eq-penalized GMM} to estimate the ETT, where the penalization term $\mathcal{R}(\theta)$ is chosen as $\mathcal{R}(\theta) = \| \alpha_{-0} \|_{2}^{2}$ and $\lambda$ are chosen as the same value in Table \ref{tab:Lambda}.


Table \ref{Tab-2} summarizes the sensitivity analysis results.  We focus the results associated with $W_1$ because the other two cases can be interpreted in a similar manner. First, at $\alpha_W=0$, we have the same result as the EPS COCA estimates in Table \ref{Tab-Zika Full}. 
Second, the 95\% pointwise confidence intervals for the effect estimate include the null effect once the sensitivity parameter $\alpha_w$ is equal to 0.74. 
Third, the effect estimate is positive once the sensitivity parameter $\alpha_w$ is greater than 0.99.
Lastly, the lower bound of the 95\% confidence interval for the crude estimate $(2.953,3.816)$ overlaps with the 95\% pointwise confidence intervals for the effect estimate once $\alpha_w$  is greater than 4.19.

\begin{table}[!htp]
\renewcommand{\arraystretch}{1.2} \centering
\small
\setlength{\tabcolsep}{7pt}
\begin{tabular}{|c|c|c|c|c|c|}
\hline
NCO                          & Statistic            & COCA & COCA (Null) & COCA (Positive) & COCA (Crude) \\ \hline
 \multirow{4}{*}{$W_1$} & $\alpha_w$ & 0.000 & 0.690 & 1.490 & 2.600 \\ \cline{2-6}
 & $\alpha_y$ & 2.677 & 1.750 & 0.325 & -0.484 \\ \cline{2-6}
 & Estimate & -2.334 & -0.718 & 0.024 & 2.330 \\ \cline{2-6}
 & 95\% CI & (-3.050,-1.618) & (-1.444,0.007) & (-0.379,0.427) & (1.672,2.989) \\ \hline
 \multirow{4}{*}{$W_2$} & $\alpha_w$ & 0.000 & 0.690 & 0.980 & 4.370 \\ \cline{2-6}
 & $\alpha_y$ & 3.060 & 1.047 & 0.273 & -4.243 \\ \cline{2-6}
 & Estimate & -2.298 & -0.646 & 0.130 & 2.288 \\ \cline{2-6}
 & 95\% CI & (-3.261,-1.334) & (-1.303,0.011) & (-0.444,0.704) & (1.473,3.103) \\ \hline
 \multirow{4}{*}{$(W_1,W_2)$} & $\alpha_w$ & 0.000 & 1.040 & 1.320 & 7.030 \\ \cline{2-6}
 & $\alpha_y$ & 2.629 & 0.829 & 0.165 & -4.611 \\ \cline{2-6}
 & Estimate & -2.462 & -0.541 & 0.006 & 2.630 \\ \cline{2-6}
 & 95\% CI & (-3.357,-1.567) & (-1.083,0.001) & (-0.457,0.469) & (2.292,2.969) \\ \hline
\end{tabular}
\caption{Summary of Sensitivity Analysis. COCA column shows the estimation result at $\alpha_w=0$. COCA (Null) column shows the estimation result at the sensitivity parameter where the 95\% confidence interval starts to contain zero. COCA (Positive) column shows the estimation result at the sensitivity parameter where the effect estimate becomes positive. 
COCA (Crude) column shows the estimation result at the sensitivity parameter where the 95\% confidence interval starts to overlap with the 95\% confidence interval for the crude estimate  $(2.953,3.816)$. $\alpha_y$ rows show the coefficient of $Y$ in $\pi_{SA}$. All results are obtained based on the penalized GMM in equation \eqref{eq-penalized GMM} with the regularization parameters are chosen as Table \ref{tab:Lambda}.}
\label{Tab-2}
\end{table}

To interpret these results, consider the scenario where $(W_1,W_2)$ are used as NCOs. Note that, at the sensitivity parameter value $\alpha_w=4.006$, the EPS COCA inference would become empirically consistent with the crude estimate. At this specific value of $\alpha_w$, the estimated coefficient for $Y$ stands at $\alpha_y = -3.250$. Compared to $\alpha_y=2.629$ at $\alpha_w=0$, this suggests that it would take a substantial violation of our primary assumption for the crude estimate to have a causal interpretation; a possibility we do not believe credible. Likewise, our sensitivity analysis suggests that it would also take incorporating a relatively large violation of our primary assumption for the COCA estimator to become consistent with the sharp null hypothesis of no causal effect $Y^{a=1}=Y^{a=0}$, therefore indicating the presence of a strong common cause of exposure and baseline outcome, leading to a near doubling of the odds of exposure to Zika virus but which does not also confound the outcome of interest; which we believe to be unlikely.    

\subsection{Over-identification Test}       \label{sec:supp:over-identification}

We formalize an over-identification test discussed in Section \ref{sec:Discussion} of the paper. For simplicity, suppose that there are two potential NCOs, denoted by $W = (W_1,W_2)$, and that $W_1$ is known a priori to be a valid NCO. One can obtain two semiparametric estimators of $\psi^*$, denoted by $\widehat{\psi}_{[1]}$ and $\widehat{\psi}_{[1,2]}$, which are constructed by either using $W_1$ or both $(W_1,W_2)$ as NCOs; see Section \ref{sec:estimator} of the main paper or the Appendix \ref{sec:supp:GMM} for details on how these estimators are constructed. From Result \ref{result-CAN} and \eqref{eq:supp:OptOmega}, $\widehat{\psi}_{[1]}$ is CAN for $\psi^*$. Likewise, if $W_2$ is also a valid NCO, $\widehat{\psi}_{[1,2]}$ is also CAN for the ETT from Result \ref{result-CAN}. However, if $W_2$ is an invalid NCO, say it violates Assumption \ref{assumption:NC}-(iii), $\widehat{\psi}_{[1,2]}$ may be no longer CAN for $\psi^*$. This implies that, we have the following result under a null hypothesis $H_0: \text{$W_2$ is a valid NCO}$:
\begin{align*}
\sqrt{N} \Big\{ \widehat{\psi}_{[1]} - \widehat{\psi}_{[1,2]} \Big\}
\stackrel{D}{\rightarrow} N \big( 0, \varsigma^2 \big) \ .
\end{align*}
A consistent estimator for $\varsigma^2$, say $\widehat{\varsigma}^2$ can be obtained in two ways. First, if parametric estimators are used, $\widehat{\varsigma}^2$ can be constructed from a consistent GMM variance estimator. Second, if semiparametric estimators are used, $\widehat{\varsigma}^2$ can be chosen as the empirical variance of the difference between the two efficient influence functions where one is associated with only $W_1$ and the other is associated with both $W_1$ and $W_2$. Therefore, a statistical test for evaluating the null hypothesis $H_0$ is given by
\begin{align*}
& \text{Reject $H_0$: $W_2$ is a valid NCO at level $\alpha \in (0,1)$ if }
T = 
\bigg| 
\frac{\widehat{\psi}_{[1,2]} - \widehat{\psi}_{[1]}}{\widehat{\varsigma} / \sqrt{N}}
\bigg| \geq z_{1-\alpha/2}
\end{align*}
where $z_{a}$ is the $100a \%$ percentile of $N(0,1)$.

In the context of the Zika virus application, we conducted the over-identification test, and the results are summarized in Table \ref{tab:Over-identification}. The findings from all four estimators indicate that there is insufficient statistical evidence to conclude that either $W_1$ or $W_2$ is an invalid NCO, as long as the other is a valid NCO.

\begin{table}[!htp]
\renewcommand{\arraystretch}{1.2} \centering
\setlength{\tabcolsep}{3pt} \footnotesize
\begin{tabular}{|cc|c|c|c|}
\hline
\multicolumn{2}{|c|}{Estimator} & Statistic & \begin{tabular}[c]{@{}c@{}}$H_0$: $W_2$ is a valid NCO\\ {\scriptsize($W_1$ is known a priori as a valid NCO)}\end{tabular} & \begin{tabular}[c]{@{}c@{}}$H_0$: $W_1$ is a valid NCO\\ {\scriptsize($W_2$ is known a priori as a valid NCO)}\end{tabular} \\ \hline
\multicolumn{2}{|c|}{\multirow{3}{*}{Semiparametric}} & Estimator & $-0.230$ & $-0.002$ \\ \cline{3-5} 
\multicolumn{2}{|c|}{} & SE & $0.316$ & $0.487$ \\ \cline{3-5} 
\multicolumn{2}{|c|}{} & Test statistic & $0.729$ & $0.004$ \\ \hline
\multicolumn{1}{|c|}{\multirow{9}{*}{Parametric}} & \multirow{3}{*}{EPS} & Estimator & $0.128$ & $0.164$ \\ \cline{3-5} 
\multicolumn{1}{|c|}{} & & SE & $0.150$ & $0.226$ \\ \cline{3-5} 
\multicolumn{1}{|c|}{} & & Test statistic & $0.853$ & $0.728$ \\ \cline{2-5} 
\multicolumn{1}{|c|}{} & \multirow{3}{*}{\begin{tabular}[c]{@{}c@{}}COCA \\ bridge\\ function\end{tabular}} & Estimator & $0.039$ & $0.152$ \\ \cline{3-5} 
\multicolumn{1}{|c|}{} & & SE & $0.186$ & $0.482$ \\ \cline{3-5} 
\multicolumn{1}{|c|}{} & & Test statistic & $0.207$ & $0.317$ \\ \cline{2-5} 
\multicolumn{1}{|c|}{} & \multirow{3}{*}{Doubly-robust} & Estimator & $-0.053$ & $0.349$ \\ \cline{3-5} 
\multicolumn{1}{|c|}{} & & SE & $0.193$ & $0.246$ \\ \cline{3-5} 
\multicolumn{1}{|c|}{} & & Test statistic & $0.275$ & $1.418$ \\ \hline
\end{tabular}
\caption{Over-identification Test Results. The critical value at level $\alpha=0.05$ is $z_{0.975} = 1.960$.}
\label{tab:Over-identification}
\end{table}

\section{Proof} \label{sec:supp:Proof}

\subsection{Proof of \eqref{EPS ID}}           \label{sec:supp:proof:EPSID}

Suppose that the EPS is correctly specified, i.e. $\pi(y, x \con \alpha^\dagger) = \Pr(A=1 \cond Y^{a=0} = y, X=x)$. Then, 
\begin{align*}
&
\EXP \bigg\{ (1-A) Y \frac{ \pi(Y, X \con \alpha^\dagger)  } { 1- \pi(Y, X \con \alpha^\dagger)  }    \bigg\}
\\
&
=
\EXP  \bigg[ \EXP \bigg\{ (1-A) Y \frac{ \pi(Y, X \con \alpha^\dagger)  } { 1- \pi(Y, X \con \alpha^\dagger)  }    
\, \bigg| \, X \bigg\}
\bigg]
\\
& =
\EXP \bigg[ \EXP \bigg\{ (1-A) Y^{a=0} \frac{ \pi(Y^{a=0} , X \con \alpha^\dagger)  } { 1- \pi(Y^{a=0} , X \con \alpha^\dagger)  }    
\, \bigg| \, X \bigg\}
\bigg]
\\
& =
\EXP \bigg[ 
\EXP \bigg\{ (1-A) Y^{a=0} \frac{ \Pr(A=1 \cond Y^{a=0},X ) }{\Pr(A=0 \cond Y^{a=0},X )}  \, \bigg| \, X  \bigg\}
\bigg]
\\
& =
\EXP \bigg\{ \Pr(A=0 \cond Y^{a=0},X) 
\EXP \big( Y^{a=0} \cond X \big) 
\frac{ \Pr(A=1 \cond Y^{a=0},X ) }{\Pr(A=0 \cond Y^{a=0},X )}    \bigg\}
\\
& =
\EXP \big\{ \EXP(A \cond Y^{a=0},X) 
\EXP \big( Y^{a=0} \cond X \big) 
\big\}
\\
& =
\EXP \big\{ 
\EXP \big( A Y^{a=0} \cond X \big) 
\big\}
\\
& =
\EXP \big( A Y^{a=0} \big) 
\end{align*}
The first equality is from the law of iterated expectations. 
The second equality is from the consistency assumption, i.e. $Y^{a=0}=Y$ if $A=0$. The third equality holds when the EPS is correctly specified. The rest identities hold from the law of iterated expectations. Similarly, we obtain
\begin{align*}
\EXP \bigg\{ (1-A) \frac{ \pi(Y, X \con \alpha^\dagger)  } { 1- \pi(Y, X \con \alpha^\dagger)  }    \bigg\}
=
\EXP (A) 
\end{align*}
Consequently, we obtain
\begin{align*}
\frac
{ \EXP \big\{ (1-A) Y \frac{ \pi(Y,X \con \alpha^\dagger)  } { 1- \pi(Y,X \con \alpha^\dagger)  }    \big\} } 
{ \EXP \big\{ (1-A) \frac{ \pi(Y,X \con \alpha^\dagger)  } { 1- \pi(Y,X \con \alpha^\dagger)  }    \big\} }
=
\frac{  \EXP \big( A Y^{a=0} \big) } { \EXP (A) }
=
\EXP \big(  Y^{a=0} \cond A=1 \big)
\end{align*}
Therefore, the GMM estimator using $g_{\PS}$ leads to $\psi_0^\dagger=\EXP \big(  Y^{a=0} \cond A=1 \big)$ if the EPS is correctly specified.

\subsection{Proof of Result \ref{result-1} and \eqref{EPS  equation}}               \label{sec:supp:proof:EPSeq}
The proof is straightforward from the following algebra:
\begin{align*}
&
\EXP 
\bigg\{ 
\frac{ \Pr(A=1 \cond Y,X=x) }{ \Pr(A=0 \cond Y,X=x) }
\, \bigg| \, W=w, X=x, A=0
\bigg\}
\\
& \stackrel{(A)}{=}
\EXP 
\bigg\{ 
\frac{ \Pr(A=1 \cond Y^{a=0}, X=x) }{ \Pr(A=0 \cond Y^{a=0}, X=x) }
\, \bigg| \, W=w, X=x, A=0
\bigg\}
\\
& =
\int 
\frac{ \Pr(A=1 \cond Y^{a=0}, X=x) }{ \Pr(A=0 \cond Y^{a=0}, X=x) }
f (Y^{a=0}=y \cond W=w, X=x, A=0 ) \, dy
\\
& =
\int 
\frac{ \Pr(A=1 \cond Y^{a=0}, X=x) }{ \Pr(A=0 \cond Y^{a=0}, X=x) }
\frac{ f (Y^{a=0}=y,  W=w, X=x, A=0 ) }{ f (W=w, X=x, A=0 ) }
\, dy
\\
& =
\int 
\frac{ \Pr(A=1 \cond Y^{a=0}, X=x) }{ \Pr(A=0 \cond Y^{a=0}, X=x) }
\frac{ f ( W=w , A=0  \cond Y^{a=0}=y, X=x ) f(Y^{a=0}=y, X=x) }{ f (W=w, X=x, A=0 ) }
\, dy
\\
& \stackrel{(B)}{=}
\int 
\frac{ \Pr(A=1 \cond Y^{a=0}, X=x) }{ \Pr(A=0 \cond Y^{a=0}, X=x) }
\frac{ \bigg\{ 
\begin{array}{l}
f ( W=w  \cond Y^{a=0}=y, X=x ) \\[-0.25cm]
\times \Pr ( A=0  \cond Y^{a=0}=y, X=x ) f(Y^{a=0}=y, X=x)
\end{array}
\bigg\} }{ f (W=w, X=x, A=0 ) }
\, dy
\\
& =
\int 
\frac{ f ( W=w  \cond Y^{a=0}=y, X=x ) \Pr ( A=1  \cond Y^{a=0}=y, X=x ) f(Y^{a=0}=y, X=x) }{ f (W=w, X=x, A=0 ) }
\, dy
\\
& \stackrel{(B)}{=}
\int 
\frac{ f ( W=w , A=1 \cond Y^{a=0}=y, X=x ) f(Y^{a=0}=y, X=x) }{ f (W=w, X=x, A=0 ) }
\, dy
\\
& =
\int 
\frac{ f ( W=w , A=1 , Y^{a=0}=y, X=x ) }{ f (W=w, A=0, X=x ) }
\, dy
\\
& =
\frac{ f ( W=w , A=1, X=x ) }{ f (W=w, A=0, X=x ) }
\\
& = 
\frac{ \Pr(A=1 \cond W=w, X=x) }{\Pr(A=0 \cond W=w, X=x) }
\end{align*}
Equality (A) holds from the consistency assumption, i.e. $Y^{a=0}=Y$ if $A=0$. Equalities (B) holds from Assumption \ref{assumption:NC}. The rest identities are trivial.

\subsection{Proof of Result \ref{result-2}}

We have
\begin{align*}
\EXP \big\{ g(Y) (1-A) \omega^*(Y,X) \big\}
&
=
\EXP \big\{ g(Y^{a=0}) (1-A) \omega^*(Y^{a=0},X) \big\}
\\
&
=
\EXP \big\{ g(Y^{a=0}) \Pr(A=0 \cond Y^{a=0},X)  \omega^*(Y^{a=0},X) \big\}
\\
&
=
\EXP \big\{ g(Y^{a=0}) \Pr(A=1 \cond Y^{a=0},X)  \big\}
\\
&
=
\EXP \big\{ g(Y^{a=0}) A \big\} \ .
\end{align*}
Therefore, by taking $g(y)=y$ and $g(y)=1$, we obtain
\begin{align*}
    \frac{ \EXP \big\{ g(Y) (1-A) \omega^*(Y,X) \big\} }{ \EXP \big\{ (1-A) \omega^*(Y,X) \big\} }
    =
    \frac{ \EXP \big(AY^{a=0} \big) } { \EXP \big(A \big) }
    =
    \EXP \big(Y^{a=0} \cond A=1 \big) \ .
\end{align*}

\subsection{Condition \eqref{Outcome COCA} Under Binary $Y$ and $W$} \label{sec:supp:proof:Binary}

If $W$ is a binary variable, any arbitrary function of $W$ is represented as $b(W) = b_0 + b_1 \cdot W $ where $b_0$ and $b_1$ are finite numbers. Therefore, Condition \eqref{Outcome COCA} is written as
\begin{align}
&
1 = E \big\{ b(W) \cond Y= 1, A=0 \big \} 
= b_0 + b_1 \Pr \big(W=1 \cond Y=1,A=0 \big)
\ ,
\label{eq:supp:binary1}
\\
&
0 = E \big\{ b(W) \cond Y= 0, A=0 \big \} 
= b_0 + b_1 \Pr \big(W=1 \cond Y=0,A=0 \big)
\ .
\label{eq:supp:binary0}
\end{align}
Therefore, the system of equations satisfy
\begin{align*}
&
\eqref{eq:supp:binary1} - \eqref{eq:supp:binary0}
&&
\Rightarrow
&&
1 = b_1 \big\{ \Pr \big(W=1 \cond Y=1,A=0 \big) - \Pr \big(W=1 \cond Y=0,A=0 \big) \big\}
\\
&
&&
\Rightarrow
&&
b_1 =  
\frac{1}{\Pr \big(W=1 \cond Y=1,A=0 \big) - \Pr \big(W=1 \cond Y=0,A=0 \big) }
\end{align*}
and
\begin{align*}
&
\Pr(W=1 \cond Y=0,A=0) 
\times
\eqref{eq:supp:binary1} - 
\Pr(W=1 \cond Y=1,A=0)
\times
\eqref{eq:supp:binary0}
\\
&
\Rightarrow
\quad 
\Pr(W=1 \cond Y=0,A=0) 
=
\big\{ \Pr(W=1 \cond Y=0,A=0)  - \Pr(W=1 \cond Y=1,A=0)  \big\} b_0
\\
&
\Rightarrow
\quad 
b_0 
=
- \frac{ 
\Pr(W=1 \cond Y=0,A=0) 
}{\Pr(W=1 \cond Y=1,A=0)  - \Pr(W=1 \cond Y=0,A=0)  }
\end{align*}
provided that $\Pr \big(W=1 \cond Y=1,A=0 \big) - \Pr \big(W=1 \cond Y=0,A=0 \big) \neq 0$. 
Therefore,
\begin{align*}
b(W=0) = b_0 
& = - \frac{ 
\Pr(W=1 \cond Y=0,A=0) 
}{\Pr(W=1 \cond Y=1,A=0)  - \Pr(W=1 \cond Y=0,A=0)  }
\\
b(W=1) = b_0 + b_1 
& = \frac{ 
- 
\Pr(W=1 \cond Y=0,A=0) +1
}{\Pr(W=1 \cond Y=1,A=0)  - \Pr(W=1 \cond Y=0,A=0)  }
\\
&
= \frac{ 
\Pr(W=0 \cond Y=0,A=0)
}{\Pr(W=1 \cond Y=1,A=0)  - \Pr(W=1 \cond Y=0,A=0)  }
\end{align*}
which is equivalent to
\begin{align*}
b(W)
=
\frac{ -(1-W) \Pr\big(  W=1|Y=0,A=0\big) + W  \Pr\big(  W=0|Y=0,A=0\big)  }{\Pr\big(
W=1|Y=1,A=0\big)  -\Pr\big(  W=1|Y=0,A=0\big)  } \ .
\end{align*}

If $\Pr \big(W=1 \cond Y=1,A=0 \big) - \Pr \big(W=1 \cond Y=0,A=0 \big) =0$, it implies that the right hand sides of \eqref{eq:supp:binary0} and \eqref{eq:supp:binary1} are the same whereas the left hand sides of those are different as 1 and 0, respectively. In this case, there is no function $b(W)$ satisfying \eqref{Outcome COCA}.

\subsection{Proof of Result \ref{result-3} and \eqref{Outcome bridge id}} \label{sec:supp:proof:OutcomeBridge}

Suppose that $b^*$ satisfies \eqref{Outcome COCA} and Assumption \ref{assumption:NC}. Then, 
\begin{align*}
&
\EXP \left\{  b^* (W,X)  \cond A=1\right\}
\\
& =   \EXP \left[  E\left\{  b^* (W,X) \cond A=1,Y^{a=0},X\right\} \cond A=1\right] 
\\
& \stackrel{(A)}{=} 
\EXP \left[  E\left\{  b^* (W,X) \cond A=0,Y^{a=0},X\right\} \cond A=1\right] 
\\
& = \iint E\left\{  b^* (W,X)  \cond A=0,Y^{a=0}=y, X=x\right\}  
f (Y^{a=0}=y,X=x \cond A=1 ) 
\, d(y,x)
\\
& \stackrel{(B)}{=} \iint E\left\{  b^* (W,X)  \cond A=0,Y=y,X=x\right\}  
f (Y^{a=0}=y,X=x \cond A=1 ) 
\, d(y,x)
\\
& \stackrel{(C)}{=} \iint y f (Y^{a=0}=y,X=x \cond A=1 ) 
\, d(y,x)
\\
& =E\big(  Y^{a=0} \cond A=1\big) \ .
\end{align*}
Equalities (A), (B), and (C) hold from Assumption \ref{assumption:NC}, consistency assumption, and \eqref{Outcome COCA}, respectively.

Suppose that the COCA confounding bridge function is correctly specified, i.e. $y = \EXP \big\{ b(W, X \con \eta^\dagger) \cond Y=y, X=x, A=0 \big\}$. Then, 
\begin{align*}
\EXP \big[   A \big\{ b(W,X \con \eta^\dagger) - \EXP(Y^{a=0} \cond A=1) \big\} \big]
&
=
\Pr(A=1)
\big[
\EXP \big\{   b(W,X \con \eta^\dagger) \cond A=1 \big\}
- \EXP(Y^{a=0} \cond A=1)
\big]
\\
& =  0 \ .
\end{align*}
Therefore, the GMM estimator using $g_{\OutReg}$ leads to $\psi_0^\dagger=\EXP \big(  Y^{a=0} \cond A=1 \big)$ if the COCA confounding bridge function is correctly specified.

\subsection{Proof of Result \ref{result-DR}}  \label{sec:supp:proof:DR}
Suppose $\EXP \left\{  b\big(  W, X \con \eta^\dagger \big)  \cond A=0,Y=y\right\}  =y$ then
\begin{align*}
& E\left[  Ab\big(  W, X \con \eta^\dagger \big)  +\frac{\big(  1-A\big)  \pi\big(
Y, X \con \mathbf{\alpha}\big)  }{ 1-\pi\big(  Y, X \con \mathbf{\alpha}\big)
}\left\{  Y-b\big(  W, X \con \eta\big)  \right\}  \right] 
\\
& =E\left[  A E\left\{  b\big(  W, X \con \eta^\dagger \big)   \cond A=1,Y^{a=0}, X\right\}
+\frac{\big(  1-A\big)  \pi\big(  Y , X\con \mathbf{\alpha}\big)  }{
1-\pi\big(  Y , X\con \mathbf{\alpha}\big)   }\left[  Y-E\left\{  b\big(
W , X \con \eta\big)  \cond A=0 ,Y^{a=0}, X\right\}  \right]  \right]
\\
& =E\left[  AE\left\{  b\big(  W, X \con \eta^\dagger \big)   \cond A=0, X,Y^{a=0}\right\}
+\frac{\big(  1-A\big)  \pi\big(  Y , X\con \mathbf{\alpha}\big)  }{
1-\pi\big(  Y , X\con \mathbf{\alpha}\big)   }\left[  Y^{a=0}-E\left\{  b\big(
W , X \con \eta\big)  \cond A=0 ,Y^{a=0}, X\right\}  \right]  \right]
\\
& =E\bigg\{  AY^{a=0}+\frac{\big(  1-A\big)  \pi\big(  Y, X \con \mathbf{\alpha
}\big)  }{  1-\pi\big(  Y, X \con \mathbf{\alpha}\big)  
}\underset{=0}{\underbrace{\big(  Y^{a=0}-Y^{a=0}\big)  }} \bigg\}  \\
& =P\big(  A=1\big)  \times E\big(  Y^{a=0} \cond A=1\big) \ .
\end{align*}
The first equality holds from the law of iterated expectations. 
The second equality holds from Assumption \ref{assumption:NC} and the consistency assumption. 
The third equality holds from $\EXP \left\{  b\big(  W, X \con \eta^\dagger \big)  \cond A=0,Y=y\right\}  =y$. 
The last equality is trivial.

Next suppose that $\pi\big(  y, x \con \mathbf{\alpha}^\dagger\big)  =\Pr\big(
A=1 \cond Y^{a=0}=y, X=x\big)  $ then
\begin{align*}
& E\left[  Ab\big(  W, X \con \eta\big)  + \frac{\big(  1-A\big)  \pi\big(
Y, X \con \mathbf{\alpha}^\dagger\big)  }{   1-\pi\big(  Y, X \con \mathbf{\alpha}^\dagger\big)
}\left\{  Y-b\big(  W, X \con \eta\big)  \right\}  \right]  \\
& =E\left[  Ab\big(  W, X \con \eta\big)  +\frac{\big(  1-A\big)  \Pr\big(
A=1 \cond Y^{a=0}, X\big)  }{  1-\Pr\big(  A=1 \cond Y^{a=0}, X\big)  
} 
\left\{  Y^{a=0}-b\big(  W, X \con \eta\big)  \right\}  \right]
\\
& =
E\left[  Ab\big(  W, X \con \eta\big)  
+
\frac{  \Pr\big(
A=0 \cond Y^{a=0}, X\big)   \Pr\big(
A=1 \cond Y^{a=0}, X\big)  }{  1-\Pr\big(  A=1 \cond Y^{a=0}, X\big)  
} 
\EXP \big\{   Y^{a=0}-b\big(  W, X \con \eta\big)    \cond A=0, Y^{a=0},X
\big\}
\right]
\\
&
=
E\left[  Ab\big(  W, X \con \eta\big)  
+
\Pr\big(
A=1 \cond Y^{a=0}, X\big) 
\EXP \big\{   Y^{a=0}-b\big(  W, X \con \eta\big)    \cond A=1, Y^{a=0},X
\big\}
\right]
\\
& =E\left[  Ab\big(  W, X \con \eta\big)  +A\left\{  Y^{a=0}-b\big(
W, X \con \eta\big)  \right\}  \right]  \\
& =P\big(  A=1\big)  \times E\big(  Y^{a=0} \cond A=1\big) \ .
\end{align*}
The first equality holds from $\pi \big( y, x \con \mathbf{\alpha}^\dagger \big)  =\Pr\big(
A=1 \cond Y^{a=0}=y, X=x\big)$ and the consistency assumption. 
The second equality holds from the law of iterated expectations. 
The third equality holds from Assumption \ref{assumption:NC}. 
The remaining equalities are trivial. Therefore, the GMM estimator using $g_{\DR}$ leads to $\psi_0^\dagger=\EXP \big(  Y^{a=0} \cond A=1 \big)$ if the EPS or the COCA confounding bridge function is correctly specified.

\subsection{Proof of Result \ref{result-IF}}

It is straightforward to verify that the efficient influence function for $\psi_1^* = E (Y \cond A=1)$ is $A (Y - \psi_1^*)/\Pr(A=1)$. Therefore, it suffices to find an (efficient) influence function of $\psi_0^*$. In the remainder of the proof, we establish that (i) the following $\InfFt_0$ is an influence function of the counterfactual mean $\psi_0^* = E(Y^{a=0} \cond A=1)$ under $\mathcal{M}$ and (ii) it is the efficient influence function for $\psi_0^*$ at the submodel $\mathcal{M}_{\text{sub}}$: 
\begin{align*}
\InfFt_0(O \con \pi^*, b^*) = 
\frac{ 1 }{\Pr \big(  A=1\big)}  
\bigg[ \frac{\big(  1-A\big)  \pi^*\big(  Y,X \big)
}{  1-\pi^*\big(  Y,X \big)     }\left\{  Y-b^*\big(
W,X\big)  \right\} 
+
A \big\{  b^*\big(W, X \big) - \psi_0^* \big\} 
\bigg]
\ .
\end{align*}

\subsubsection{Proof of (i)} 

Let $\mathcal{M}(t)$ be the one-dimensional parametric submodel of $\mathcal{M}$ and $f(O \con t) \in \mathcal{M}(t)$ be the density at $t$. We suppose that the true density is recovered at $t^*$, i.e., $f^*(O) := f(O \con t^*)$. 
Let $s(\cdot \cond \cdot \con t): = \partial \log f(\cdot \cond \cdot \con t) / \partial t$ be the score function evaluated at $t$, and let $s^*(\cdot \cond \cdot)$ be the score function evaluated at $t^*$. 
Let $E\ETA \big\{ g(O) \big\}$ be the expectation operator at $f(O \con t)$. 

Let $b(W,X \con t)$ be the COCA confounding bridge function satisfying
\begin{align*}
&
0
=
\EXP\ETA \big\{ Y - b(W,X \con t) \cond Y, A=0 , X \big\} \ .
\end{align*}
Let $\psi_0(t) = E\ETA \big\{  b\big(  W, X \con t \big) \cond A=1 \big\} =  E\ETA \big\{ A b\big(  W, X \con t \big) \big\} / E\ETA \big(A \big) $ be the counterfactual mean evaluated at $t$. Let $\psi_N(t) = E\ETA \big\{ A b\big(  W, X \con t \big) \big\} $ and $\psi_D(t) = E\ETA \big(A \big)$. Note that $\psi_0^* = 
E \big\{  b^*\big(  W, X\big) \cond A=1 \big\}$ and $\psi_N^* = E \big\{ A b^*\big(  W, X\big) \big\}$ and $\psi_D^* = E (A)$. 
Taking the derivative of the above restrictions with respect to $t$ and evaluating at $t^*$, the score functions $s^*$ satisfies:
\begin{align}
\EXP \Big[
\big\{ Y - b^* (\bW,\bX) \big\} s^* (\bW \cond Y, A=0, \bX)         - \nabla_t b (W,X \con t^*)	
\, \Big| \, Y, A=0, \bX \Big] = 0 \ .
\label{eq:tangent-1}
\end{align}

The pathwise derivative of $\psi_0(t)$ is
\begin{align*}
\frac{\partial \psi_0(t) }{\partial t}
& =
\frac{ \EXP \ETA
\big\{
A \nabla_{t} b(W,X \con t)
+
A b(W,X \con t) s(W,A,X \con t)
\big\}
}{ \EXP \ETA(A) }
-
\frac{ \psi_0(t)  }{ \psi_D(t) }
\EXP \ETA \big\{ A \cdot s(A \con t) \big\}
\\
&
=
\frac{ \EXP \ETA
\big\{
A \nabla_{t} b(W,X \con t)
+
A b(W,X \con t) s(Y,W,A,X \con t)
\big\}
}{ \psi_D(t) }
-
\frac{ \psi_0(t)  }{\psi_D(t) }
\EXP \ETA \big\{ A \cdot s(Y,W,A,X \con t) \big\}
\\
&
=
\frac{ \EXP \ETA
\big[
A \nabla_{t} b(W,X \con t)
+
A \big\{  b(W,X \con t) - \psi_0(t) \big\} s(Y,W,A,X \con t)
\big]
}{ \psi_D(t) }
\ ,
\end{align*}
which is evaluated at $t^*$ as follows:
\begin{align*}
\frac{\partial \psi_0(t) }{\partial t}
\Bigg|_{t=t^*}
& 
=
\frac{ \EXP
\big[
A \nabla_{t} b(W,X \con t^*)
+
A \big\{  b^*(W,X) - \psi_0^* \big\} s^*(O)
\big]
}{ \Pr(A=1) }
\ ,
\end{align*}

We find that
\begin{align*}
&
E \big\{ A \nabla_{t} b(W,X \con t^*) \big\}
\\
&
=
E \Big[ E \big\{ A \nabla_{t} b(W,X \con t^*) \cond \potY{0}{},X \big\} \Big]
\\
&
\stackrel{(\dagger)}{=}
E \Big[ E \big( A \cond \potY{0}{},X \big)  
E \big\{ \nabla_{t} b(W,X \con t^*) \cond \potY{0}{},X \big\} \Big]
\\
&
=
E \bigg[ \Pr \big( A=0 \cond \potY{0}{},X \big)  
\frac{ \pi^*(\potY{0}{},X) }{1-\pi^*(\potY{0}{},X)}
E \big\{ \nabla_{t} b(W,X \con t^*) \cond \potY{0}{},X \big\} \bigg]
\\
&
=
E \bigg[ E \bigg\{ \frac{(1-A) \pi^*(\potY{0}{},X)}{1-\pi^*(\potY{0}{},X)} \nabla_{t} b(W,X \con t^*) \, \bigg| \, \potY{0}{},X \bigg\} \bigg]
\\
&
=
E \bigg\{ \frac{(1-A) \pi^*(Y,X)}{1-\pi^*(Y,X)} \nabla_{t} b(W,X \con t^*) \bigg\}
\\
&
=
E \bigg[ \frac{(1-A) \pi^*(Y,X)}{1-\pi^*(Y,X)} 
E \big\{ \nabla_{t} b(W,X \con t^*) \cond Y,A=0,X \big\}
\bigg]
\\ 
&
\stackrel{(*)}{=}
E \bigg[ \frac{(1-A) \pi^*(Y,X)}{1-\pi^*(Y,X)} 
E \big[ \big\{ Y - b^*(W,X) \big\} s^*(W \cond Y, A=0, X)  \cond Y,A=0,X \big]
\bigg]
\\ 
&
=
E \bigg[ \frac{(1-A) \pi^*(Y,X)}{1-\pi^*(Y,X)} 
\big\{ Y - b^*(W,X) \big\} s^*(W \cond Y, A=0, X)
\bigg]
\\ 
&
\stackrel{(\$)}{=}
E \bigg[ \frac{(1-A) \pi^*(Y,X)}{1-\pi^*(Y,X)} 
\big\{ Y - b^*(W,X) \big\} 
\Big[
E \big\{ s^*(O) \cond Y,W,A=0,X \big\}
-
E \big\{ s^*(O) \cond Y,A=0,X \big\}
\Big]
\bigg]
\\ 
&
=
E \bigg[ \frac{(1-A) \pi^*(Y,X)}{1-\pi^*(Y,X)} 
\big\{ Y - b^*(W,X) \big\} 
s^*(O)
\bigg]
\\
&
-
E \bigg[ \frac{(1-A) \pi^*(Y,X)}{1-\pi^*(Y,X)} 
\big[ Y - E \big\{ b^*(W,X) \cond Y,A=0,X \big\} \big]
E \big\{ s^*(O) \cond Y,A=0,X \big\}
\bigg]
&
\\ 
&
\stackrel{(\#)}{=}
E \bigg[ \frac{(1-A) \pi^*(Y,X)}{1-\pi^*(Y,X)} 
\big\{ Y - b^*(W,X) \big\} 
s^*(O)
\bigg] \ .
\end{align*}
Equalities with no marks are straightforward to show from the law of iterated expectations. 
Equalities with $(\dagger)$ are established from Assumption \ref{assumption:NC}: $W \indep A \cond (\potY{0}{},X)$.
The equality with $(*)$ is established from \eqref{eq:tangent-1}.
The equality with $(\$)$ holds from the property of the score function.
The equality with $(\#)$ holds from the definition of the COCA confounding bridge function. As a consequence, we find
\begin{align*}
\frac{\partial \psi_0(t) }{\partial t}
\Bigg|_{t=t^*}
& 
=
\frac{ \EXP
\big[
A \nabla_{t} b(W,X \con t^*)
+
A \big\{  b^*(W,X) - \psi_0^* \big\} s^*(O)
\big]
}{ \Pr(A=1) }
\\
&
=
\EXP \big\{ \InfFt_0(O \con \pi^*, b^*) \cdot s^*(O) \big\} \ ,
\end{align*}
implying that $\psi_0^*$ is a pathwise differentiable parameter \citep{Newey1990} with the corresponding influence function $\InfFt_0(O \con \pi^*,b^*)$.

\subsubsection{Proof of (ii)}

It suffices to show that the influence function $\InfFt_0(O \con \pi^*, b^*)$ belongs to the tangent space of $\mathcal{M}$ at the submodel $\mathcal{M}_{\text{sub}}$. Note that the model imposes a restriction in \eqref{eq:tangent-1} on the score function. The restriction implies that the tangent space of $\mathcal{M}$ consists of the functions $S(O) \in \mathcal{L}_{2,0}(O)$ satisfying
\begin{align} \label{eq:tangent-2}
\EXP \Big[
\big\{ Y - b^* (\bW,\bX) \big\} S (\bW \cond Y, A=0, \bX)        
\, \Big| \, Y, A=0, \bX \Big] 
\in \text{Range}(T_1) \ .
\end{align}
Under \HL{Surjectivity}, we have $\text{Range}(T_1) = \mathcal{L}_2(Y,A=0,X)$. Therefore, any $S(O) \in \mathcal{L}_{2,0}(O)$ satisfies the restriction \eqref{eq:tangent-2} at the submodel $\mathcal{M}_{\text{sub}}$. This implies that $\InfFt_0(O \con \pi^*, b^*)$ belongs to the tangent space of $\mathcal{M}$ at the submodel $\mathcal{M}_{\text{sub}}$. 

\subsection{Proof of Result \ref{result-CAN}}

In what follows, we simply denote $\InfFt^*(\bO) = \InfFt(\bO \con \omega^*, b^*)$ and $\AVER(V) = \sum_{i=1}^{N} V_i / N$. If we show that $\widehat{\psi}^{(k)}$ has the asymptotic representation as
\begin{align}						\label{eq-ssestimator}
\big| \mathcal{I}_k \big|^{1/2}
\Big\{
\widehat{\psi}^{(k)} - \psi^*
\Big\}
=
\frac{1}{\big| \mathcal{I}_k \big|^{1/2}}
\sum_{ i \in \mathcal{I}_k } \InfFt^*(\bO_i) + o_P(1) \ , 
\end{align}
then we establish $\widehat{\psi} = K^{-1} \sum_{k=1}^{K} \widehat{\psi}^{(k)}$ has the asymptotic representation as 
\begin{align}					\label{eq-fullestimator}
\sqrt{N} \Big( \widehat{\psi} - \psi^* \Big)
=
\frac{1}{\sqrt{N}}
\sum_{i=1}^{N}  \InfFt^*(\bO_i) + o_P(1) \ .
\end{align}
Therefore, the asymptotic normality result holds from the central limit theorem. Thus, we focus on showing that  \eqref{eq-ssestimator} holds.

Recall that $\widehat{\psi}^{(k)}$ is
\begin{align}								\label{eq-psihat0}
\widehat{\psi}^{(k)} 
& =
\Big\{ \AVER  \big( A \big) \Big\}^{-1}
\Big[
\AVER\SSS
\underbrace{			
\Big[
A \big\{ Y - \widehat{b}\LSS(W,X) \big\}
-
(1-A) \widehat{\omega}\LSS(Y,X) \big\{ Y - \widehat{b}\LSS(W,X) \big\}
\Big]}_{=: \widehat{\zeta}\LSS }	
\Big]
\end{align}
Therefore, we find the left hand side of \eqref{eq-ssestimator} is 
\begin{align*}
\big| \mathcal{I}_k \big|^{1/2}
\Big\{
\widehat{\psi}^{(k)} - \psi^*
\Big\}
& 	=
\frac{1}{|\mathcal{I}_k|^{1/2}}
\sum_{i \in \mathcal{I}_k}
\bigg\{
\frac{ \widehat{\zeta}_i\LSS }{\AVER \big( A \big)}
-
\frac{ A_i \psi^*}{\AVER\SSS \big( A \big)}
\bigg\}
\\
&	=
\underbrace{	
\frac{1}{ 2}
\bigg[  
\frac{ \Pr(A=1) }{ \AVER(A) } - \frac{ \Pr(A=1) }{ \AVER\SSS (A) } 
\bigg]}_{o_P(1)}
\underbrace{ 
\frac{1}{\big| \mathcal{I}_k \big|^{1/2}}		
\sum_{i \in \mathcal{I}_k}
\frac{\widehat{\zeta}_i\LSS + A_i \psi^* }{\Pr(A=1)}
}_{O_P(1)}
\\
& \quad 
+
\underbrace{
\frac{1}{2}
\bigg[ 
\frac{ \Pr(A=1) }{ \AVER(A) } + \frac{ \Pr(A=1) }{ \AVER\SSS (A) } 
\bigg]}_{=1+o_P(1)}
\underbrace{
\frac{1}{ \big| \mathcal{I}_k \big|^{1/2}}
\sum_{i \in \mathcal{I}_k}
\frac{\widehat{\zeta}_i\LSS - A_i \psi^* }{\Pr(A=1)} }_{\text{\HT{AN}: } O_P(1)}
\\
& =
\frac{1}{ \big| \mathcal{I}_k \big|^{1/2}}
\sum_{i \in \mathcal{I}_k}
\frac{\widehat{\zeta}_i\LSS - A_i \psi^* }{\Pr(A=1)} +o_P(1) \ .
\end{align*}
The second line holds because $\AVER(A)=\Pr(A)+o_P(1)$ and $ \AVER\SSS(A) = \Pr(A=1) + o_P(1)$ from the law of large numbers, and  $\widehat{\zeta}_{i} \LSS $ is bounded. The third row holds because term \HL{AN} is asymptotically normal, which will be shown later. 

Let $\zeta^*$ be the numerator of the uncentered influence function, i.e., 
\begin{align*}
\zeta^* 
=
A \big\{ Y -   b^*\big(W, X \big)  \big\}
-
\big(  1-A\big) \omega^*(Y,X) 
\left\{  Y-b^*\big( W,X\big)  \right\}  \ .
\end{align*}
Note that we have $\InfFt^*(\bO) = (\zeta^* - A \psi^*)/\Pr(A=1)$.


Let $\EMP_{\mathcal{I}_k} (V) = | \mathcal{I}_k |^{-1/2} \sum_{i \in \mathcal{I}_k} \big\{ V_i - \EXP(V_i) \big\} $ be the empirical process of $V_i$ centered by $\EXP(V_i)$. Similarly, let $\EMP_{\mathcal{I}_k}\LSS \big( \widehat{V}\LSS \big) = | \mathcal{I}_k |^{-1/2} \sum_{i \in \mathcal{I}_k} \big\{ \widehat{V}_i\LSS - \EXP\LSS(\widehat{V}\LSS) \big\} $ be the empirical process of $\widehat{V}\LSS$ centered by $\EXP\LSS(\widehat{V}\LSS)$ where $\EXP\LSS (\cdot)$ is the expectation after considering random functions obtained from $\mathcal{I}_k^c$ as fixed functions. The empirical process of $\widehat{\zeta}\LSS - A \psi^*$ is
\begin{align} 
\big| \mathcal{I}_k \big|^{-1/2}
\sum_{i \in \mathcal{I}_k} \big\{ \widehat{\zeta}\LSS  - A \psi^* \big\}
= 	&
\  \EMP_{\mathcal{I}_k} \big( \zeta^*  - A \psi^* \big)
\label{Term1}
\\
& 
+
\big| \mathcal{I}_k \big|^{1/2}
\cdot
\EXP\LSS
\big\{ \widehat{\zeta}\LSS - \zeta^* \big\}
\label{Term3}
\\
& +
\EMP_{\mathcal{I}_k}\LSS
\big(
\widehat{\zeta}\LSS - \zeta^*
\big)
\label{Term2} \ ,
\end{align}
where
\begin{align*}
\EMP_{\mathcal{I}_k} \big( \zeta^*  - A \psi^* \big)
& =
\EMP_{\mathcal{I}_k} 
\Big[
A \big\{ Y-  b^*\big(W, X \big) - \psi^* \big\} 
-
\big(  1-A\big) \omega^*(Y,X) 
\left\{  Y-b^*\big(
W,X\big)  \right\} 
\Big]
\\
&
=
\big| \mathcal{I}_k \big|^{-1/2}
\sum_{i \in \mathcal{I}_k} \Big\{ \zeta_i^* - \Pr(A=1) \psi^* \Big\}
\\
\EMP_{\mathcal{I}_k}\LSS
\big(
\widehat{\zeta}\LSS - \zeta^*
\big)
&
=
\EMP_{\mathcal{I}_k}\LSS
\left[
\begin{array}{l}
A \big\{ Y -  \widehat{b}\LSS(W,X) \big\}
-
(1-A) \widehat{\omega}\LSS(Y,X) \big\{ Y - \widehat{b}\LSS(W,X) \big\}
\\
-
A \big\{ Y - b^*\big(W, X \big) \big\}
+
\big(  1-A\big) \omega^*(Y,X) 
\left\{  Y-b^*\big(
W,X\big)  \right\} 
\end{array}		
\right] \ .
\end{align*}
From the derivation below, we find that \eqref{Term3} and \eqref{Term2} are $o_P(1)$, indicating that \eqref{Term1} is asymptotically normal, and consequently, \HL{AN} is $O_P(1)$. Moreover, we establish \eqref{eq-ssestimator}, concluding the proof as follows:
\begin{align*}
\big| \mathcal{I}_k \big|^{1/2}
\Big\{
\widehat{\psi}^{(k)} - \psi^*
\Big\}
& =
\big| \mathcal{I}_k \big|^{-1/2}
\sum_{i \in \mathcal{I}_k}
\frac{\widehat{\zeta}_{i} \LSS - A_i \psi_0^* }{\Pr(A=1)}
+
o_P(1)
\\
& =
\big| \mathcal{I}_k \big|^{-1/2}
\sum_{i \in \mathcal{I}_k}
\frac{ \zeta_{i}^*  - A_i \psi^* }{\Pr(A=1)}
+
o_P(1)
=
\big| \mathcal{I}_k \big|^{-1/2}
\sum_{i \in \mathcal{I}_k} \InfFt^*(\bO_i)
+
o_P(1) \ .
\end{align*}

\noindent \textbf{Showing that \eqref{Term3} is $o_P(1)$}

First, we introduce the following equality:
\begin{align}
&
E \big\{ A g(W,X) \big\}
\nonumber
\\
&
=    
E \big[ \Pr(A=1 \cond \potY{0}{},X) E \big\{ g(W,X) \cond \potY{0}{},A=1,X \big\}  \big]
\nonumber
\\
&
=    
E \big[ E (1-A \cond \potY{0}{},X ) \omega^*(\potY{0}{},X) 
E \big\{ g(W,X) \cond \potY{0}{},A=0,X \big\}  \big]
\nonumber
\\
& 
=
E \big\{ (1-A) \omega^*(\potY{0}{},X) g(W,X) \big\}
\nonumber
\\
& 
=
E \big\{ (1-A) \omega^*(Y,X)   g(W,X) \big\} \ .
\label{eq-proof-aux1}
\end{align}

We find 
\begin{align*}
& 
\Big\|
\big| \mathcal{I}_k \big|^{1/2}
\EXP\LSS
\big\{ \widehat{\zeta}\LSS - \zeta^* \big\}
\Big\|
\\
&
=
\big| \mathcal{I}_k \big|^{1/2}
\left\|
\EXP\LSS
\left[
\begin{array}{l}
(1-A) \widehat{\omega}\LSS(Y,X) \big\{ Y - \widehat{b}\LSS(W,X) \big\}
+ A \widehat{b}\LSS(W,X)
\\
-
\big(  1-A\big) \omega^*(Y,X) 
\left\{  Y-b^*\big(
W,X\big)  \right\} 
-
A    b^*\big(W, X \big)
\end{array}
\right] 
\right\|
\\
&
=
\big| \mathcal{I}_k \big|^{1/2}
\bigg\|
\EXP\LSS
\bigg[
(1-A) \widehat{\omega}\LSS(Y,X) \big\{ b^*(W,X) - \widehat{b}\LSS(W,X) \big\}
+ A \big\{ \widehat{b}\LSS(W,X) - b^*\big(W, X \big) \big\}
\bigg]
\bigg\|
\\
&
\stackrel{(\dagger)}{=}
\big| \mathcal{I}_k \big|^{1/2}
\left\|
\EXP\LSS
\left[
\begin{array}{l}
(1-A)\widehat{\omega}\LSS(Y,X) \big\{ b^*(W,X) - \widehat{b}\LSS(W,X) \big\}
\\
+ (1-A) \omega^*(Y,X) \big\{ \widehat{b}\LSS(W,X) - b^*\big(W, X \big) \big\} 
\end{array}
\right]
\right\|
\\
&
=
\big| \mathcal{I}_k \big|^{1/2}
\bigg\|
\EXP\LSS
\bigg[ 
(1-A)
\Big\{ \widehat{\omega}\LSS(Y,X) - \omega^*(Y,X) \Big\}
\Big\{ b^*(W,X) - \widehat{b}\LSS(W,X) \Big\}
\bigg]
\bigg\|
\end{align*}
Equality $(\dagger)$ holds from \eqref{eq-proof-aux1} by choosing $g = b^* - \widehat{b}\LSS $. Therefore, $
\big| \mathcal{I}_k \big|^{1/2}
\EXP\LSS
\big\{ \widehat{\zeta}\LSS - \zeta^* \big\}$ is upper bounded as
\begin{align*}
    & 
    \Big\|
\big| \mathcal{I}_k \big|^{1/2}
\EXP\LSS
\big\{ \widehat{\zeta}\LSS - \zeta^* \big\}
\Big\|
\\
&
\leq
    \big| \mathcal{I}_k \big|^{1/2}
\EXP\LSS
\bigg[ 
\Big\| \widehat{\omega}\LSS(Y,X) - \omega^*(Y,X) \Big\| \cdot 
\Big\| 
    \EXP\LSS\big[ 
    (1-A)
    \big\{ b^*(W,X) - \widehat{b}\LSS(W,X) \big\}
    \cond Y,X 
    \big]
\Big\|
\bigg]
\\
&
\leq
    \big| \mathcal{I}_k \big|^{1/2}
\Big\| \widehat{\omega}\LSS(Y,X) - \omega^*(Y,X) \Big\|_{P,2}
\Big\| 
    \EXP\LSS\big[ 
    (1-A)
    \big\{ b^*(W,X) - \widehat{b}\LSS(W,X) \big\}
    \cond Y,X 
    \big]
\Big\|_{P,2}
\end{align*}
and 
\begin{align*}
    & 
    \Big\|
\big| \mathcal{I}_k \big|^{1/2}
\EXP\LSS
\big\{ \widehat{\zeta}\LSS - \zeta^* \big\}
\Big\|
\\
&
\leq
    \big| \mathcal{I}_k \big|^{1/2}
\EXP\LSS
\bigg[ 
\Big\| \widehat{b}\LSS(W,X) - b^*(W,X) \Big\|
\Big\| 
    \EXP\LSS\big[ 
    (1-A)
    \big\{ \omega^*(Y,X) - \widehat{\omega}\LSS(Y,X) \big\}
    \cond W,X 
    \big]
\Big\|
\bigg]
\\
&
\leq
    \big| \mathcal{I}_k \big|^{1/2}
\Big\| \widehat{b}\LSS(W,X) - b^*(W,X) \Big\|_{P,2}
\Big\| 
    \EXP\LSS\big[ 
    (1-A)
    \big\{ \omega^*(Y,X) - \widehat{\omega}\LSS(Y,X) \big\}
    \cond W,X 
    \big]
\Big\|_{P,2}  \ .
\end{align*}
This concludes that
\begin{align*}
&
\Big\|
\big| \mathcal{I}_k \big|^{1/2}
\EXP\LSS
\big\{ \widehat{\zeta}\LSS - \zeta^* \big\}
\Big\|
\\
&
\leq 
\big| \mathcal{I}_k \big|^{1/2}
\cdot 
\min 
\left[ 
\begin{array}{l}
    \Big\| \widehat{\omega}\LSS - \omega^* \Big\|_{P,2}  
    \Big\| 
    \EXP\LSS\big[ 
    (1-A)
    \big\{ b^*(W,X) - \widehat{b}\LSS(W,X) \big\}
    \cond Y,X 
    \big]
\Big\|_{P,2}
    ,
    \\
    \Big\| \widehat{b}\LSS - b^* \Big\|_{P,2}
\Big\| 
    \EXP\LSS\big[ 
    (1-A)
    \big\{ \omega^*(Y,X) - \widehat{\omega}\LSS(Y,X) \big\}
    \cond W,X 
    \big]
\Big\|_{P,2} 
\end{array}
\right]
\\
&
= o_P(1) \ .
\end{align*}
where the second inequality is from Assumption \ref{reg}-(iii).

\bigskip
 
\noindent \textbf{Showing that \eqref{Term2} is $o_P(1)$}

Next, we establish that Term \eqref{Term2} is $o_P(1)$.  Suppose $g(\bO)$ is a mean-zero function. Then, $\EXPk \big\{ \EMPk(g) \big\} = 0$. Therefore, it suffices to show that $\VARk \big\{ \EMPk(g) \big\} = o_P(1)$, i.e., 
\begin{align*}
\VARk \big\{ \EMPk(g) \big\}
=
\VARk \bigg\{
\frac{1}{\sqrt{|\mathcal{I}_k|}} \sum_{i \in \mathcal{I}_k} g(\bO_i)
\bigg\}
=
\VARk \big\{ g(\bO) \big\}
=
\EXPk \big\{ g(\bO)^2 \big\}
=
o_P(1) \ .
\end{align*}
The variance of \eqref{Term2} is
\begin{align}
& 
\VARk \big\{
\eqref{Term2}
\big\}
\nonumber
\\
& =
\VARk \left[
\EMPk
\left[
\begin{array}{l}
A \big\{ Y -  \widehat{b}\LSS(W,X) \big\}
-
(1-A) \widehat{\omega}\LSS(Y,X) \big\{ Y - \widehat{b}\LSS(W,X) \big\}
\\
-
A \big\{ Y -    b^*\big(W, X \big) \big\}
+
\big(  1-A\big) \omega^*(Y,X) 
\left\{  Y-b^*\big(
W,X\big)  \right\} 
\end{array}		
\right]
\right]
\nonumber
\\
&
=
\EXPk 
\left[
\left[
\begin{array}{l}
(1-A) \widehat{\omega}\LSS(Y,X) \big\{ Y - \widehat{b}\LSS(W,X) \big\}
+ A \widehat{b}\LSS(W,X)
\\
-
\big(  1-A\big) \omega^*(Y,X) 
\left\{  Y-b^*\big(
W,X\big)  \right\} 
-
A    b^*\big(W, X \big)
\end{array}		
\right]^2
\right]
\nonumber
\\
&
\stackrel{(*)}{\precsim }
\EXPk 
\left[
\begin{array}{l}
\big[ (1-A) \big\{ \widehat{\omega}\LSS(Y,X) - \omega^*(Y,X) \big\} Y \big]^2
\\
+
\big[ (1-A) \widehat{\omega}\LSS(Y,X) \widehat{b}\LSS(W,X) - 
(1-A)
\omega^*(Y,X) b^*(W,X) \big]^2
\\
+
\big[ A \big\{ \widehat{b}\LSS(W,X) - b^*(W,X) \big\} \big]^2
\end{array}
\right]
\nonumber
\\
&
\stackrel{(*)}{\precsim }
\EXPk 
\left[
\begin{array}{l}
\big[ (1-A) \big\{ \widehat{\omega}\LSS(Y,X) - \omega^*(Y,X) \big\} Y \big]^2
\\
+
\big[ (1-A) \big\{ \widehat{\omega}\LSS(Y,X)  - \omega^*(Y,X) \big\} 
\widehat{b}\LSS(W,X) \big]^2
\\
+
\big[ (1-A) \omega^*(Y,X)
\big\{ \widehat{b}\LSS(W,X) - b^*(W,X) \big\} \big]^2 
\\
+
\big[ A \big\{ \widehat{b}\LSS(W,X) - b^*(W,X) \big\} \big]^2
\end{array}
\right]
\quad 
\begin{array}{l}
\text{\HT{Term-Q1}}
\\
\text{\HT{Term-Q2}}
\\
\text{\HT{Term-Q3}}
\\
\text{\HT{Term-Q4}}
\end{array}
\nonumber
\\
&
\stackrel{(\dagger)}{\precsim}
\big\| \widehat{\omega}\LSS(Y,X) - \omega^*(Y,X) \big\|_{P,2}^2
+
\big\| \widehat{b}\LSS(Y,X) - b^*(Y,X) \big\|_{P,2}^2 
\nonumber
\\
&
=
o_P(1) \ .
\label{eq-variance0}
\end{align}
Inequality $(*)$ holds from the Cauchy-Schwartz inequality $(\sum_{j=1}^{M} a_j)^2 \leq M (\sum_{j=1}^{M} a_j^2)$. Inequality $(\dagger)$ is established in items (a) and (b) below. The last convergence rate is obtained from Assumption \ref{reg}. 

\begin{itemize}[leftmargin=0cm,itemsep=0cm]
\item[(a)] \HL{Term-Q1}
\begin{align*}
\text{\HL{Term-Q1}}
&
=
\EXPk \big[ \big[ (1-A) \big\{ \widehat{\omega}\LSS(Y,X) - \omega^*(Y,X) \big\} Y \big]^2 \big]
\\
&
=
\EXPk \big[ (1-A) \big\{ \widehat{\omega}\LSS(Y,X) - \omega^*(Y,X) \big\}^2 Y^2 \big]
\\
&
=
\EXPk \big[ \Pr(A=0 \cond Y,X) \big\{ \widehat{\omega}\LSS(Y,X) - \omega^*(Y,X) \big\}^2  Y^2 \big]
\\
&
\stackrel{(*)}{\precsim}
\EXPk \big\{ \big\| \widehat{\omega}\LSS(Y,X) - \omega^*(Y,X) \big\| \cdot Y^2 \big\}
\\
&
\stackrel{(\dagger)}{\precsim}
\EXPk \big\{ \big\| \widehat{\omega}\LSS(Y,X) - \omega^*(Y,X) \big\|^2 \big\}  \cdot
\EXPk \big(  Y^4 \big)
\\
& 
\precsim
\big\| \widehat{\omega}\LSS(Y,X) - \omega^*(Y,X) \big\|_{P,2}^2
\ .
\end{align*}
Inequality $(*)$ holds from Assumption \ref{reg}, which implies $\Pr(A=0 \cond Y,X) \in [c,1-c]$ for $c>0$ and $\big\| \widehat{\omega}\LSS(Y,X) - \omega^*(Y,X) \big\|_{\infty} \leq 2C$. Inequality $(\dagger)$ holds from the Cauchy-Schwartz inequality.

\item[(b)] \HL{Term-Q2}-\HL{Term-Q4}
\begin{align*}
\text{\HL{Term-Q2}}
&
=
\EXPk \big[ \big[ (1-A) \big\{ \widehat{\omega}\LSS(Y,X)  - \omega^*(Y,X) \big\} 
\widehat{b}\LSS(W,X) \big]^2 \big]
\\
&
\precsim
\EXPk \big\{ \big\| \widehat{\omega}\LSS(Y,X) - \omega^*(Y,X) \big\|^2 \big\} 
\\
&
\precsim
\big\| \widehat{\omega}\LSS(Y,X) - \omega^*(Y,X) \big\|_{P,2}^2 \ .
\\[0.5cm]
\text{\HL{Term-Q3}}
&
=
\EXPk \big[ \big[ (1-A) \omega^*(Y,X)
\big\{ \widehat{b}\LSS(W,X) - b^*(W,X) \big\} \big]^2 \big]
\\
&
\precsim
\EXPk \big\{ \big\| \widehat{b}\LSS(W,X) - b^*(W,X) \big\|^2 \big\} 
\\
&
\precsim
\big\| \widehat{b}\LSS(W,X) - b^*(W,X) \big\|_{P,2}^2 \ .
\\[0.5cm]
\text{\HL{Term-Q4}}
&
=
\EXPk \big[ \big[ A \big\{ \widehat{b}\LSS(W,X) - b^*(W,X) \big\} \big]^2 \big]
\\
&
\precsim
\EXPk \big\{ \big\| \widehat{b}\LSS(W,X) - b^*(W,X) \big\|^2 \big\} 
\\
&
\precsim
\big\| \widehat{b}\LSS(W,X) - b^*(W,X) \big\|_{P,2}^2 \ .
\end{align*}

\end{itemize}

\noindent\textbf{Consistency of the Variance Estimator}

The proposed variance estimator is
\begin{align*}
&
\widehat{\sigma}^2
=
\frac{1}{K} \sum_{k=1}^{K} \widehat{\sigma}^{2,(k)}
\ , \
&&
\widehat{\sigma}^{2,(k)}
=
\frac{1}{|\mathcal{I}_k|}
\sum_{i \in \mathcal{I}_k}
\Bigg\{
\frac{ \widehat{\zeta}_i \LSS - A_i 
\widehat{\psi}\LSS  }{ \AVER(A) }  \Bigg\}^{2} \ ,
\end{align*}
where $\widehat{\zeta}\LSS$ is defined in \eqref{eq-psihat0}. Therefore, it suffices to show that $\widehat{\sigma}^{2,(k)} =\sigma^2 + o_P(1)$. Note that $\widehat{\sigma}^{2,(k)} - \sigma^2$ is    
\begin{align}
\widehat{\sigma}^{2,(k)} - \sigma^2
& 
=
\big\{ \AVER(A) \big\}^{-2}
\AVER\SSS
\Big[
\big\{ \widehat{\zeta}\LSS - A \widehat{\psi}  \big\}^2
\Big] -
\sigma^2
\nonumber
\\
&
=
\big\{ \Pr(A=1) \big\}^{-2}
\AVER\SSS
\Big[
\big\{ \widehat{\zeta}\LSS - A \widehat{\psi}  \big\}^2
\Big] -
\sigma^2 + o_P(1)
\nonumber
\\
&
=
\big\{ \Pr(A=1) \big\}^{-2}
\Big[
\AVER\SSS
\Big[
\big\{ \widehat{\zeta}\LSS - A \widehat{\psi}  \big\}^2
\Big] -
\AVER\SSS
\Big\{
\big( \zeta^* - A \psi^*  \big)^2
\Big\}
\Big]
\nonumber
\\
& \hspace*{2cm}
+		 
\Big[
\big\{ \Pr(A=1) \big\}^{-2}
\AVER\SSS
\Big\{
\big( \zeta^* - A \psi^*  \big)^2
\Big\}
-
\sigma^2
\Big] + o_P(1)
\nonumber
\\
& = 
\big\{ \Pr(A=1) \big\}^{-2}
\Big[
\AVER\SSS
\Big[
\big\{ \widehat{\zeta}\LSS - A \widehat{\psi}  \big\}^2
\Big] -
\AVER\SSS
\Big\{
\big( \zeta^* - A \psi^*  \big)^2
\Big\}
\Big]
+ o_P(1)
\label{eq-variance1} \ .
\end{align}
The third and fifth lines hold from the law of large numbers. Therefore, it is sufficient to show that \eqref{eq-variance1} is also $o_P(1)$. From some algebra, we find the term in  \eqref{eq-variance1} is
\begin{align*}
&
\frac{1}{|\mathcal{I}_k|^{-1}}
\sum_{i \in \mathcal{I}_k}
\Big[ 
\big\{
\widehat{\zeta}_i\LSS
-
A \widehat{\psi} 
\big\}^2
-
\big\{
\zeta_i^*
-
A
\psi^*
\big\}^2
\Big]
\\
& =
\frac{1}{|\mathcal{I}_k|}
\sum_{i \in \mathcal{I}_k}
\big[
\big\{
\widehat{\zeta}_i\LSS
-
A
\widehat{\psi}
\big\}
-
\big\{
{\zeta}_i^*
-
A
\psi^*
\big\}
\big]\big[
\big\{
\widehat{\zeta}_i\LSS
-
A
\widehat{\psi} 
\big\}
+
\big\{
\zeta_i^*
-
A
\psi^*
\big\}
\big]
\\
& = 
\frac{1}{|\mathcal{I}_k|}
\sum_{i \in \mathcal{I}_k}
\big[
\big\{
\widehat{\zeta}_i\LSS
-
A
\widehat{\psi}
\big\}
-
\big\{
{\zeta}_i^*
-
A
\psi^*
\big\}
\big] \big[
\big\{
\widehat{\zeta}_i\LSS
-
A
\widehat{\psi}
\big\}
-
\big\{
{\zeta}_i^*
-
A
\psi^*
\big\}
+ 
2 
\big\{
{\zeta}_i^*
-
A
\psi^*
\big\}
\big]
\\
& = 
\frac{1}{|\mathcal{I}_k|}
\sum_{i \in \mathcal{I}_k}
\Big[
\big\{
\widehat{\zeta}_i\LSS
-
A
\widehat{\psi}
\big\}
-
\big\{
{\zeta}_i^*
-
A
\psi^*
\big\}
\Big]^2
+
\frac{2 }{|\mathcal{I}_k|}
\sum_{i \in \mathcal{I}_k}
\Big[
\big[
\big\{
\widehat{\zeta}_i\LSS
-
A
\widehat{\psi}
\big\}
-
\big\{
{\zeta}_i^*
-
A
\psi^*
\big\}
\big]
\big\{
{\zeta}_i^*
-
A
\psi^*
\big\}
\Big]			\ .
\end{align*}
Let $\widehat{\delta}_i\LSS =\big\{
\widehat{\zeta}_i\LSS
-
A
\widehat{\psi}
\big\}
-
\big\{
{\zeta}_i^*
-
A
\psi^*
\big\}$. From the H\"older's inequality, we find the absolute value of \eqref{eq-variance1} is upper bounded by
\begin{align*}
\big\| \eqref{eq-variance1} \big\|
& \precsim
\AVER\SSS \Big[  \big\{ \widehat{\delta} \LSS \big\}^2 \Big]
+
2
\AVER\SSS \Big[  \big\{ \widehat{\delta} \LSS \big\}^2 \Big]
\cdot 
\AVER\SSS \Big\{  \big( \zeta^* - A \psi^* \big)^2 \Big\} \ .
\end{align*}
Since $\AVER\SSS \big\{  \big( \zeta^* - A \psi^* \big)^2 \big\} = \Pr(A=1)^2 \sigma^2 + o_P(1) = O_P(1)$, \eqref{eq-variance1} is $o_P(1)$ if $\AVER\SSS \big[  \big\{ \widehat{\delta} \LSS \big\}^2 \big] = o_P(1)$. From some algebra, we find
\begin{align*}
\AVER\SSS \Big[  \big\{ \widehat{\delta} \LSS \big\}^2 \Big]
& \leq
\frac{2}{|\mathcal{I}_k|}
\sum_{i \in \mathcal{I}_k} 
\big\{
\widehat{\zeta}_{i}\LSS
-
\zeta_{i}^*
\big\}^2
+
2 \big( \widehat{\psi}  - \psi^* \big)^2
\\
& =
\EXP\LSS\Big[
\big\{
\widehat{\zeta}\LSS
-
\zeta^*
\big\}^2 \Big] + o_P(1)
+
2 \big( \widehat{\psi}  - \psi^* \big)^2
=
o_P(1) \ . 
\end{align*}
The first line holds from $(V_1+AV_2)^2 \leq 2V_1^2 + 2V_2^2$. The second line holds from the law of large numbers applied to $\big\{ \widehat{\zeta}\LSS
-
\zeta^*
\big\}^2$. The last line holds from \eqref{eq-variance0} and $\widehat{\psi} = \psi^* + o_P(1)$, which is from the asymptotic normality of the estimator.  This concludes the proof.

\newpage

\bibliographystyle{apa}
\bibliography{COCA.bib}

\end{document}